\documentclass{emulateapj}
\usepackage{psfig}
\usepackage{here}

\begin{document}


\lefthead{Chakrabarti et al.}
\righthead{}

\title{An Evolutionary Model For Submillimeter Galaxies}

\author{Sukanya Chakrabarti\altaffilmark{1,2}, Yeshe Fenner\altaffilmark{1}, Lars Hernquist\altaffilmark{1}, T.J. Cox\altaffilmark{1}, Philip F. Hopkins\altaffilmark{1}}

\altaffiltext{1,2}{
Harvard-Smithsonian Center for Astrophysics, 60 Garden Street, Cambridge, MA 02138 USA, schakrabarti@cfa.harvard.edu}
\altaffiltext{2} {National Science Foundation Postdoctoral Fellow}

\begin{abstract}

We calculate multi-wavelength spectral energy distributions (SEDs) (spanning optical to millimeter wavelengths) from
simulations of major galaxy mergers with black hole feedback which produce submillimeter bright galaxies (SMGs), using a self-consistent three-dimensional radiative transfer code.  These calculations allow us to predict multiwavelength correlations for this important class of galaxies.  We review star formation rates, the time evolution of the $850~\micron$ fluxes, along with the time evolution of the $M_{\rm BH}-M_{\rm star}$ relation of the SMGs formed in the mergers.   We reproduce correlations for local AGN observed in Spitzer Space Telescope's IRAC bands, and make definitive predictions for infrared X-ray correlations that should be testable by combining observations by Spitzer and the upcoming Herschel mission with X-ray surveys.  To aid observational studies, we quantify the far-infrared X-ray correlations.  Our dynamical approach allows us to directly correlate observed clustering in the data as seen in IRAC color-color plots with the relative amount of time the system spends in a region of color-color space.  We also find that this clustering is positively correlated with the stars dominating in their contribution to the total bolometric luminosity.  The merger simulations also allow us to directly correlate the $850~\micron$ flux with the ratio of the black hole luminosity to the total luminosity, which is an inherent and testable feature of our model.  We present photo albums spanning the lifetime of SMGs, from their infancy in the pre-merger phase to the final stage as an elliptical galaxy, as seen in the observed $3.6~\micron$ and $450~\micron$ band to visually illustrate some of the morphological differences between mergers of differing orbital inclination and progenitor redshift.  We compare our SEDs from the simulations to observations of SMGs and find good agreement.  We find that SMGs are a broader class of systems than starbursts or quasars.  We introduce a simple, heuristic classification scheme on the basis of the $L_{\rm IR}/L_{\rm x}$ ratios of these galaxies, which may be interpreted qualitatively as an evolutionary scheme, as these galaxies evolve in $L_{\rm IR}/L_{\rm x}$ while transiting from the pre-merger stage, through the quasar phase, to a merger remnant.

\end{abstract}



\section{Introduction}

The luminous ($L_{\rm IR} \ga 1\times 10^{12} L_{\odot}$), high-redshift, dusty galaxies which were 
discovered in large numbers by the Submillimeter Common-User Bolometer Array (SCUBA), now designated as submillimeter galaxies (SMGs), generate a significant fraction of the cosmic energy output (Smail et al. 1997; Ivison et al. 1998; Blain et al. 2002).  As such, SMGs are key cosmological players, believed to be responsible for more than half of the star formation at $z \sim 2$ (Blain et al. 2002).  

These systems have been studied now in a diverse range of wavelengths, spanning the range from X-rays to radio wavelengths.  UV/optical spectroscopic redshifts (Chapman et al. 2003b, 2005, Swinbank et al. 2004) yield a median redshift of $\sim 2$ for this population.  The presence of AGN in these systems has been confirmed in X-ray surveys (Alexander et al. 2005a,b), however the contribution of the AGN to the bolometric luminosity remains unclear.  Borys et al.'s (2005) analysis of rest-frame X-ray and near-IR data suggests that SMGs fall nearly two orders of magnitude below the local $M_{\rm star}-M_{\rm BH}$ relation, when Eddington-limited accretion is assumed to calculate the black hole masses.  HST imaging indicates that these are large irregular galaxies (Chapman et al. 2003a), with some preliminary evidence for extended starbursts (Chapman et al. 2004), assuming that the radio emission is taken to trace the starburst.  Blain et al. (2004b) and 
Chapman et al. (2004) stress the need to understand the SEDs of
these systems, specifically, in regards to ascertaining whether a
population of hot SMGs, which would be undetected in current sub-mm
surveys, would contribute significantly to the infrared emission
of $z\sim 2$ galaxies.  The most direct observational probe of the rest-frame far-infrared of high redshift sub-mm selected galaxies is the set of SHARC-2 $350~ \micron$ observations of $z\sim 2$ systems obtained by Kovacs et al. (2006).    CO observations indicate that these galaxies have large gas reservoirs ($M_{\rm gas} \sim 10^{10}-10^{11} M_{\odot}$) (Greve et al. 2004, Neri et al. 2003).  SMGs have also been studied through high resolution ($\sim 1''$) millimeter imaging and CO observations (Genzel et al. 2003, Tacconi et al. 2006) to reveal large gas masses; Tacconi (et al. 2006) infer from their observations that their sample of SMGs do not have extended starbursts and suggest that SMGs are likely to be scaled-up, more gas-rich versions of local 
Ultra Luminous Infrared Galaxies (ULIRGs), the dusty infrared-bright galaxies with $L_{8~\micron-1000~\micron} \ga 1\times 10^{12} L_{\odot}$ discovered in large numbers by IRAS (Soifer et al. 1984; 1987).  SMGs have been proposed as candidates for the progenitors of the most massive spheroids in the local universe (Lilly et al. 1999).  There is also some indication that SMGs are a clustered population (Blain et al. 2004a), with similar correlation lengths for SMGs and quasars (Croom et al. 2005). 

These recent intriguing set of observations have prompted the development of various models to fit the spectral energy distributions (SEDs) of SMGs, and to infer an evolutionary scheme for this important class of galaxies.  On the interpretative end, Farrah et al. (2002) suggest, on the basis of axially symmetric radiative transfer calculations which incorporate a model of evolving HII regions, that the 
starburst dominates in its contribution to wavelengths longer than rest-frame far-IR.  Rowan-Robinson (2000) argues that high-redshift Hyper Luminous Infrared Galaxies (HLIRGs) have star formation rates
greater than $1000~M_{\odot}/\rm yr$, and are candidates for primeval galaxies undergoing 
a burst of star formation, rather than merging systems.  Efstathiou \& Rowan-Robinson (2003) find 
that ``cirrus'' axisymmetric models, i.e., extended emission from dusty envelopes with low optical depths, 
better describe SCUBA sources, and that they may be more akin to optically selected high
redshift galaxies than obscured starbursting galaxies.  Baugh et al. (2005) have been able
to reproduce the observed galaxy number counts at $850~\micron$, assuming a top-heavy
IMF within the context of a semi-analytic prescription for galaxy evolution, coupled with the axisymmetric
dust radiative transfer code developed by Silva et al. (1998).
 
High-resolution imaging of local ULIRGs has revealed the complex morphologies
of these systems (Soifer et al. 2000, Goldader et al. 2002, Scoville et al. 2000), 
indicating a merger-driven origin.  This scenario is supported by simulations demonstrating
that tidal interactions during a major merger cause gas inflows by gravitational torques (e.g. Barnes \& Hernquist 1991, 1996), leading to nuclear starbursts (e.g., Mihos \& Hernquist 1994, 1996).  A recent analysis of spectroscopic data in conjunction with high-resolution infrared imaging (Dasyra et al. 2006) indicates that the majority of ULIRGs are formed in nearly equal mass major mergers.   
Unless high redshift ULIRGs are more quiescent than the local systems, it is
unlikely that axisymmetric models will prove to be representative
of SMGs.  The complex dynamical behavior of merging galaxies has been successfully modeled
in numerical simulations incorporating feedback from central black holes (Springel et al. 2005a,b; Di Matteo et al. 2005),
and shown to reproduce the relation between mergers and quasar populations in optical and
x-ray observations (Hopkins et al. 2006b,c).  Chakrabarti et al. (2006a) 
employed these simulations of major mergers with black hole and starburst-driven feedback (Cox et al. 2006a)
to calculate the infrared emission of local ULIRGs and LIRGs
using a self-consistent fully three-dimensional radiative transfer code.
These calculations reproduced observed trends such as the empirical 
warm-cold IRAS classification (de Grijp et al. 1985), wherein energetically active AGN are found to be correlated with high $F_{25~\micron}/F_{60~\micron}$ colors ($F_{25~\micron}/F_{60~\micron} \ga 0.2$), and demonstrated
that these trends are directly driven by feedback processes.

We build upon these recent developments here by calculating panchromatic SEDs, spanning
optical to millimeter wavelengths, of simulations of gas-rich major
mergers by using a self-consistent three-dimensional radiative 
transfer code, which treats the scattering, absorption, and reemission of 
photons from dust grains (Chakrabarti \& Whitney, 2006, in preparation).
Stellar spectra are specified by Starburst 99 calculations (Leitherer
et al. 1999, Vazquez \& Leitherer 2005). 
We conduct a multiwavelength analysis of correlations from rest-frame near-infrared to
sub-mm bands, as well as infrared X-ray correlations.  We also compare our simulated
SEDs with recent far-IR observations of SMGs by Kovacs et al. (2006).    
Our goal in this paper is to highlight the 
physical and dynamical basis for the multiwavelength correlations that 
we predict.   In particular, our multiwavelength dynamical approach allows us to cast trends in color-color
space directly in terms of color evolution as a function of time, or
of color evolution as a function of the relative luminosities from
the black hole and the stars.  While we investigate the emergent SEDs
of SMGs across a wide wavelength range in this paper by analyzing
a number of simulations (and different realizations
of these simulations by varying the IMF and dust opacity
curve, as we describe in \S 2), we defer a derivation of the infrared
luminosity function to a future paper, as this depends on a 
full cosmological model of merger rates, and is outside the
scope of this paper (Chakrabarti et al. 2006c, in preparation).

The organization of this paper is as follows.  In \S 2, we review 
the merger simulations and the translation from the Smooth
Particle Hydrodynamics (SPH) information to the spatial grid that
we use to do the radiative transfer calculations.  In \S3, we
review our radiative transfer methodology and the dust model 
we have used, along with our adopted model for PAH emission.  \S 4 presents our results, beginning in \S 4.1 which gives the star formation rates, the evolution 
of the $850~\micron$ fluxes, and the $M_{\rm BH}-M_{\rm star}$ relation
for SMGs.  In \S 4.2, we present the simulated IRAC color-color plot in the rest-frame,
 and explain the clustering in this plot.  In \S 4.3, we make a number 
of predictions infrared X-ray correlations, which can be tested empirically
by obtaining a large sample of observations in a narrow wavelength range.  
We present photo albums spanning the lifetime of SMGs during the 
active phase in \S 4.4.  In \S 5, we compare our SEDs from the simulations
to observed data, and present the IRAC color-color plot 
in the observed frame for galaxies at $z=2$, and show a 
plot depicting the variation of the $850~\micron$ fluxes as a 
function of X-ray luminosity.  We introduce a simple classification scheme
for SMGs on the basis of the $L_{\rm IR}/L_{\rm x}$ ratios which also
corresponds to an evolutionary scheme.  We conclude in \S 6. 

\section{Merger Simulations \& Specification of Interstellar Medium}

\begin{deluxetable*}{lcccc}[!h]
\tablewidth{0pt}
\tablecolumns{4}
\tablecaption{SMG Simulations}
\tablehead{\colhead{Simulation} & \colhead{Progenitor Redshift} & \colhead{Orbital Orientation}  &  \colhead{$V_{\rm vir}(\rm km/s)$}}
\startdata
e160(vc3e) &  0   &  e   & 160 \\
h160(A3)   &  0   &  h   & 160  \\
h226(A4)   &  0   &  h   & 226  \\
h320(A5)   &  0   &  h   & 320   \\
h500(A6)   &  0   &  h   & 500   \\
e226(A4e)  &  0   &  e   & 226   \\
e320(A5e)  &  0   &  e   & 320   \\
e500(A6e)  &  0   &  e   & 500   \\
z3e270(z3A4.5e) & 3 &  h   & 270    \\
z3h500(z3A6) &  3   &  h   & 500    \\
\enddata
\end{deluxetable*}

We employ a new version of the parallel TreeSPH code GADGET-2
(Springel 2005), which uses an entropy-conserving formulation of
smoothed particle hydrodynamics (Springel \& Hernquist 2002), and
includes a sub-resolution, multiphase model of the dense interstellar
medium (ISM) to describe star formation (Springel \& Hernquist 2003a,
henceforth SH03; see also Springel \& Hernquist 2003b).  
Black holes are represented by ``sink''
particles that accrete gas, with an accretion rate estimated using a
Bondi-Hoyle-Lyttleton parameterization, with an upper limit equal to
the Eddington rate (Springel et al. 2005b). The bolometric luminosity
of the black hole is then $\L_{\rm bol}=\epsilon_{\rm r}\dot{M}c^{2}$,
where $\epsilon_{\rm r}=0.1$ is the radiative efficiency.  We further
allow a small fraction ($\sim 5\%$) of $\L_{\rm bol}$ to couple dynamically
to the gas as thermal energy. This fraction is a free parameter,
determined in Di Matteo et al. (2005) by matching the $M_{\rm
BH}$-$\sigma$ relation.  We do not attempt to resolve the gas
distribution immediately around the black hole, but instead assume
that the time-averaged accretion can be estimated from the gas on the
scale of our spatial resolution, which is a few tens of parsecs in
the best cases.

We adopt the ISM prescription developed by Chakrabarti et al. (2006a)
in their study of the infrared properties of local ULIRGs.
Owing to the limited resolution of current numerical simulations, the
interstellar medium (ISM) must be treated in a volume-averaged manner.
In particular, the simulations discussed here adopt the multiphase
model of SH03, in which individual SPH particles represent a region of
the ISM that contains cold clouds embedded in a diffuse hot medium.
Because the SPH calculation uses only the volume-averaged pressure,
temperature, and density to evolve the hydrodynamics, this model does
not provide specific information regarding the cold clumps. Since
these cold clumps harbor the dust that produces the infrared emission,
we must adopt a model which determines their number, density, 
size, and location.  

As the first step to specify the properties of the cold, dense gas, we
assume that each SPH particle contains one dense molecular cloud at its
center.  Because we know the temperature and density of the volume-average
SPH particle, assuming a temperature (say, $10^{3} \rm ~K$, as was assumed in
SH03) for this cold cloud and that it exists in
pressure equilibrium with the hot phase we readily attain its density.
However, observations of molecular clouds indicate that their pressure is
dominated by turbulent motions (Blitz et al. 2006; Solomon et al. 1997), rather than thermal energy, and thus
neglecting this feature yields very dense clouds, with small volume
filling factors.  In the work presented here, we take the turbulent
pressure to be $\sim 100$ times that of the thermal pressure; the turbulent
velocity as estimated from line widths is of order 10 times the thermal
sound speed in star forming regions, which would lead to a factor of $\sim 100$ 
for the ratio of the turbulent to thermal pressure (Plume et al. 1997).  With this
assumption, the density is much lower than without the turbulent support
and its volume filling factor is greatly increased.  (Our approach is similar to that employed
by Narayanan et al. [2006] in their study of the evolution of
the molecular gas in mergers.)  

We have considered a diverse
range of simulations in this paper, of varying mass, orbital inclination,
and progenitor redshift.
We give a summary of the simulations
analyzed in this paper in Table 1, including the progenitor redshift of the
two galaxies, the orbital orientation, which is denoted ``h'' for a co-planar
merger, and ``e'' for an arbitrary inclination, and the virial velocity.  The names that we use to refer to the simulations along with names used in previous papers (i.e., Hopkins et al. 2005b) given in parentheses.  A detailed description of these models is given in Cox et al. (2006a).  The orbital orientation denoted ``e'' corresponds to $\theta_{\rm 1}=30,\phi_{\rm 1}=60$ for the first disk, and $\theta_{\rm 2}=-30, \phi_{\rm 2}=45$ for the second disk (also in Table 1 of Cox et al. 2006a).   We construct the disk progenitors at redshift $z$ in a manner similar to Robertson et al. (2006) which follows the Mo, Mao \& White (1998) formalism in determining the scale lengths of the disks, and the disk concentrations are scaled following Bullock et al. (2001).  

\section{Radiative Transfer Methodology}

We use a self-consistent three-dimensional Monte Carlo radiative equilibrium code (Chakrabarti \& Whitney, 2006, in prep) to calculate the emergent SEDs and images from the merger simulations as a function of evolutionary state.  We review some of the main points here.  This code incorporates the Monte Carlo radiative equilibrium routine developed by Bjorkman \& Wood (2001) to solve for the equilibrium temperature of the dust grains, in a manner similar to that implemented by Whitney et al. (2003).  Individual photons are tracked directly in the Monte Carlo scheme.  When a photon is absorbed by a grid cell, it may raise the temperature of that grid cell so that it emits infrared radiation; the emissivity of this grid cell depends on the temperature.  The Bjorkman \& Wood (2001) algorithm corrects the temperature in a grid cell by sampling the new photon frequency from the difference of two emissivity functions - that from the previous temperature and that computed with the new temperature.  This is not an explicitly iterative scheme; a sufficiently large number of photons must be absorbed by the grid cells so that they eventually relax to an equilibrium temperature.  Since the density within a grid cell is constant, one can easily integrate the optical depth through the grid to account for the attenuation of the photons.  This code also treats scattering, representing the scattering phase function as a Henyey-Greenstein phase function (see Whitney et al. 2003; Whitney \& Wolff 2002 for a detailed description of this).  Since the scattering efficiency is proportional to $1/\lambda^{4}$, it is particularly relevant for the shorter wavelengths.  Whitney et al. (2003) carried out convergence studies to compare their results to a set of benchmark calculations developed for spherically symmetric codes.  Chakrabarti \& Whitney (2006, in prep) have carried out both photon and grid resolution convergence studies to verify that photon numbers of order $10^{8}$ yield converged results from mm to near-IR wavelengths for the three-dimensional grids we have used here, which follow the clump size distribution (which result from our ISM prescription discussed in \S 2) and will at a minimum, resolve (i.e., by $\sim 2$ cells) the median of the clump size distribution; the grid resolution was varied by an order of magnitude to find converged results.  For fine grid structures, it is necessary to increase the photon number proportionately to ensure that each grid cell receives enough photons to reach an equilibrium temperature (i.e., the photon and grid resolution studies are not independent).  Off-nuclear sources of radiation are allocated photons in proportion to the fraction of bolometric luminosity they contribute.  The radiative equilibrium temperature calculation needs to be referenced to the total bolometric luminosity of all of the sources of radiation.  For large three-dimensional grids, it is crucial to allocate tens of millions of photons for the grid cells to reach an equilibrium temperature; otherwise, the emissivity of the dust may be artificially lowered.  This code incorporates a ``peeling-off'' algorithm (Yusef-Zadeh, Morris \& White 1984; Wood \& Reynolds 1999) to calculate the images, which are then convolved with broadband filter functions.

We model the intrinsic AGN continuum spectrum following Hopkins, Richards
\& Hernquist (2006) (HRH), which is based on optical through hard X-ray observations
(Elvis et al. 1994, George et al. 1998, Perola et al. 2002, Telfer et
al. 2002, Ueda et al. 2003, Vignali et al. 2003), with a reflection
component generated by the PEXRAV model (Magdziarz \& Zdziarski 1995).
The HRH spectrum is similar to that developed by Marconi et al. (2004), 
but it is more representative of the shapes of typical observed quasars;
it includes a template for the observed ``hot dust'' component
 longward of $1~\micron$.  
Since we calculate the dust emission self-consistently, we enforce
a Rayleigh-Jeans cutoff beyond $1 ~\micron$ in the input AGN spectrum.  In our
models here, the presence of dust emission at wavelengths longer 
$1~\micron$ is derived from our radiative transfer calculations.
We have performed Starburst 99 calculations (Leitherer et al. 1999,
Vazquez \&  Leitherer 2005) to calculate the stellar 
spectrum and bolometric stellar luminosity, taking as input the age, mass, and metallicity of the stars
from the simulations, for both the Kroupa and Salpeter IMFs.  

We use the Weingartner \& Draine (2001) (henceforth WD01) $R_{V}=5.5$
dust opacity model, both with and without the addition of ice mantles.  
The inclusion of ice mantles (which lead to an effective
increase in the opacity of $\sim~2$ for $T\la 100~\rm K$ above which
ices sublimate, e.g., Pollack et al. 1994) is motivated by
spectroscopic studies that have identified ice spectral features
in protostellar environments and in dusty galaxies (Allamandola
et al. 1992; Spoon et al. 2002). 
The mass fraction of dust is equal to $1/105.1$
for solar abundances (WD01).  WD01's models have been shown to
reproduce the observed extinction curves for the Milky Way, as well as
regions of low metallicity, such as the LMC and the SMC.  The opacity
normalization per gram of dust, $\kappa_{\lambda_{0}}$,is equal to
$0.27 \delta$ for $\lambda_{0}=100~\micron$, where $\delta$ is the
dust-to-gas ratio relative to solar, which we take to be unity.  Dunne
\& Eales (2001) and Klaas et al. (2001) found that the dust-to-gas
ratio for a large sample of ULIRGs is comparable to Milky Way values,
when they fit two-temperature blackbodies to the far-IR SEDs.
Wilson et al. (in preparation) also find similar results
when they use multi-temperature blackbodies to fit the
SEDs.  Previous work, based on fitting single temperature blackbodies had
found lower dust-to-gas ratios (Dunne et al. 2000) by a
factor of two.  Further, this dust model has been used successfully
in fitting the observed SEDs of ULIRGs and SMGs by Chakrabarti \& McKee (2005) 
and Chakrabarti \& McKee (2006, in preparation).  
We adopt a phenomenological model of PAH emission here.  
If there is a UV photon of sufficient magnitude
to excite PAH emission, then we add on an observed PAH template
(adopted from Dopita et al. 2005) to the SED, where the PAH strength is scaled to the $200~\micron$ flux as in Dale \& Helou (2002).  

\begin{figure*}[!ht] \begin{center}
\centerline{\psfig{file=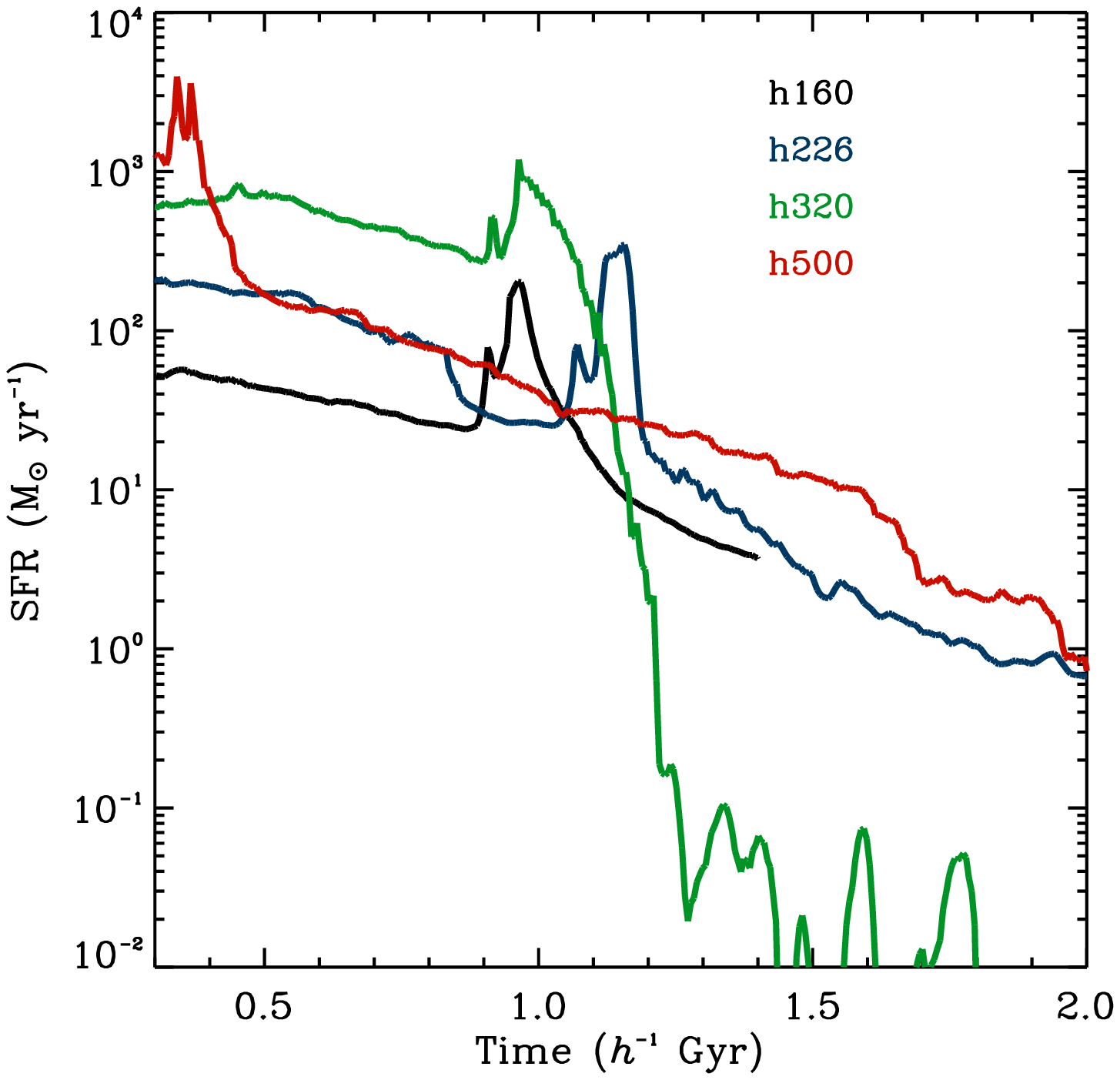,width=2.2in}
\psfig{file=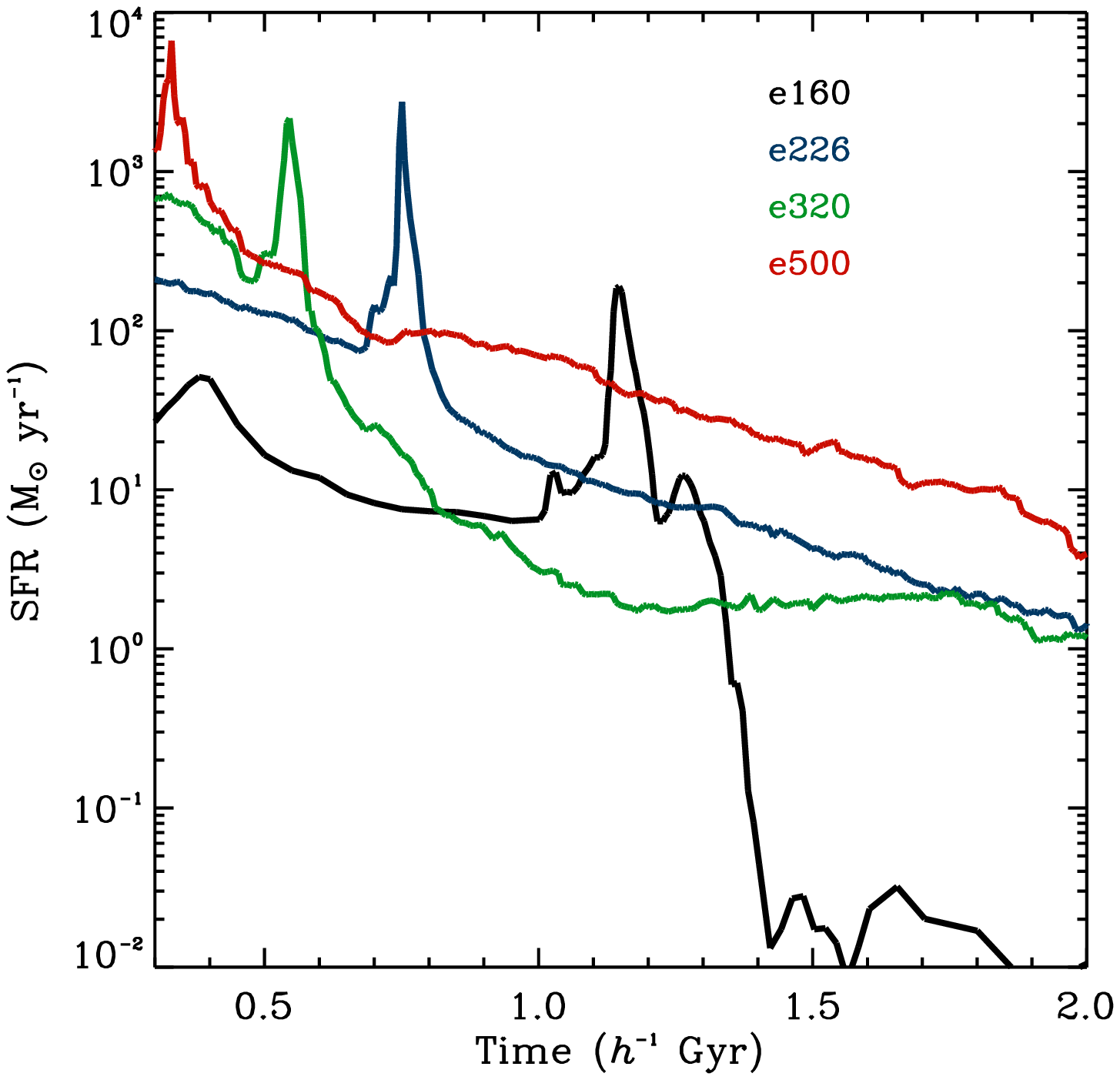,width=2.2in}
\psfig{file=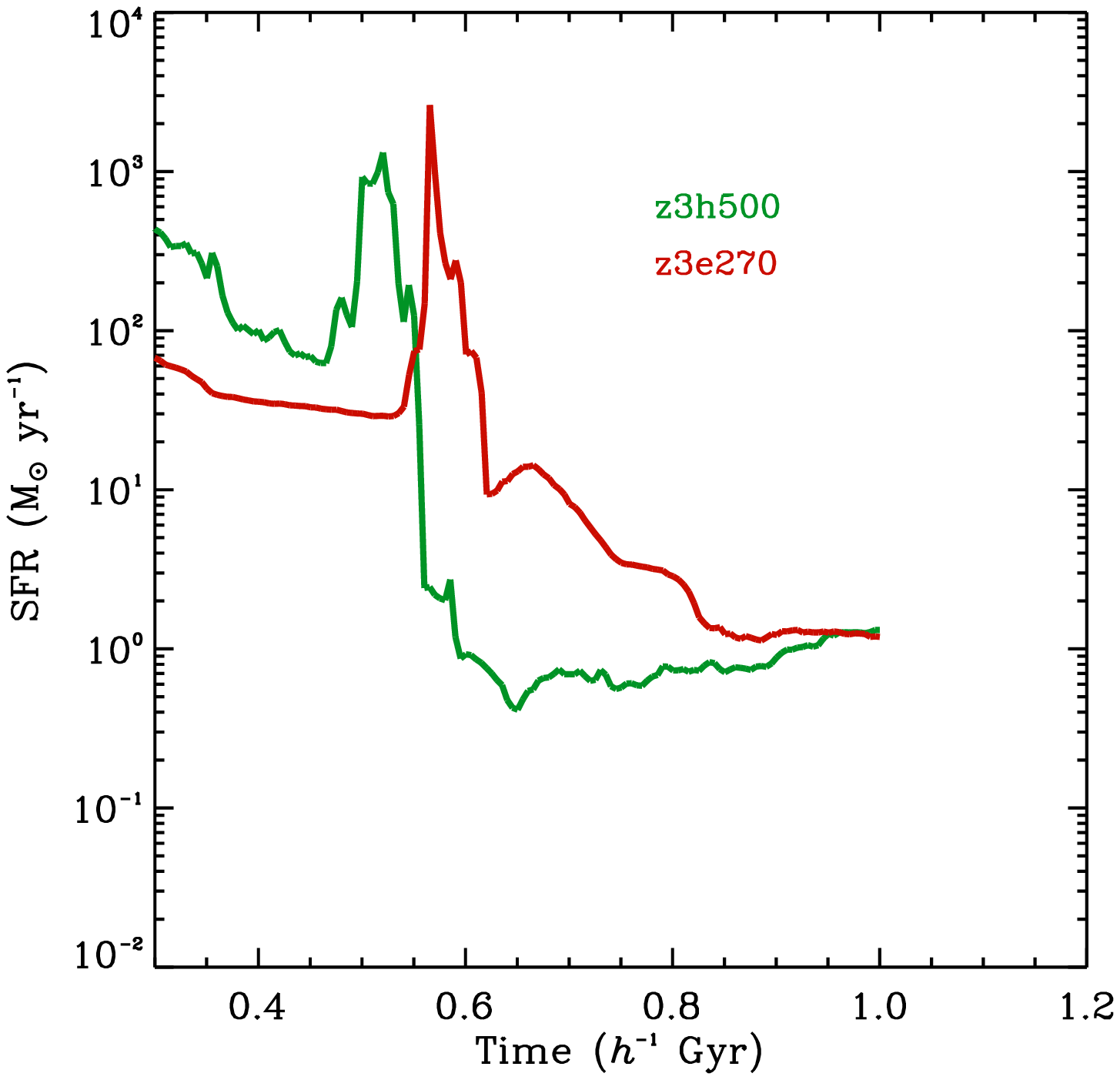,width=2.2in}}
\end{center}
\caption{Star Formation Rate (in $M_{\odot}/\rm yr$) vs. time for the merger simulations analyzed in this paper.  Left panel shows the $z=0$ co-planar mergers of varying virial velocity; the number refers to the virial velocity of the simulations in $\rm km/s$; middle panel shows the $z=0$ non co-planar mergers; right panel shows the scaled $z=3$ mergers (``e'' denotes arbitrary inclination and ``h'' is co-planar)}\label{fig:1}
\end{figure*}

\section{Results \& Analysis}

We plot in Figure 1 the star formation rates for the simulations analyzed
in this paper.  SH03 implement an algorithm for star formation that is 
similar to the Kennicutt (1998) relation, i.e., $\Sigma_{\rm SFR} \propto \Sigma_{\rm gas}^{1.4}$.
For progenitors at a fixed redshift and gas fraction, the star formation rate
is proportional to (essentially a linear power of) the virial velocity, since $M_{\rm vir}=V_{\rm vir}^3 / (10\,G\,H(z))$
and $R_{\rm vir}=V_{\rm vir}/(10\,G\,H(z))$ which gives $\Sigma_{\rm SFR} \propto V_{\rm vir}^{1.4}$ for
progenitors at the same redshift, with fixed gas fraction.   
The sequence of plots in Figure 1 (a) and (b) shows that the peak star formation rate scales
with virial velocity for progenitors at a given redshift. 
Disk progenitors at redshift $z$ are constructed in a manner similar to that described 
by Robertson et al. (2006); the size of the disks are set by the Mo, Mao \& White (1998) formalism wherein the disk contains a fraction of the total galaxy angular momentum equal to its mass fraction.  
The disk scale length is then determined by the galaxy spin (which is set to a constant that is motivated
by results from cosmological N-body simulations) and the dark matter halo concentration, $C_{\rm vir}$, which scales essentially as $1/(1+z)$ (with a weak dependence on the virial mass, e.g., equation 3 in Robertson et al. 2006).  This dependence of the concentration on the redshift offsets the $1/H(z)$ dependence of $R_{\rm vir}$ 
(i.e., when substituting this dependence of the concentration in eq. 28 of Mo et al. 1998) so that the disk scale radius decreases by about 20 \% to $z=3$ while there is a factor of $\sim 5$ loss in mass.  Therefore, the star formation will depend primarily on the mass of the disk (and will decrease proportionately) for progenitors at redshift $z$.  The scaled $z=3$ simulations, for a given 
virial velocity (e.g. z3h500 ($z=3$) in Figure 1c compared to h500 in Figure 1a) have lower
star formation rates, as they are less massive.  We also discuss later in \S 4.4 that the simulations with
progenitors at $z=3$ have more compact morphologies than the
simulations with disk properties of $z=0$ simulations.  Inferred star formation rates
of SMGs vary from the extremely high star formation rates of cirrus models,
$\rm SFR \ga 1000 M_{\odot}/\rm yr$ (Efstathiou \& Rowan-Robinson 2003; Rowan-Robinson 2000), 
to about $500 M_{\odot}/\rm yr$ if a factor of two is used to account for
the AGN contribution to the infrared luminosity, along with Kennicutt's (1998)
relation for the star formation rate and the infrared luminosity (Genzel et al. 2003);
inferring star formation rates without including the contribution of the AGN leads
to star formation of the order of several thousand solar mass per year (Ivison et al.
1998).  In our simulations, during the active phase, the star formation rates are on
average $\sim 500-1000 M_{\odot}/\rm yr$ and produce bright SMGs as we discuss below.

\begin{figure}[!b] \begin{center}
\centerline{\psfig{file=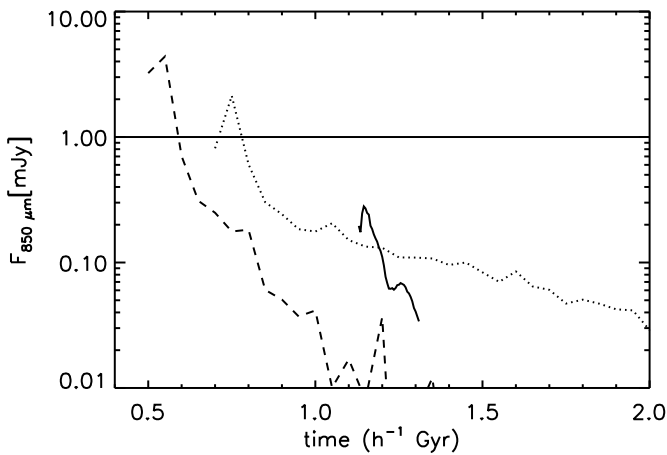,width=3.7in}}
\end{center}
\caption{$F_{850~\micron}$ (in observed frame for galaxies at $z=2$, $ D_{\rm L} \approx 15.5$ Gpc) vs. time for the SMGs of varying virial velocity (which corresponds to varying galaxy mass for galaxies with progenitors at the same redshift) in our sample; solid line is the e160 simulation, dotted line is e226, dashed line is e320.  Solid horizontal line marks $F_{850~\micron} > 1~\rm mJy$, the empirical definition of SMGs.  These all have the same orbital inclination.}\label{fig:2}
\end{figure}

Henceforth, we will illustrate specific features of the
merger simulations for representative samples of simulations.  
We show in Figure 2 the observed $850~\micron$ flux as a function of time
for e160, e226, and e320, which span the range from virial velocities of
$160~\rm km/s$ to $320~\rm km/s$.  The $850~\micron$ flux is shown
for galaxies at a luminosity distance of 
$ D_{\rm L} \approx 15.5$ Gpc (this corresponds to
the rest-frame $283~\micron$ flux at $z=2$).  SMGs are an empirically 
defined class of galaxies - based on the observed $850~\micron$ flux; 
systems with observed $F_{850~\micron} \ga 1~\rm~mJy$ are 
designated as sub-mm bright (Ivison et al. 2000; Blain et al. 1998).  The solid
horizontal line demarcates this 
criterion ($F_{850~\micron} \ga 1~\rm~mJy$) on our flux versus time plots.
(This empirical definition of sub-mm bright may well change owing to improved
sensitivity of future sub-mm surveys, i.e., SCUBA-2) As is clear, the peak $850~\micron$ flux
increases as a function of virial velocity (which corresponds to galaxy
mass for galaxies with progenitors at the same redshift).  The more massive
galaxies show a more rapid and pronounced decline in their sub-mm flux as
a function of time as these systems are losing gas content
both owing to AGN feedback clearing out the dust and gas in a more
violent fashion (Hopkins et al. 2006d, Hopkins \& Hernquist 2006),
as well as to forming stars at a faster rate.

\begin{figure}[!b] \begin{center}
\centerline{\psfig{file=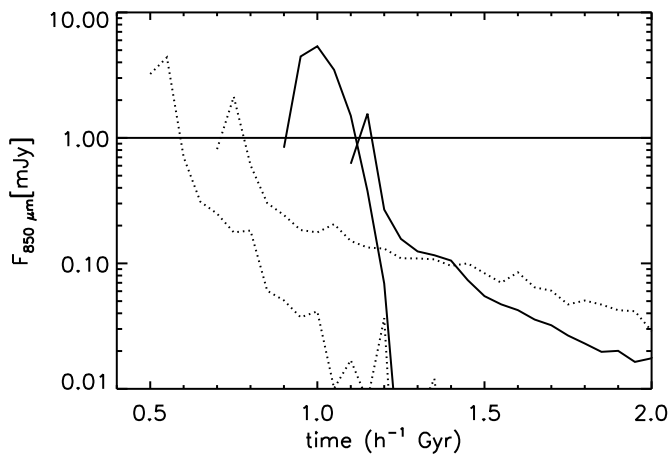,width=3.7in}}
\end{center}
\caption{$F_{850~\micron}$ (in observed frame for galaxies at $z=2$, $D_{\rm L} \approx 15.5$ Gpc) vs. time for the SMGs of varying orbital inclination for h226, e226, h320, and e320. 
The solid line shows the co-planar cases: h226 and h320, and the dotted line the non co-planar cases e226 and e320.}\label{fig:3}
\end{figure}

\begin{figure} \begin{center}
\centerline{\psfig{file=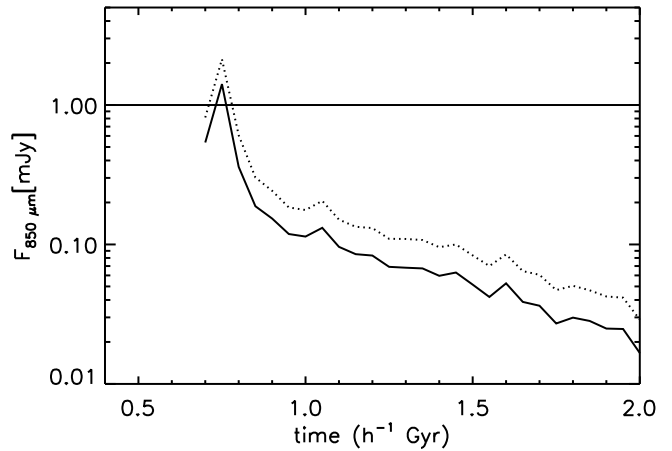,width=3.7in}}
\end{center}
\caption{$F_{850~\micron}$ (in observed frame for galaxies at $z=2$, $ D_{\rm L} \approx 15.5$ Gpc) vs. time for the SMGs of varying dust opacity shown here for e226, with and without the inclusion of ice mantles.  Solid line has the WD01 opacity, while the dotted line includes ice mantles.}\label{fig:4}
\end{figure}

Figure 3 shows the observed $850~\micron$ flux as a function of time
for galaxies with varying orbital inclination for e226, h226, e320, and h320.
While there is no clear difference between the peak fluxes of the
galaxies shown here of varying orbital inclination, the co-planar
mergers (shown here as the solid lines) do have a sharper decline in
their $850~\micron$ flux as a function of time.  Finally, Figure 4
contrasts the time evolution of the $850~\micron$ flux when the WD01 
opacity (solid line) is used instead of the WD01 model with inclusion of ice mantles (dotted line).  
  Since the observed $850~\micron$ flux is nearly optically thin, the 
effect of adding ice mantles corresponds to increasing the $850~\micron$ flux.
Figure 5 shows the e226 and e320 simulations with the Salpeter (solid line) and Kroupa (dashed line) IMFs.  Using the Kroupa IMF leads to a slight increase in the observed $850~\micron$ flux (by about $\sim 30\%$).  Contrasting Figure 1 with Figures 2-5 demonstrates that the observed $850~\micron$ flux does 
track the star formation rate quite closely.  We return to this point in a forthcoming paper where we analyze a larger set of simulations and give fitting relations between the $850~\micron$ flux and relevant parameters, such as the star formation rate.

\begin{figure} \begin{center}
\centerline{\psfig{file=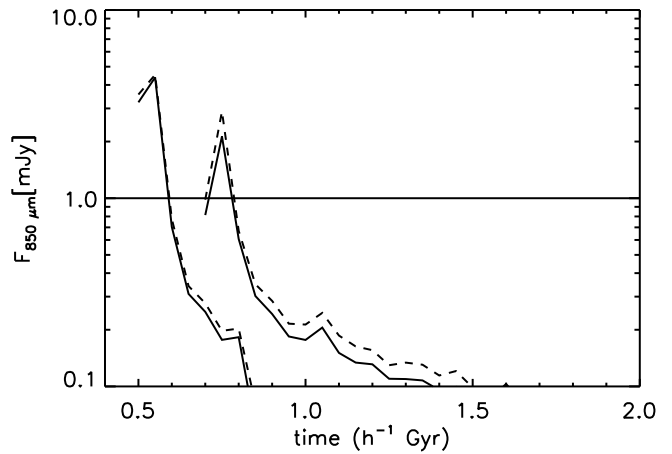,width=3.7in}}
\end{center}
\caption{$F_{850~\micron}$ (in observed frame for galaxies at $z=2$, $ D_{\rm L} \approx 15.5$ Gpc) vs. time for for e226 and e320, shown here for Salpeter (solid line) and Kroupa (dashed line) IMFs.}\label{fig:5}
\end{figure}

\begin{figure} \begin{center}
\centerline{\psfig{file=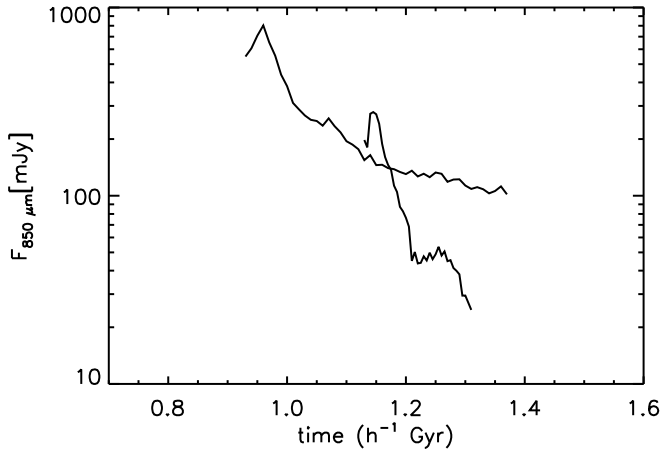,width=3.7in}}
\end{center}
\caption{$F_{850~\micron}$ vs. time for galaxies designed to be representative of the local population (e160 and h160), placed at $D_{\rm L}=77~\rm Mpc$.}\label{fig:6}
\end{figure}

Figure 6 shows the time evolution of the observed $850~\micron$ fluxes for the e160 and h160 simulations - which are designed to have virial velocities representative of local ULIRGs (the e160 simulation has 40\% gas fraction initially, while the h160 simulation has 100 \% at the start of the simulation).  These simulations have sub-mm fluxes similar to those of local LIRGs and ULIRGs.  Averaging the $850~\micron$ fluxes in Table 3 of Dunne \& Eales (2001) gives $183~\rm mJy$, with Arp 220 topping the list at $832~\rm mJy$.   Such objects at $D_{\rm L} \sim 77~\rm Mpc$ would be bright in the sub-mm for a longer fraction of their lifetime than at $z=2$, as has been shown previously in this section.    We discuss in Chakrabarti et al. (2006c, in preparation) that the time variation of the $850~\micron$ fluxes leads to a simple interpretation for the observed $850~\micron$ number counts, as well as the clustering properties of SMGs, similar to the characterization of the quasar luminosity function by Hopkins et al. (2005a-c, 2006a), and the weak luminosity dependence of clustering properties of quasars, as discussed by Lidz et al. (2006).  Although we do not derive the sub-mm luminosity functions in this paper, it is clear that simulations of gas-rich major mergers naturally lead to the production of SMGs.

\begin{figure}[!b] \begin{center}
\centerline{\psfig{file=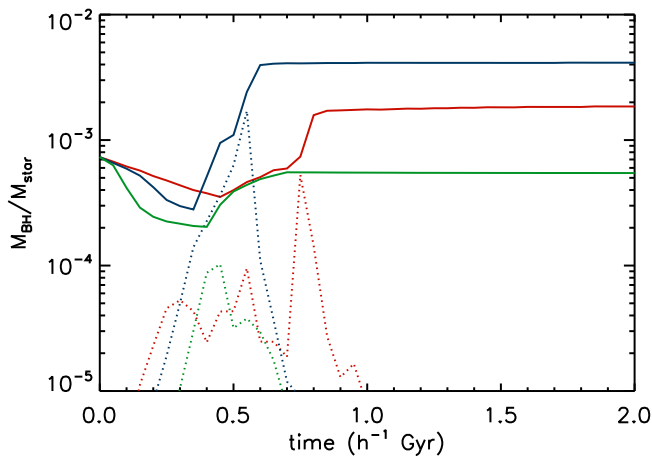,width=3.7in}}
\end{center}
\caption{$M_{\rm BH}/M_{\rm star}$ versus time for e226 (red line), e320 (shown in blue),and e500 (shown in
green).  The dotted lines correspond to deriving the black hole mass for Eddington limited accretion.This shows that deriving black hole masses under the assumption of Eddington-limited accretion underestimates the black hole masses by a factor of $\sim 5$ during the SMG phase ($t\sim 0.5 \rm h^{-1}~Gyr$).}\label{fig:7}
\end{figure}

We show in Figure 7 the evolution of $M_{\rm BH}/M_{\rm star}$ as a function of time
for e226 (red line), e320 (blue line) and e500 (green line).  At late times, we see that
these simulations tend to the observed $M_{\rm star}-M_{\rm BH}$ relation for 
spheroids, i.e., the mass of the black hole is $\sim 0.1 \%$ of the spheroid 
mass (Magorrian et al. 1998).  Borys et al. (2005) discuss the relationship
between the stellar and black hole mass for SMGs at a median redshift of 2.  Their 
X-ray observations indicate that about half of their sample of 13 SMGs host AGN.
They derive the relation between the black hole mass, from the observed X-ray luminosity under the assumption 
of Eddington-limited accretion, and stellar mass, from the rest-frame
K band observations, to find that the SMGs fall
about two orders of magnitude below the local relation.  The SMG phase 
for the simulations shown in Figure 7 occurs around $t \sim 0.5 \rm~ h^{-1} Gyr$ - such systems would
indeed lie below the local relation, but not by two orders of magnitude.
We show the change in $M_{\rm BH}/M_{\rm star}$ when we obtain the black hole
masses assuming Eddington-limited growth (the dotted lines in this figure)
- which would then lead to the appearance of such objects falling about
two orders of magnitude below the local relation.  While we find that 
the black hole masses would grow perhaps by a factor of $\sim 5$ between 
the SMG phase at $z=2$ and the present-day, it is possible that Borys et al.'s (2005)
assumption of Eddington-limited growth has led to an overestimate of this
factor by more than an order of magnitude.  

\subsection{IRAC Correlations}

\begin{figure}[!b] \begin{center}
\centerline{\psfig{file=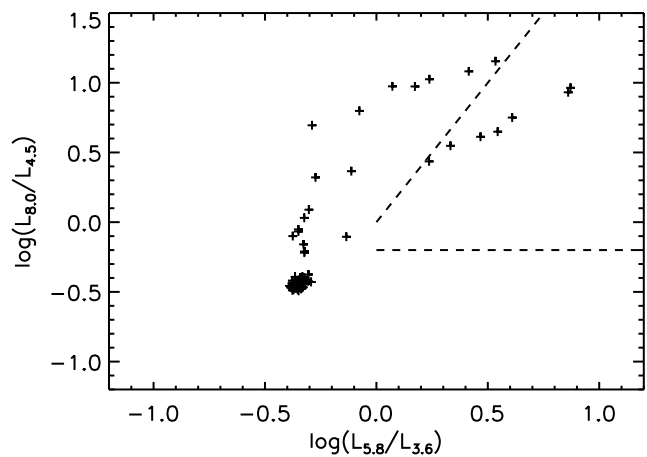,width=3.7in}}
\end{center}
\caption{Correlations in rest-frame IRAC bands, shown for all the simulations analyzed in this paper.  Dashed lines demarcate color criteria to pick out obscured AGN sample by Lacy et al. (2004).}\label{fig:8}
\end{figure}

We plot in Figure 8 an IRAC color-color plot in the rest-frame.  This IRAC color-color plot was proposed by Lacy et al. (2004) to identify AGN; the AGN are the sources that fall within the dashed lines.  We recover the general trends of the observed IRAC color-color plot from Spitzer Space Telescope's First Look Survey (FLS) - namely, the clustering of the fluxes in the lower left hand corner of the plot (both colors being bluer), as well as a small fraction of these sources falling on the redder sequence.  We show a much smaller sample here than that shown by Lacy et al. (2004), which depicts tens of thousands of sources, while we consider ten simulations, at about 20 different snapshots during the active phase on average.  Specifically, of the 16,000 objects shown in the 
Lacy plot, 2000 are likely to contain AGN, while 20 were robustly identified to be
obscured AGN; about eight out of about of about two hundred points fall in the dashed region from our
simulated data.  (For each time snapshot in the simulation, the median flux over viewing angles constitutes a single data point).

\begin{figure*} \begin{center}
\centerline{\psfig{file=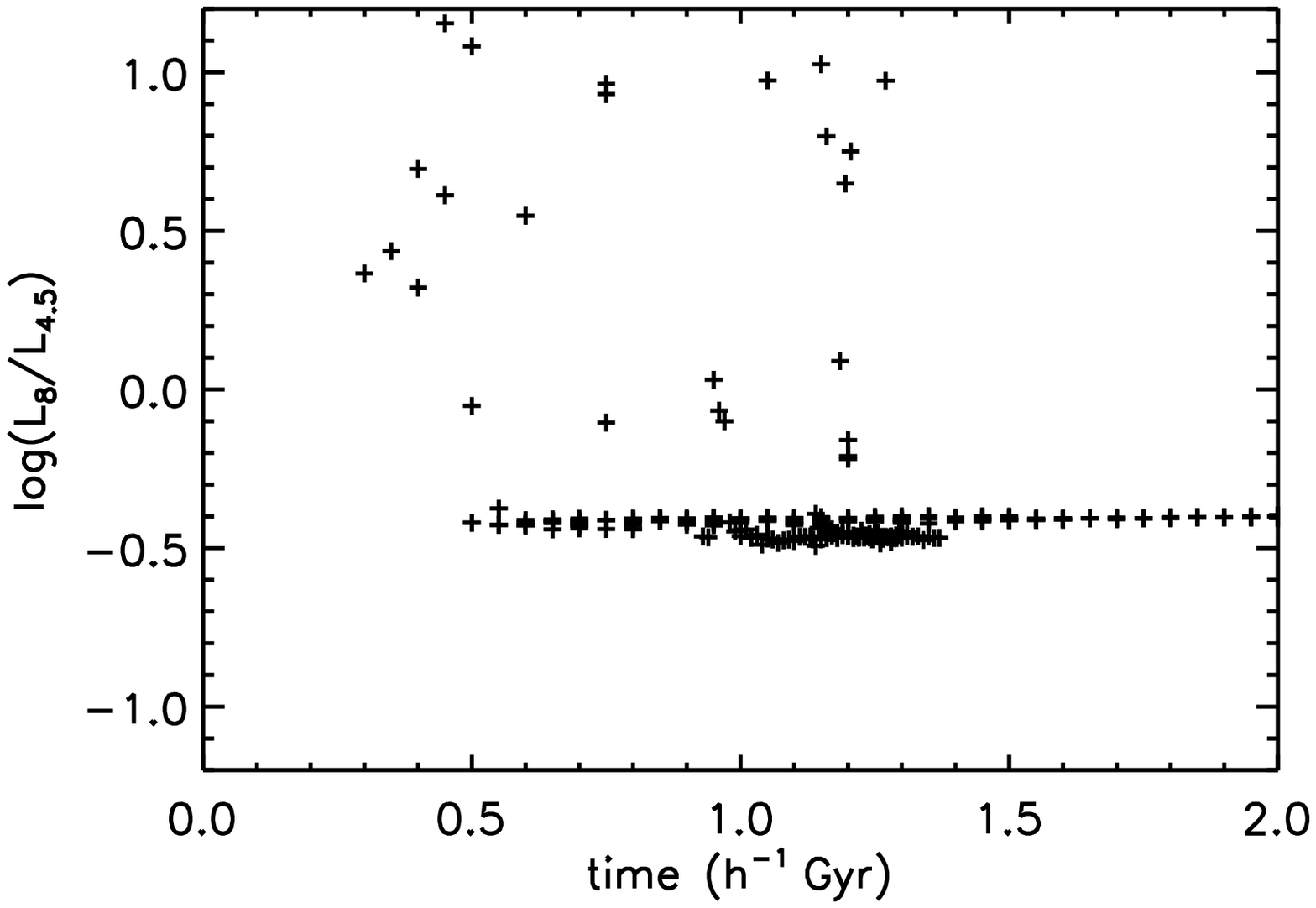,width=3.in}
\psfig{file=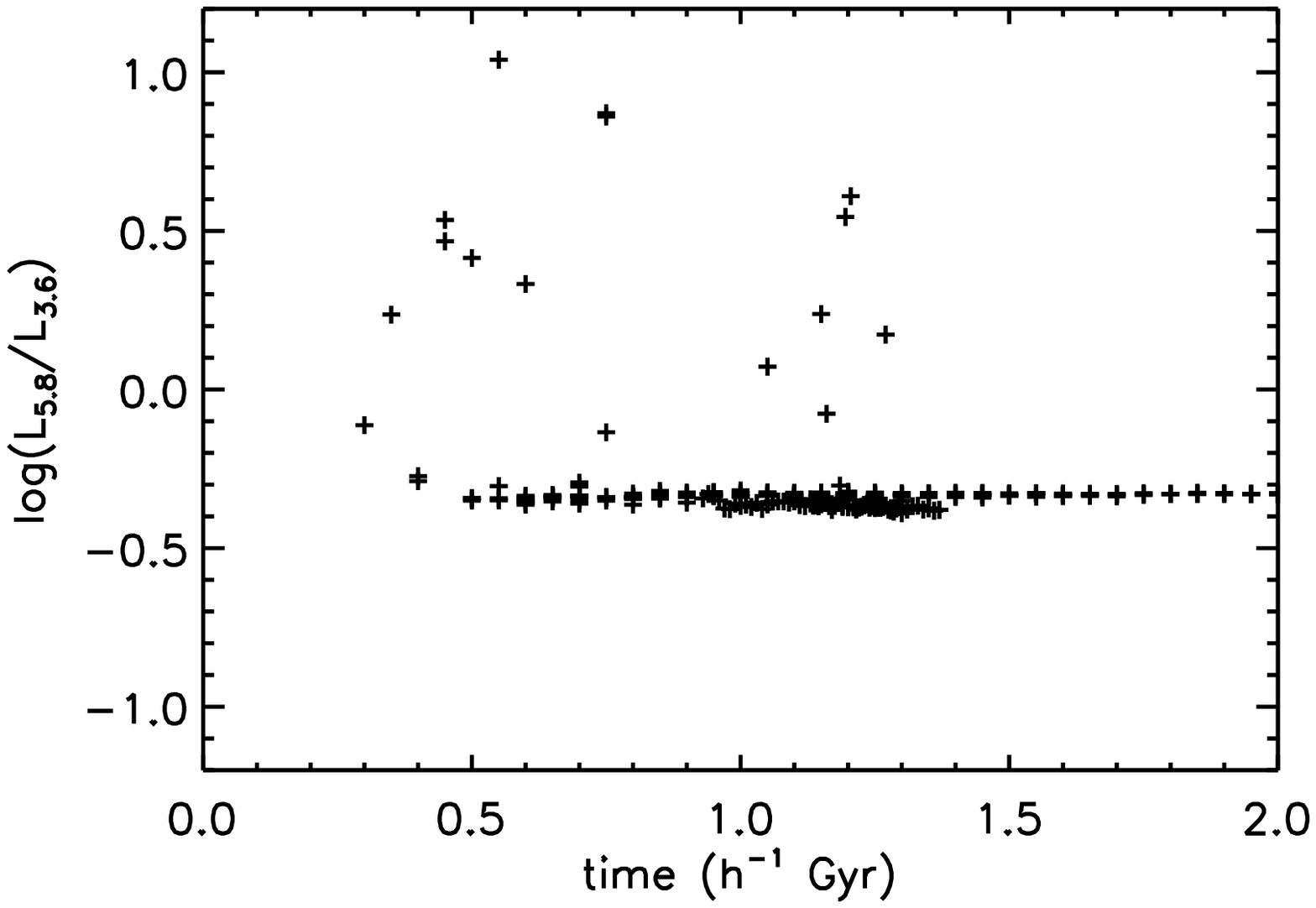,width=3.in}}
\end{center}
\caption{Rest-frame IRAC Color-Color plots as a function of time. The clustering in the lower left hand corner of the IRAC color-color plot owes to the simulated galaxies spending most of their time in that phase.}\label{fig:9}
\end{figure*}

The power of the dynamical approach afforded by calculating fluxes from the simulations is that we can unfold this color-color plot to a color vs. time plot for each axis, as shown in the following Figure 9.  As this plot shows, the clustering that is present in the observed color-color plot, which we reproduce also in the simulation IRAC color-color plot, is naturally explained in our models by the SMG spending more of its lifetime in that region of color space. 
 The low $F_{8~\micron}/F_{4.5~\micron}$ and $F_{5.8~\micron}/F_{3.6~\micron}$ colors correspond to the Rayleigh-Jeans tail of attenuated starlight (modulated by the albedo) influencing the IRAC colors.  (It is a simple exercise to show that the Rayleigh-Jeans tail of stellar spectra leads to $\rm log(F_{8~\micron}/F_{4.5~\micron})$ and $\rm log(F_{5.8~\micron}/F_{3.6~\micron})$ roughly of order -0.5, independent of the details of the stellar spectrum.)  Figure 10 plots the IRAC colors as a function of $L_{\rm BH}/L_{\rm stars}$ - this shows that if the IRAC colors fall within the dashed region demarcated by Lacy et al. (2004), the source is certainly an energetically active AGN; i.e., the bolometric luminosity from the black hole is greater than or comparable to the luminosity from the stars.  However, there are also cases, when the AGN's luminosity is greater than or comparable to the stars, and yet it is not in the AGN-demarcated region of color-color space.  For instance, most of the points which have blue colors in $F_{5.8~\micron}/F_{3.6~\micron}$ and red colors in $F_{8~\micron}/F_{4.5~\micron}$ do have $L_{\rm BH} \ga L_{\rm stars}$, as well as a few of the points that have blue $F_{5.8~\micron}/F_{3.6~\micron}$ and blue $F_{8~\micron}/F_{4.5~\micron}$.  (For the most part, though, sources that have blue colors for both axes are dominated by the stellar luminosity.)  This variation seems to owe to the effects of larger and more extended starbursts in some of the co-planar mergers.  Therefore, while we confirm Lacy's et al. (2004) IRAC correlations in general, our interpretation of the blue-red plume, as well as sources that lie close to the demarcating line for AGN, is different from their analysis.  We suggest that some of these sources may also have energetically active AGN, which could be detected in deep X-ray surveys.  

\begin{figure*} \begin{center}
\centerline{\psfig{file=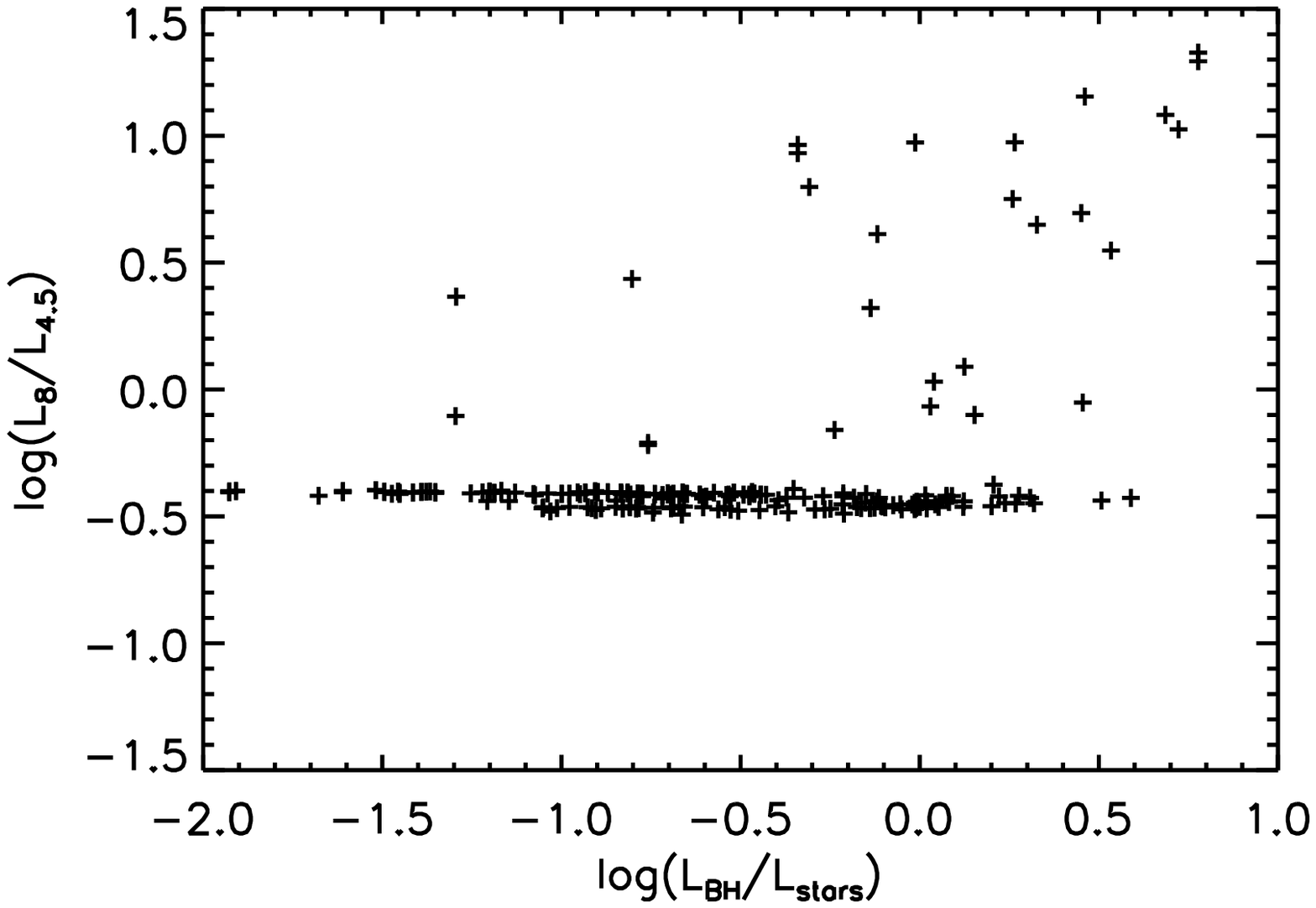,width=3.in}
\psfig{file=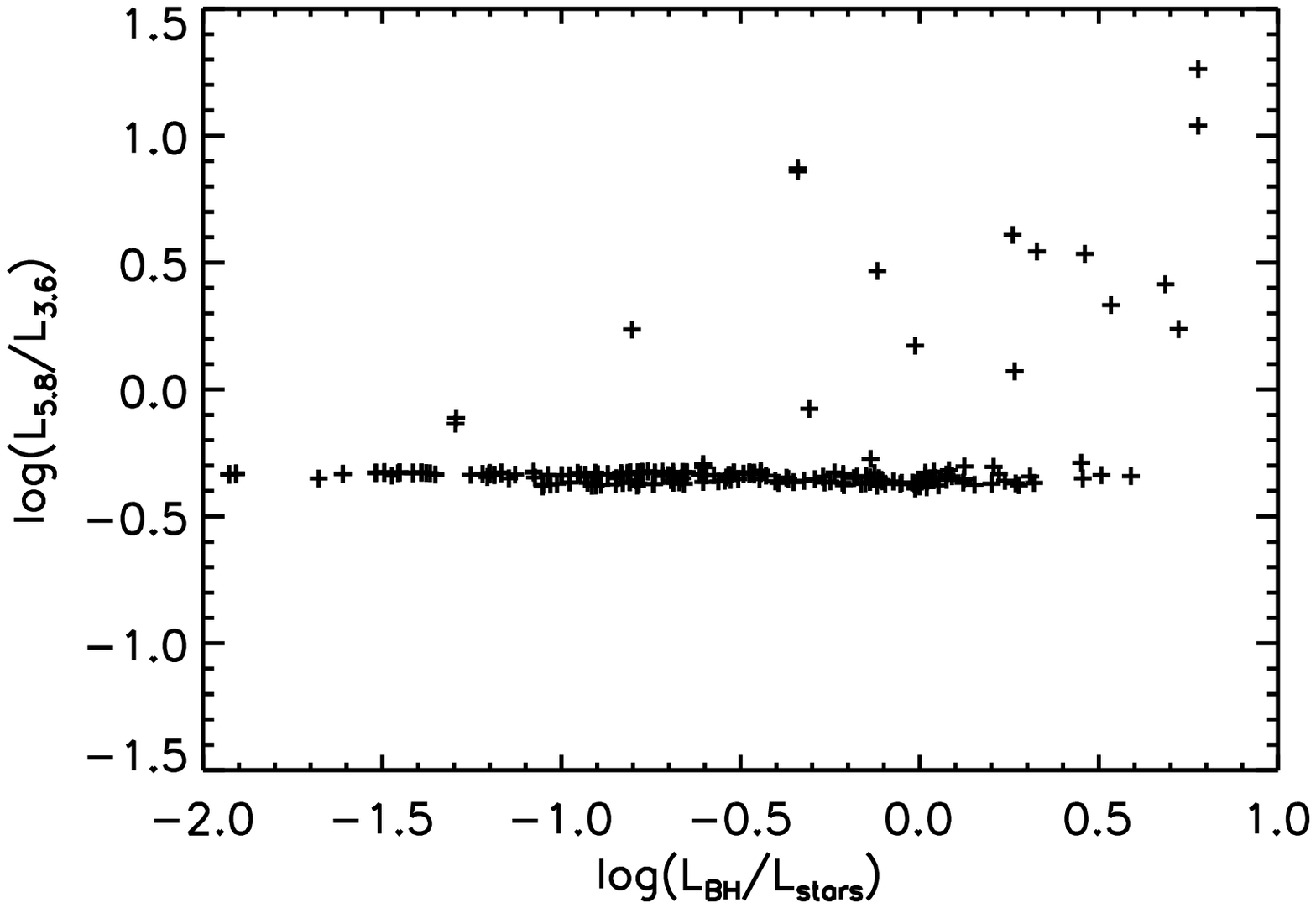,width=3.in}}
\end{center}
\caption{Rest-frame IRAC Color-Color plots as a function of $L_{\rm BH}/L_{\rm stars}$. The clustering in the lower left hand corner of the IRAC color-color plot is positively correlated with the stars dominating in their contribution to the total bolometric luminosity.  The redder colors are positively correlated with the black hole being energetically active, i.e., $L_{\rm BH} \ga L_{\rm stars}$.  There are however cases when the colors are blue and $L_{\rm BH} \ga L_{\rm stars}$ (see discussion in \S 4.1).}\label{fig:10}
\end{figure*}

We emphasize that this interpretation of the IRAC color-color plot is most appropriate when viewed in the rest-frame, as the clustering properties and relative fraction in the AGN demarcated region can then be directly correlated with $\it{dynamical}$ properties of these galaxies, such as the relative amount of time the system spends in some region of color-color space and the relative contribution from the black hole luminosity.   One can interpret the FLS IRAC color-color plot roughly in the manner outlined above as well since Lacy et al. (2004) note that the median photometric redshift of their candidate AGN is $\sim 0.3$.  We discuss later in \S 4.5 that the continuum is not as sensitive to redshifting the SED as the PAH features - hence the continuum in the observed frame is quite similar (for a range of redshifts at least, as we discuss in \S 4.5) to the rest-frame color-color plot.  Nonetheless, the dynamical interpretation of trends in color-color space is most easily understood in terms of rest-frame quantities.

The recent, comprehensive work by Barmby et al. (2006) corroborates our suggestion regarding the blue-red plume containing energetically active AGN.  Specifically,
Barmby et al. (2006) find that the infrared properties of X-ray sources in the Extended Groth Strip are very diverse - about 
40\% of the X-ray detected sources have red power-law SEDs in the $3.6-8.0~\micron$ IRAC bands,
while another 40\% have blue SEDs.  Thus, while the presence of sources in the
AGN demarcated region in the IRAC color-color plot is a sure sign of an energetically active
AGN, the converse is not necessarily true (as Figure 10 has shown).  The diversity of infrared
properties of X-ray detected sources is further highlighted by Rigby et al. (2004), who found
that X-ray hard AGNs are not preferentially infrared-brighter; Alonso-Herroro et al. (2004)
found that sources in the Lockman Hole with similar selection criteria have a variety of 
optical/IR spectral types.  Therefore, while AGN identification using the IRAC color-color
selection is generally effective, it may undercount the number of energetically active AGN
as it would miss the bluer region.

\begin{figure} \begin{center}
\centerline{\psfig{file=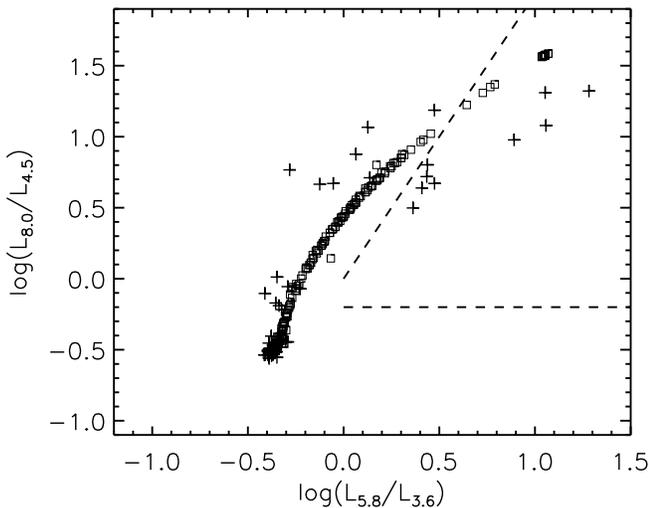,width=3.7in}}
\end{center}
\caption{Correlations in rest-frame IRAC bands, with inclusion of PAH emission model (squares designate the combined spectrum, including the continuum and PAH model).  Dashed lines demarcate color criteria to pick out obscured AGN sample by Lacy et al. (2004).}\label{fig:11}
\end{figure}

We show in Figure 11 our simulated IRAC color-color plot with inclusion of our phenomenological model of PAH emission.  The PAH region, as shown in this figure, is qualitatively in agreement with Sajina et al.'s (2005) results, who use a combined template of stellar emission, a power-law for the mid-IR spectrum, and a PAH template to simulate an IRAC color-color plot.  Interestingly, our color-color plot for the continuum (as shown in Figure 8) - which is derived from self-consistent three-dimensional radiative transfer solutions - is in better agreement with the observed FLS IRAC color-color plot shown in Lacy et al. (2004) than Sajina et al.'s (2005) results.  Lacy et al. (2004) note that the median photometric redshift of the candidate AGN is $\sim 0.3$ in the FLS color-color plot.  The location of the PAH region in the color-color plot is a sensitive function of redshift.  While in the rest-frame the PAH region extends into (and bends towards) the AGN demarcated region (as we have shown here), redshifting the PAH template to $z\ga 0.1$ causes the PAH region to bend into the bluer region - resulting in the appearance of the ``bunny-ear'' shape seen in FLS IRAC color-color plot, as has been noted by Sajina et al. (2005).  We show later in \S 4.5 an IRAC color-color plot in the observed frame for galaxies at a range of redshifts - from 
$z \approx 0.3 - 2$,
which do indeed result in the characteristic ``bunny-ear'' shape for the $z=0.3$ sample.

We note that that the scaled $z=3$ simulations do not occupy a region of color-color space (in the rest-frame) that is distinct from that of the $z=0$ simulations.  This suggests that at least qualitatively, the appearance of (in particular, the clustering of the colors) simulated galaxies on color-color plots is a generic feature of simulations of major mergers where black hole feedback has a significant and sharp effect on the evolution of the system, but which evolve otherwise passively.  Extrapolating this trend to high redshift systems suggests that simulations of high-redshift systems ($z \ga 6$) would show a similar behavior in color-color space.

\subsection{Infrared X-ray Correlations}

\begin{figure} \begin{center}
\centerline{\psfig{file=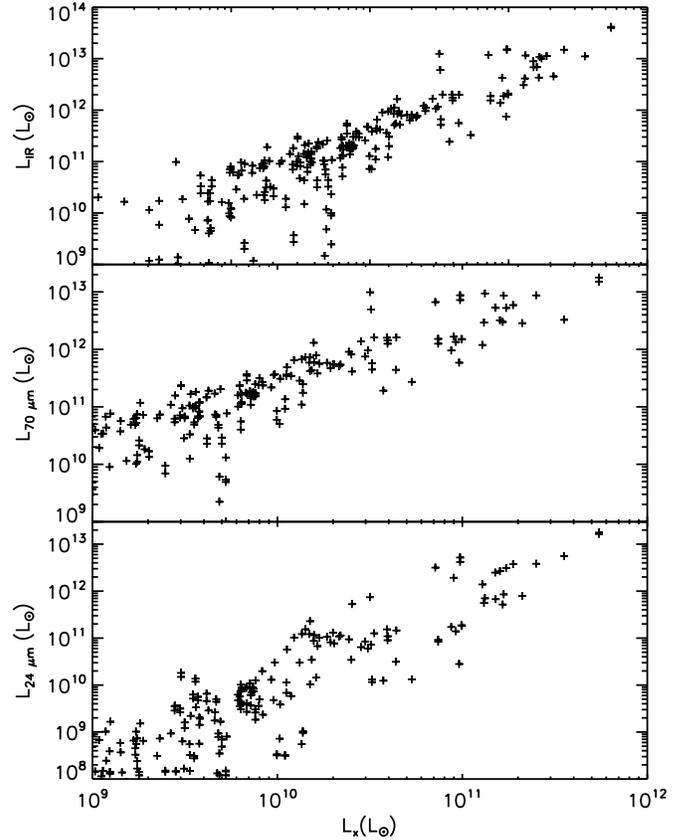,width=3.7in}}
\end{center}
\caption{(a) Top panel - predictions for infrared luminosity (integrated from $8~\micron$ to $1000~\micron$) vs. the hard X-ray, (b) Middle panel - predictions for $70 ~\micron$ rest-frame luminosity vs. the hard X-ray correlations, (c) Bottom panel - predictions for $24 ~\micron$ rest-frame luminosity vs. the hard X-ray luminosity.}\label{fig:12}
\end{figure}

A large population of obscured AGN is discerned from the hard X-ray background (Comastri et al. 1995); these systems would re-radiate a significant fraction of their luminosity into the infrared,
in principle explaining the spectral shape of the
X-ray background (see, e.g., Hopkins et al. 2006a, HRH).
Therefore, infrared X-ray correlations hold promise for being particularly useful diagnostics for discerning and
understanding the nature of both energetically active AGN, and dormant AGN approaching their accretion phase in merging galaxies.  We show in Figure 12 (a) the infrared luminosity versus the hard X-ray luminosity (from 2-10 keV) from our set of simulations.  In our model, the hard X-ray luminosity derives from the black hole luminosity, which we compute following the model of Marconi et al. (2004).  On the high X-ray luminosity end ($L_{\rm x} \ga 10^{9} L_{\odot}$), the contribution from the black hole to the X-ray luminosity is significantly higher than that from star formation; even moderately high values of the star formation rate ($\sim 100~M_{\odot}/\rm yr$) lead to x-ray luminosities of order $\sim 10^{8}~L_{\odot}/\rm yr$ (Gilfanov et al. 2004).  X-ray emission from hot gas is also negligible on the high x-ray luminosity end - from analyzing the x-ray emission from hot gas in merger simulations, Cox et al. (2006b) find that the maximum x-ray emission would not exceed $\sim 10^{8} L_{\odot}$.  

\begin{figure*}[!t] \begin{center}
\centerline{\psfig{file=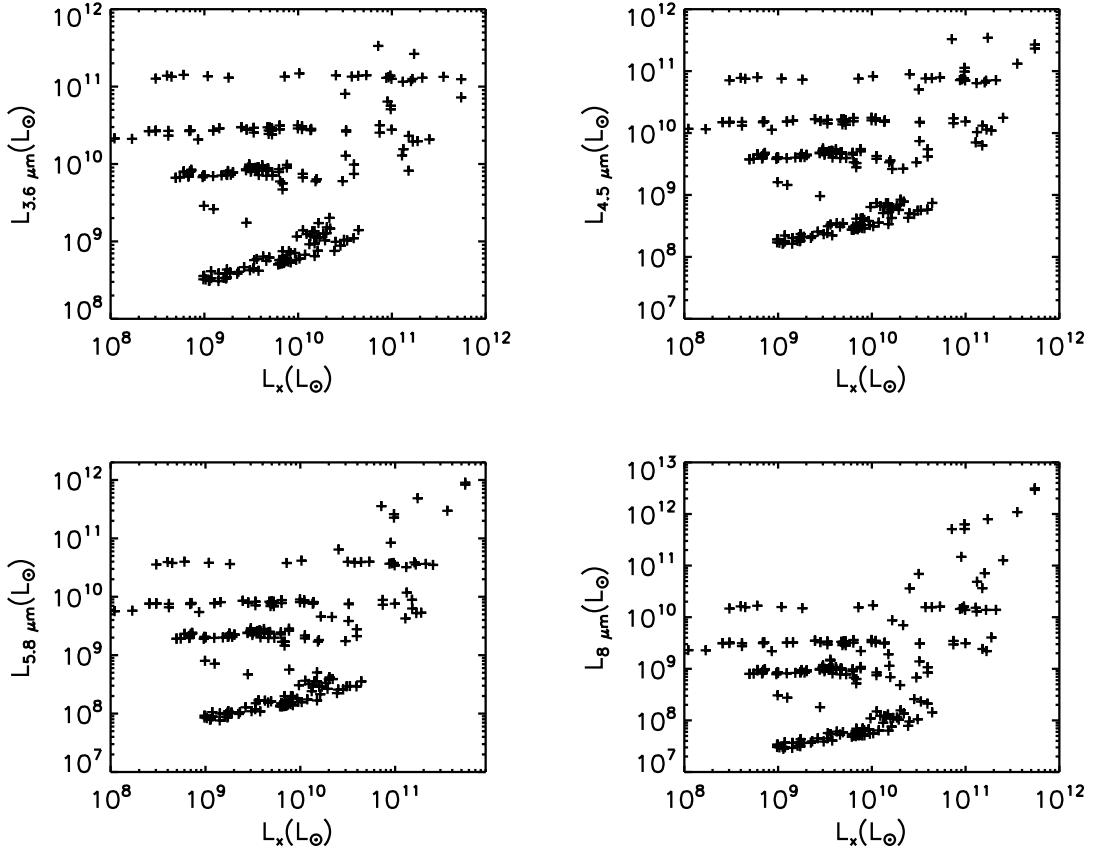,width=6in}}
\end{center}
\caption{(top-left) The rest-frame $3.6~\micron$ luminosity vs. the hard X-ray, (top-right) the rest-frame $4.5~\micron$ vs. the hard X-ray, (bottom-left) rest-frame $5.8~\micron$ vs. the hard X-ray luminosity, (bottom-right) rest-frame $8~\micron$ vs. hard X-ray luminosity.  Note that the clearest correlations are for the longer wavelength bands at higher X-ray luminosities (see discussion in \S 4.2).}\label{fig:13}
\end{figure*}

Figure 12 (a) shows that there is a clear correlation between the the total infrared luminosity (which we get from integrating the SED from $8~\micron$ to $1000~\micron$) and the hard X-ray luminosity, which extends over nearly four orders of magnitude, with an increase
in the scatter at low X-ray luminosities.  This makes intuitive sense - in our model, low X-ray luminosities correspond to the pre- (and post-) accretion phase, before (and after) the black hole has become energetically active ($L_{\rm BH} \ga L_{\rm star}$).  In the low X-ray luminosity phase, the infrared luminosity is partly powered by distributed sources of radiation; i.e., stars, and partly by a heavily enshrouded black hole.  On the most X-ray luminous end, the black hole dominates in its contribution to the total bolometric luminosity - this case is essentially analogous to a central source of luminosity powering the infrared radiation.  Hence, there should be a correlation between the luminosity of the central source and the reprocessed radiation in this phase.  These systems, at their peak infrared luminosity, usually exceed ULIRG, 
and sometimes HLIRG
($L_{\rm IR} \ga 10^{13} L_{\odot}$)
luminosities.  On average, the X-ray luminosities are of order $10^{10} L_{\odot}$, though there are a few on the very bright X-ray end ($L_{\rm x-ray} \sim 10^{11} L_{\odot}- 10^{12} L_{\odot}$) - which correspond to the phase when the black hole dominates the bolometric luminosity.  Alexander et al. (2005a) find that the X-ray luminosity of their SCUBA selected sample is $\sim 10^{11} L_{\odot}$, with only four in their sample exceeding $10^{11} L_{\odot}$.  Alexander et al. (2005a) use the locally observed radio far-IR correlation to find that the X-ray luminosity is about 250 times lower than the infrared luminosity, whereas we find that it is typically $\sim 100$ times lower.  The radio far-IR correlation has considerable scatter in luminous galaxies (Sadler et al. 2002), and its extrapolation to high redshifts is uncertain.  As such, this discrepancy may in part owe to the fact that using the radio far-IR correlation underpredicts the infrared luminosity since the contribution from the black hole has not been taken into account.  Another possibility is that (part of) their X-ray sample is Compton thick, and hence have higher intrinsic luminosities than have been accounted for.  Multi-wavelength infrared observations are needed to observationally derive the rest-frame infrared luminosities to test our prediction.  

We find that the $70~\micron$ luminosity is roughly forty times the X-ray luminosity, and displays the same behavior as the infrared luminosity (Figure 11b).  Figure 12 (c) shows that a similar behavior for the rest-frame $24 ~\micron$ flux - with the brightest $24 ~\micron$ objects being strongly correlated with the hard X-ray luminosity.  We stress that these correlations are predictions from our model, which can be directly tested by combining far-IR and X-ray data at $\it{comparable}$ sensitivities for large samples of SMGs.  This should be made possible by the upcoming Herschel mission.  Source-stacking analysis may make testing this prediction feasible, at least in a statistical sense, by using existing MIPS data.

We also show in Figures 13 (a-d) the rest-frame luminosities, derived from the continuum only, in the IRAC bands plotted versus the hard X-ray flux.  These plots display a more complex behavior than the far-infrared X-ray correlations.  There is a positive correlation only at high X-ray luminosities, with the weakest correlations, i.e., increase in $L_{\rm IRAC~bands}$ as a function of X-ray luminosity, seen for the shorter wavelengths.  The different curves owe to more massive simulations having higher luminosities; actual observations will probe nearly a continua of masses and hence will not appear discrete.  At the longer IR wavelengths, the dust opacity curve has a relatively simple (essentially power-law) form and the emission
comes from the outer regions of the envelope where the dust is cooler.  However, the rest-frame IRAC fluxes typically emanate from optically thick regions, which are often heated to high temperatures ($T \ga 100 \rm K$), with their emissivity modulated by (distinctly non power-law) spectral features in the dust opacity curve.  (For a discussion of, and analytic expressions for, the characteristic emission radii of different frequency regimes in dust envelopes, see CM05.)  Nonetheless, on the bright end, the problem does simplify for the longer wavelength IRAC bands.  The shortest wavelength is the least correlated with the X-ray luminosity - dust is rarely heated to temperatures high enough ($T \ga 900~\rm K$) to produce $3.6~\micron$ emission, except on the very bright end.  As such, this band is generally probing the attenuated Rayleigh-Jeans fall-off of the stellar light, particularly at low X-ray luminosities.  The fluxes in these bands are also influenced by scattering of photons by dust grains since the scattering efficiency increases sharply at short wavelengths and leads to an increase in the flux by factors of 3-5 (for optical depths characteristic of the simulations), especially as the effective optical depth of the envelope decreases (Chakrabarti \& Whitney, 2006, in preparation).  Moreover, they will also be influenced by PAH emission.   Yan et al. (2005) find that 60\% of their sources
between redshifts of 1.8 and 2.6, have weak or no PAH emission, while about half have prominent silicate absorption lines.  Both PAH spectral features and non-power law continuum features will heighten the relative scatter for IRAC X-ray correlations, as compared to the far-IR X-ray correlations.  Rigby et al. (2004) emphasize the diversity of $24~\micron$ properties of X-ray selected AGN, even in a narrow redshift slice ($z\sim 0.7$).  They do see some indication of the ratio of the $24~\micron$ to X-ray luminosity increasing with X-ray hardness for their subsample at $z\sim 0.7$, but there is scatter even in this subsample.  A detailed comparison with observations is beyond the scope of this paper - in a future paper, we use the SEDs from the simulations as templates to K-correct observed fluxes of sources with spectroscopic redshifts, using the redshift distribution from Chapman et al. (2005) for SMGs.      

\subsection{Photo Albums of the Lifetimes of SMGs}

We present photo albums of the time evolution of the surface brightness in the observed $3.6~\micron$ band, as well as the $450~\micron$ band during the lifetime of three SMGs from our sample.  We adopt a luminosity distance of 15.5 Gpc corresponding to $z=2$ ($\rm \Omega_{M}=0.3$, $\rm \Omega_{\Lambda}=0.7$, $H_{0}=70$) in these figures and take into account the instrumental angular resolution.  These figures should not be interpreted as what would be observed by SCUBA and Spitzer's IRAC since we have not convolved these images with realistic PSFs; doing so would not leave any visible structure.  Our goal here is merely to use these images as a visual aid in contrasting the time evolution of the surface brightness and morphologies of three SMGs of differing orbital inclination and progenitor redshift during their lifetimes.  

Figure 14 shows a co-planar merger (h320) from its pre-merger phase (a), to close to the main feedback phase as the AGN starts to clear out the surrounding dust and gas (b), to the final remnant (c).  Figure 15 (a-c) shows the corresponding images in the observed $450~\micron$ band at the same times.  The galaxy becomes brighter in the IRAC bands as it is becoming progressively fainter in the longer wavelength SCUBA band.  This simply reflects the time variation of the fluxes - as the AGN feedback clears out the obscuring material, more of the emitted energy (the SED) shifts to shorter wavelengths, and there is correspondingly less emission at longer wavelengths.  This trend would be seen also in surface brightness maps of the CO emission which probe the cooler regions of the envelope.  Another point of note is that this co-planar merger has a disk-like morphology, while the non co-planar merger, shown in the following Figure 16 and 17, is more extended and non disk-like.  However, it shows the same time evolution of the surface brightness - the brightness in the IRAC bands increases as the brightness in the SCUBA bands decreases as more of the emitted energy shifts to shorter wavelengths.  Finally, the set of images shown in Figures 18 and 19 depicts the time evolution of a simulation with a progenitor at $z=3$, z3h270, (with a virial velocity comparable to h320 and e320, the ones shown in the previous two figures).  While the trends in the time evolution of the surface brightness mentioned earlier for the $z=0$ scaled simulations are also seen here, one clear difference is that the $z=3$ scaled simulations are much more compact than the $z=0$ scaled simulations, which have emission extending to $\ga 10~\rm kpc$ at times, while the $z=3$ simulation does not show any structure on scales larger than 1 kpc in the observed $3.6~\micron$ band, though some extended long wavelength emission is seen over several kpc in the observed $450~\micron$ band.

\begin{figure*} \begin{center}
\centerline{\psfig{file=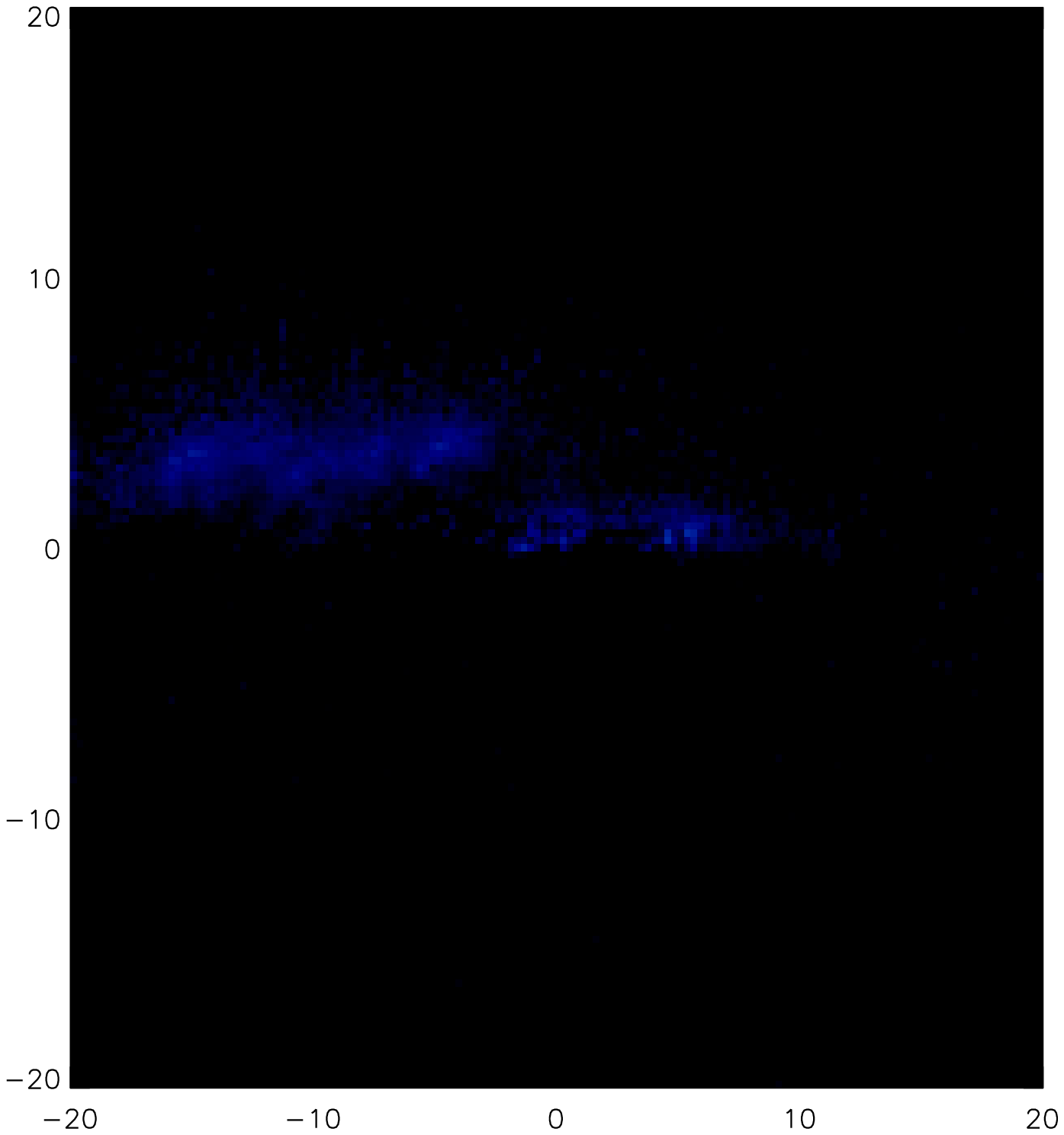,width=2.in}
\psfig{file=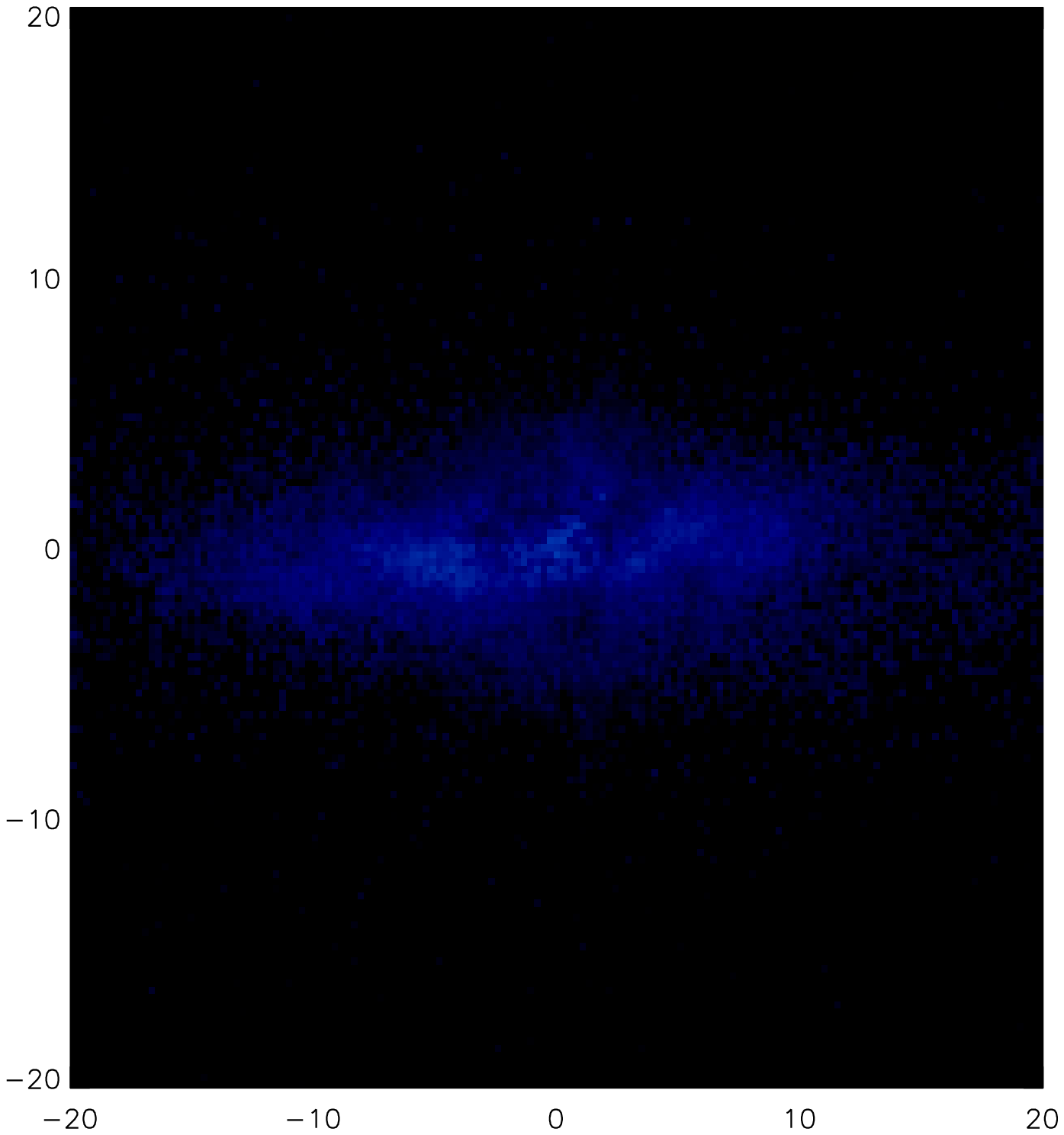,width=2.in}
{\psfig{file=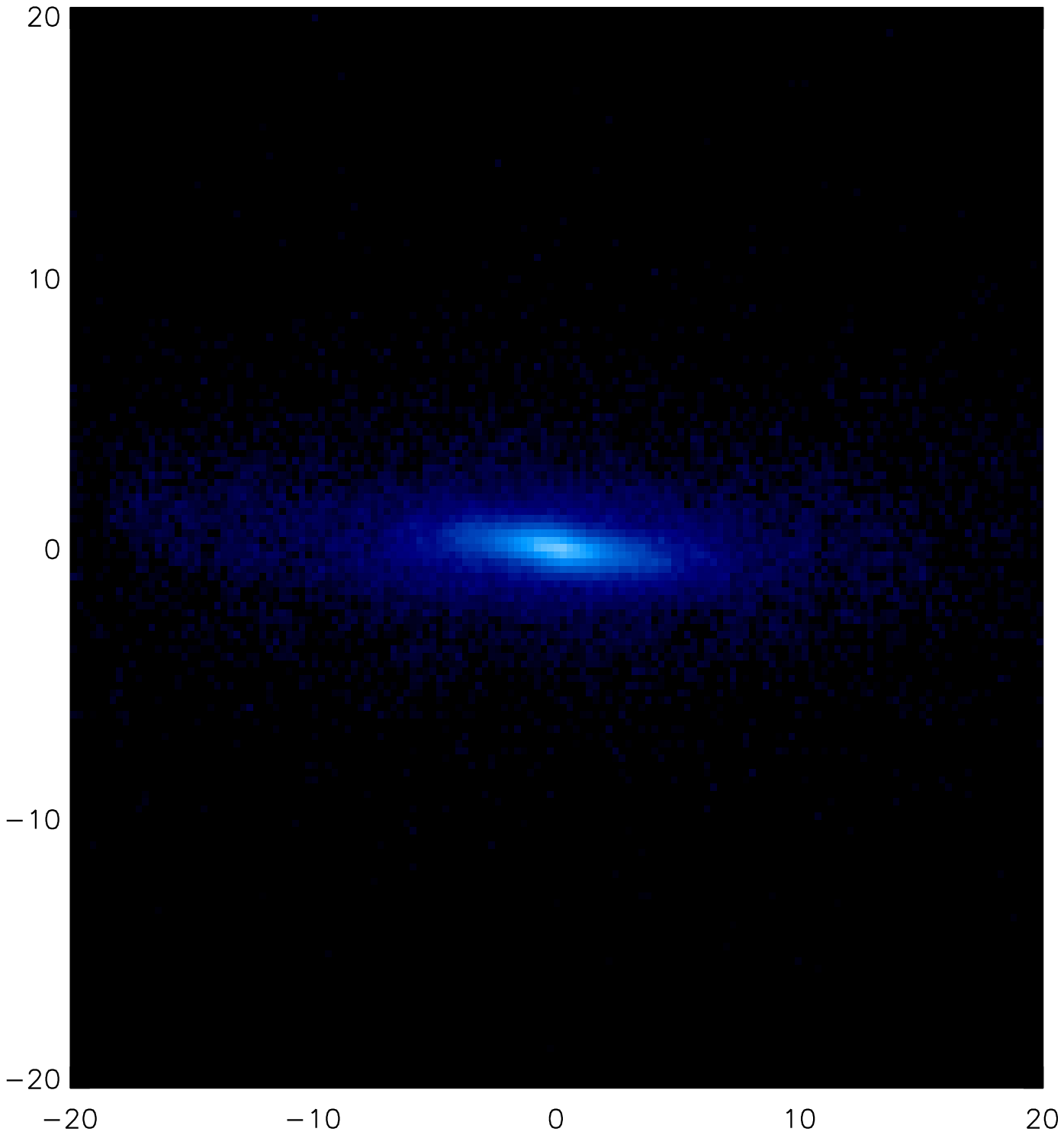,width=2.in}}}
\end{center}
\caption{Photo Album of the Lifetime of an SMG produced in a co-planar merger (h320) in observed $3.6~\micron$ band, (a) Pre-merger phase, (b) Close to Main Feedback Phase, (c) After Main Feedback Phase.}\label{fig:14}
\end{figure*}

\begin{figure*} \begin{center}
\centerline{\psfig{file=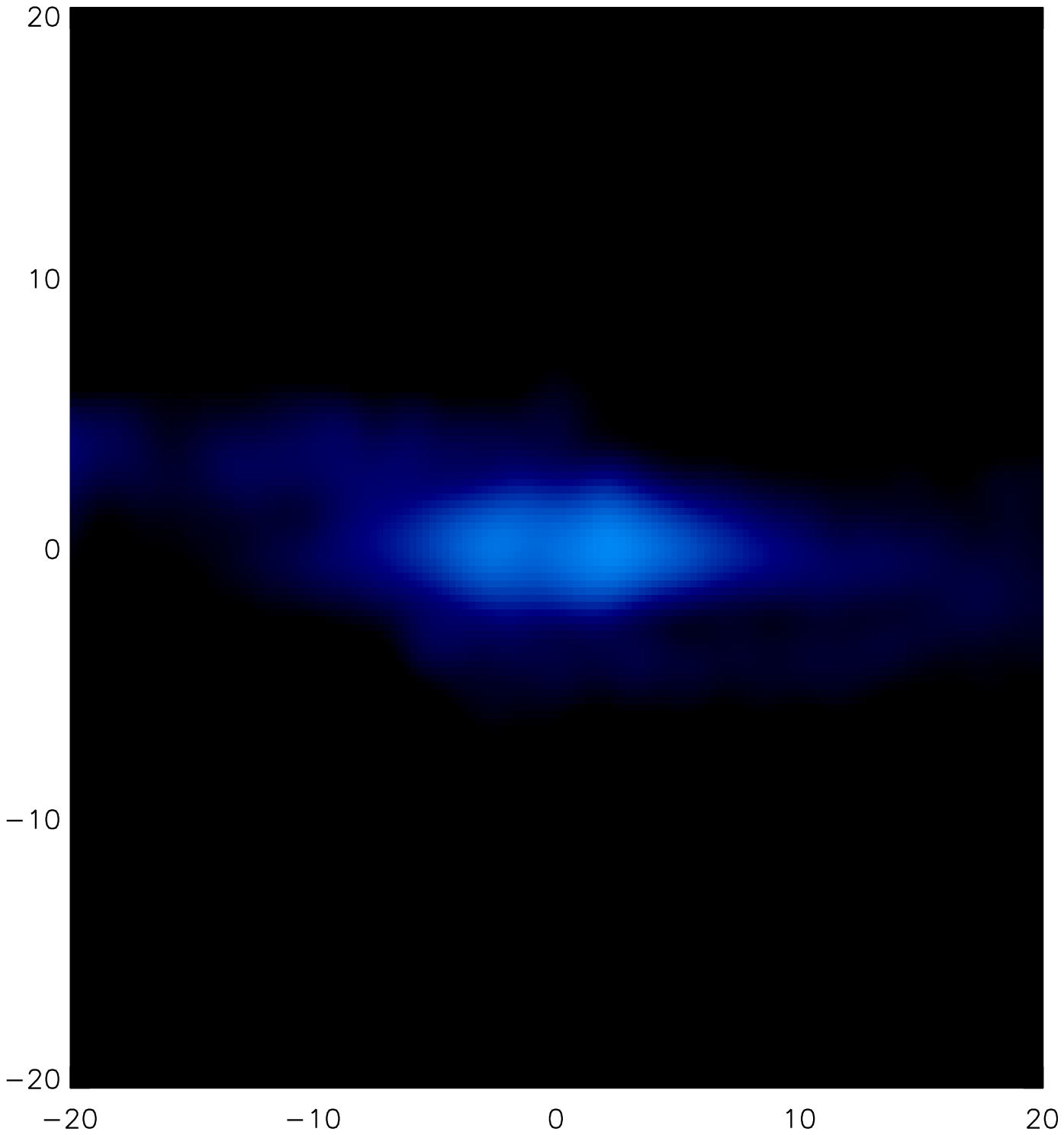,width=2.in}
\psfig{file=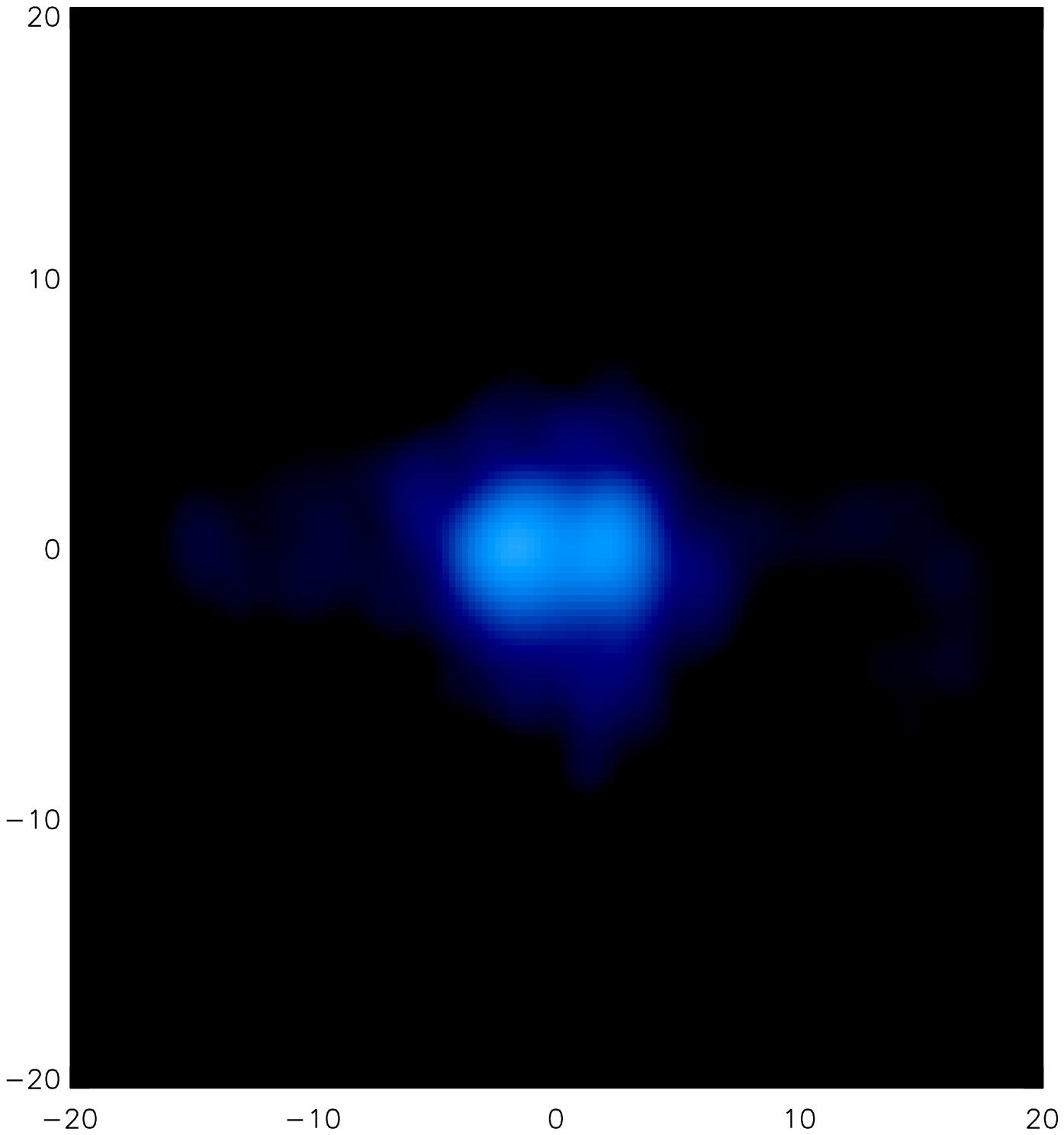,width=2.in}
{\psfig{file=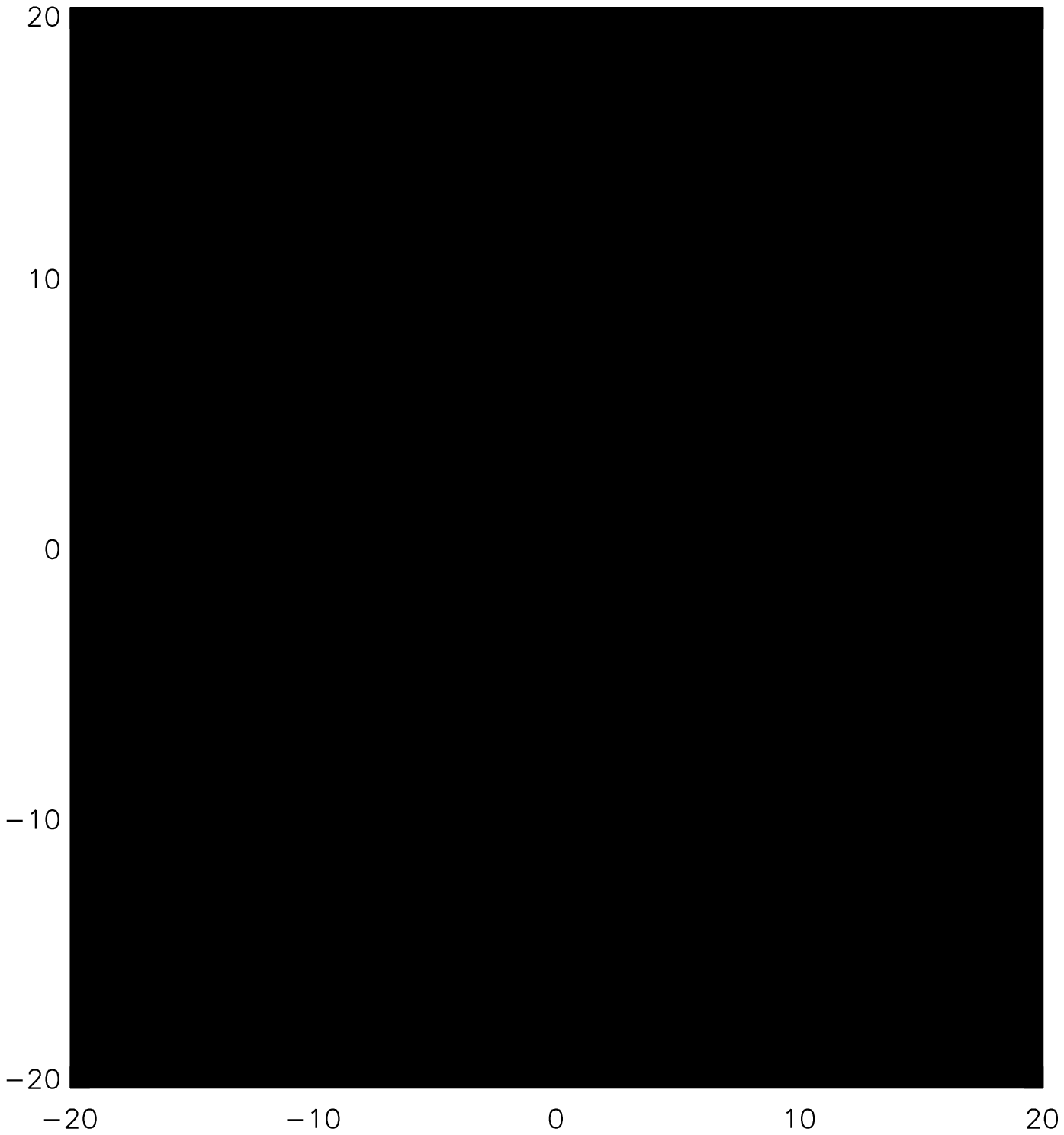,width=2.in}}}
\end{center}
\caption{Photo Album of the Lifetime of an SMG produced in a co-planar merger (h320) in observed $450~\micron$ band at same times as in $3.6~\micron$ band.  Times beyond the last phase would be black, i.e., would show no emission in SCUBA.  This shows that more of the emitted energy is distributed to the shorter wavelengths as AGN feedback clears out the obscuring material.}\label{fig:15}
\end{figure*}

\begin{figure*} \begin{center}
\centerline{\psfig{file=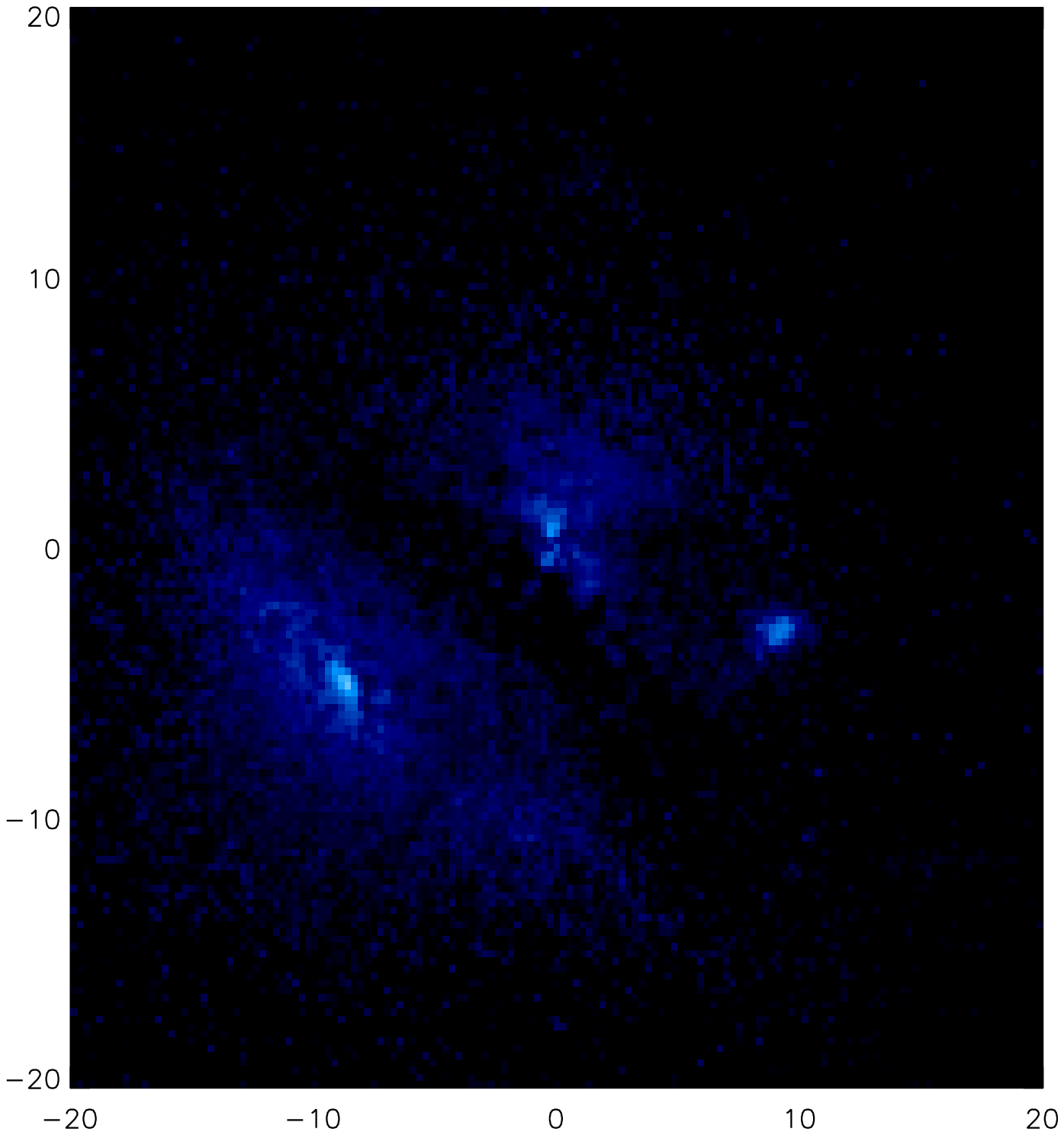,width=2.in}
\psfig{file=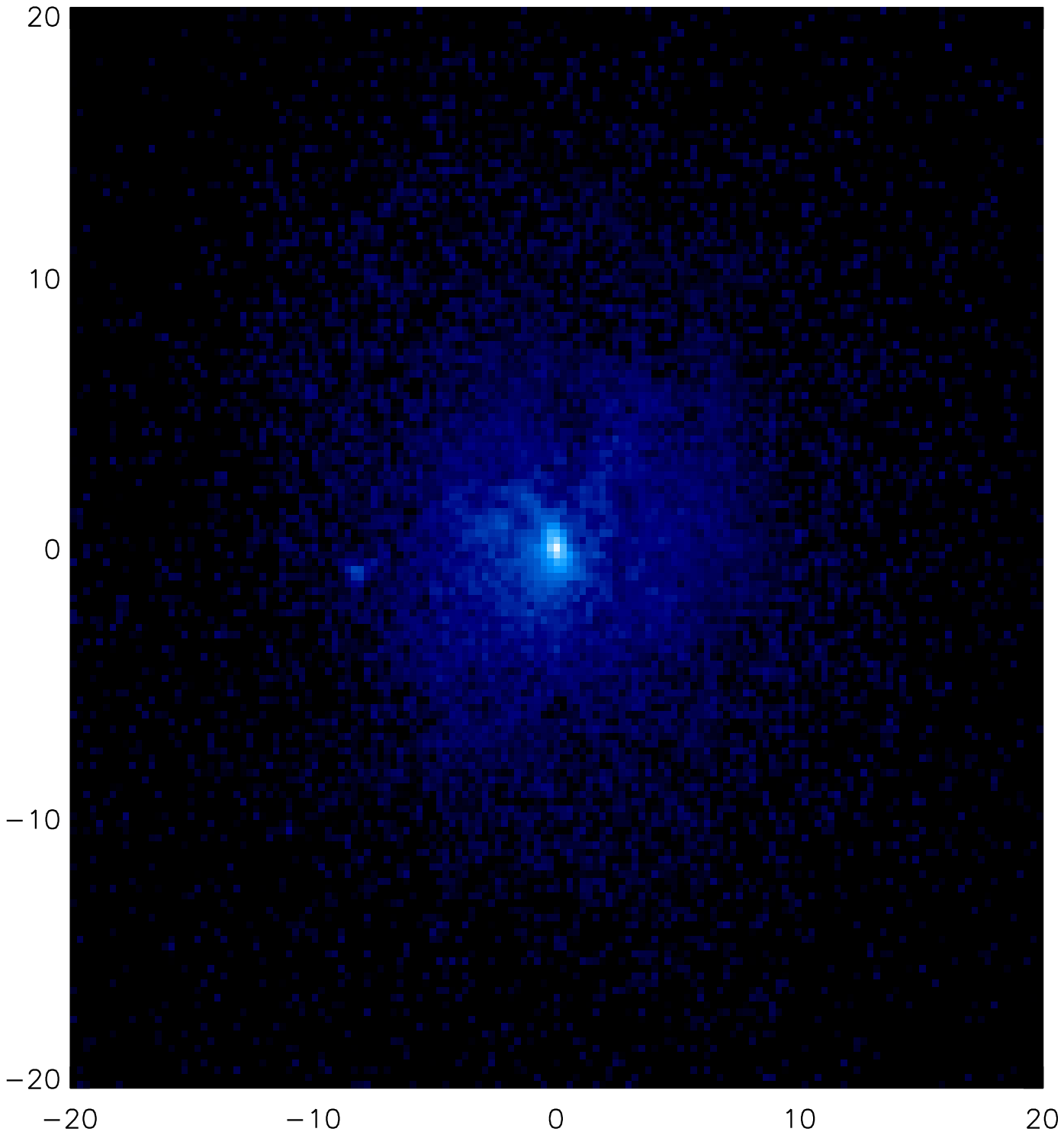,width=2.in}
{\psfig{file=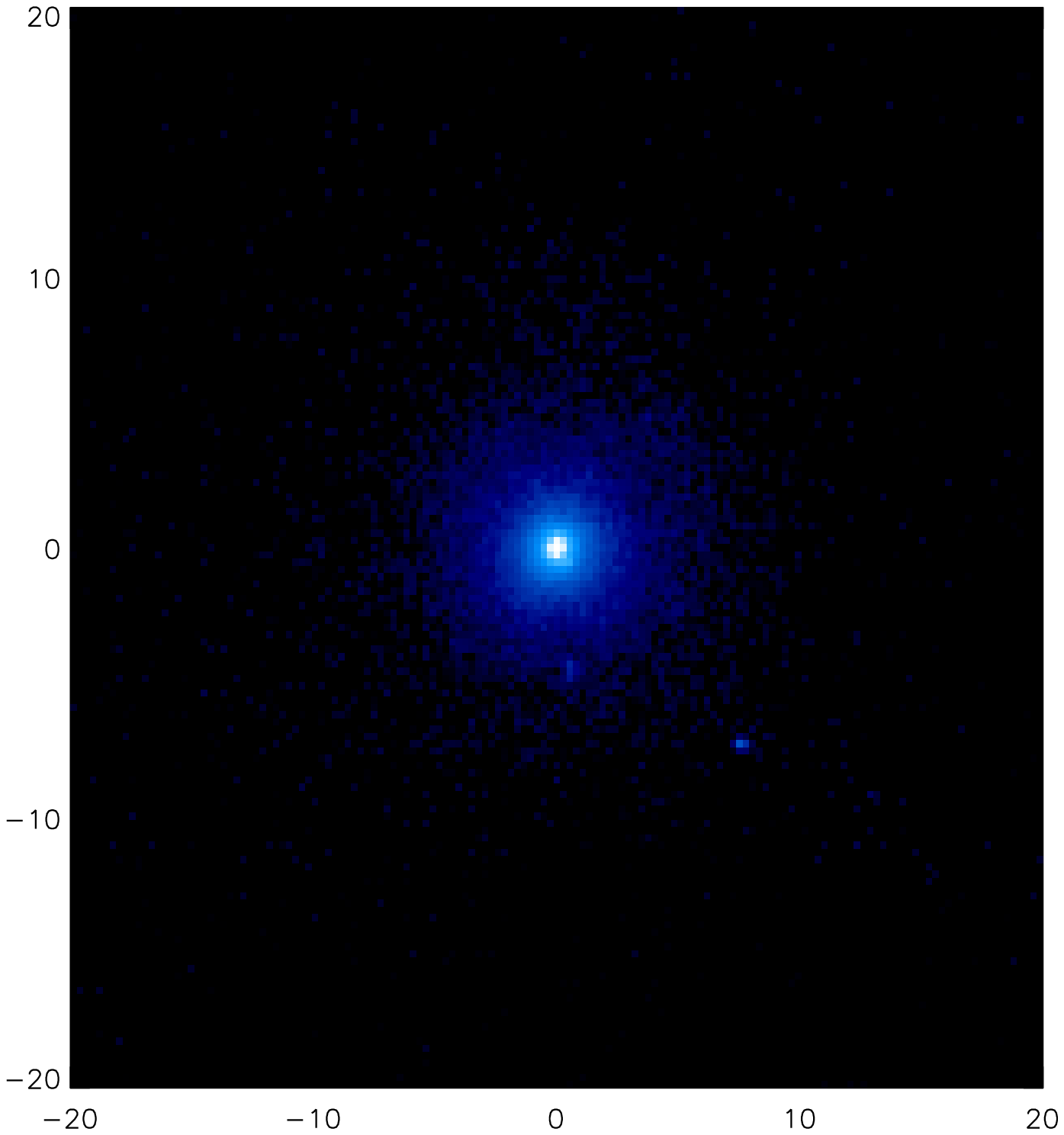,width=2.in}}}
\end{center}
\caption{Photo Album of the Lifetime of an SMG produced in a non co-planar merger (e320) in observed $3.6~\micron$ band, (a) Pre-merger phase, (b) Close to Main Feedback Phase, (c) After Main Feedback Phase.}\label{fig:16}
\end{figure*}

\begin{figure*} \begin{center}
\centerline{\psfig{file=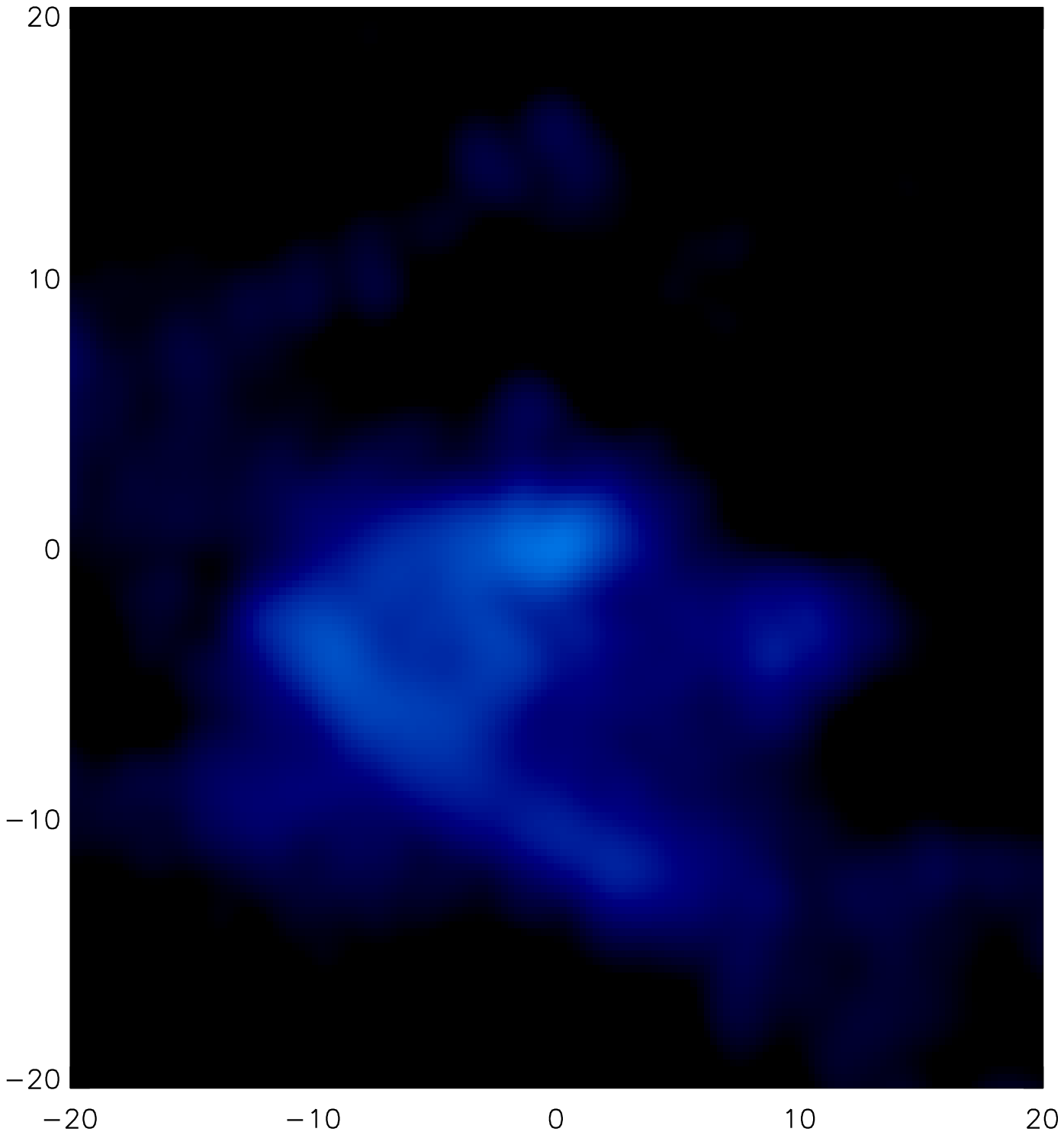,width=2.in}
\psfig{file=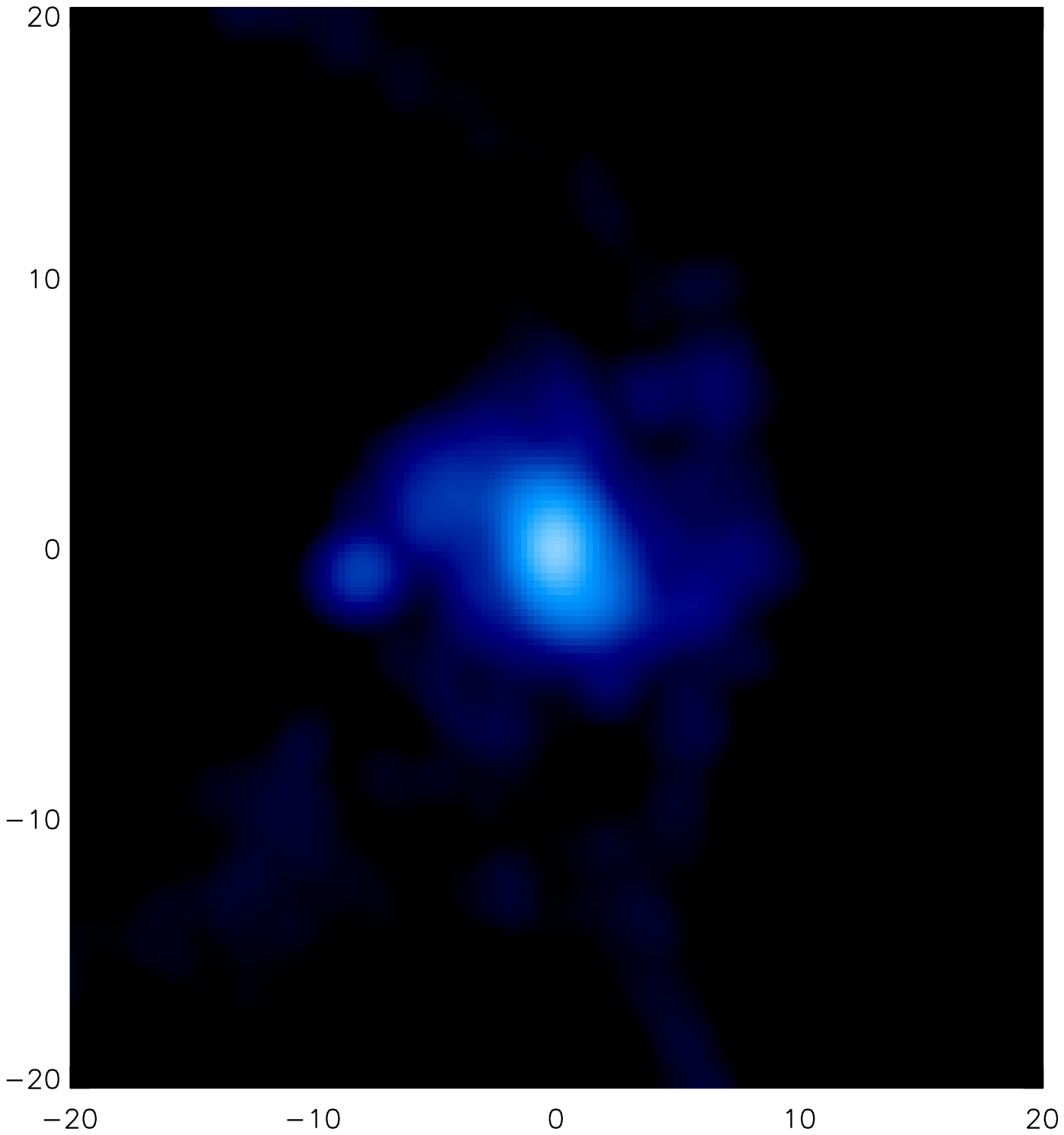,width=2.in}
{\psfig{file=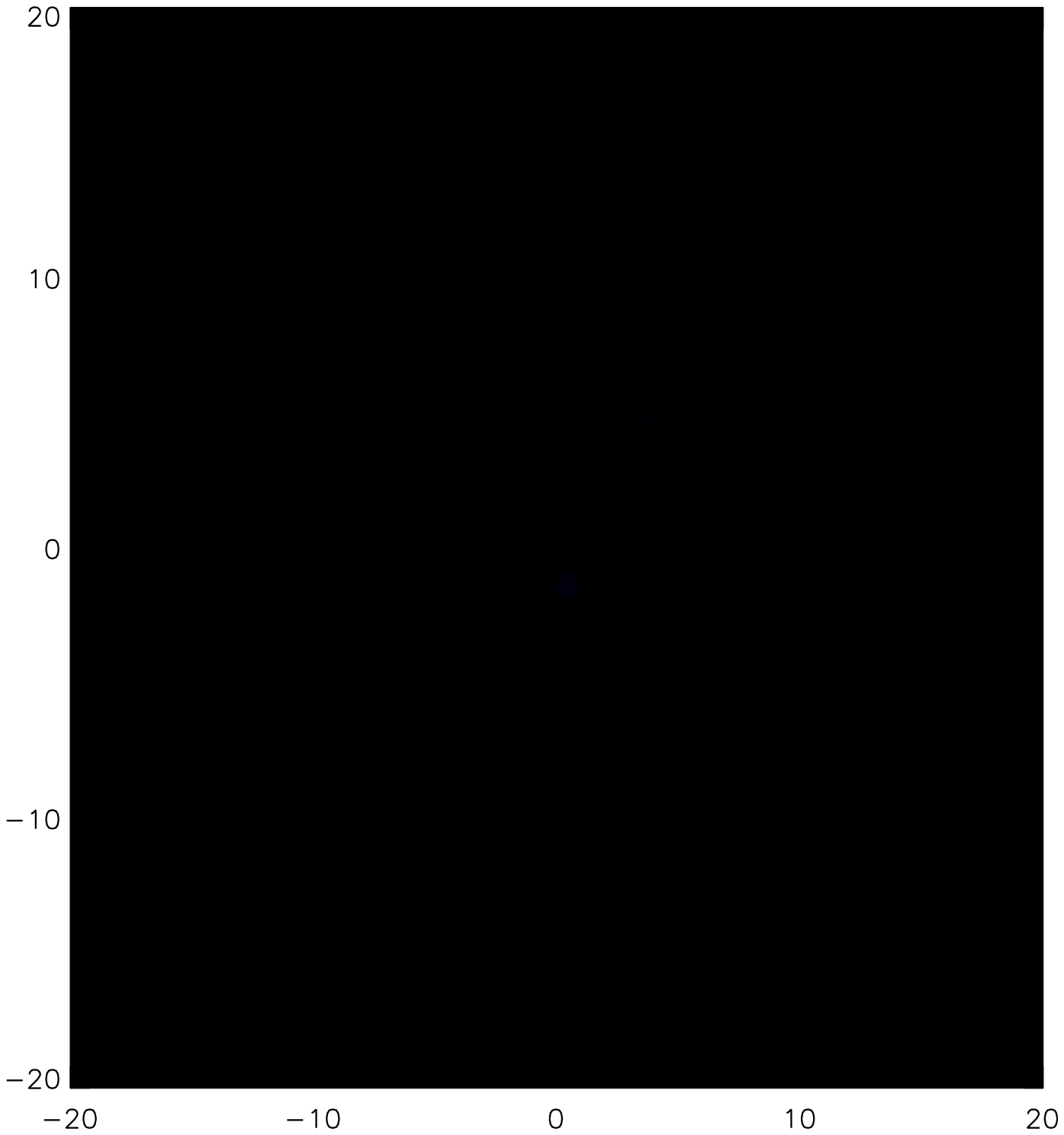,width=2.in}}}
\end{center}
\caption{Photo Album of the Lifetime of an SMG produced in a non co-planar merger (e320) in observed $450~\micron$ band at same times as in shown above in the $3.6~\micron$ band.  Times beyond the last phase would be black; i.e., would show no emission in this band.}\label{fig:17}
\end{figure*}

\begin{figure*} \begin{center}
\centerline{\psfig{file=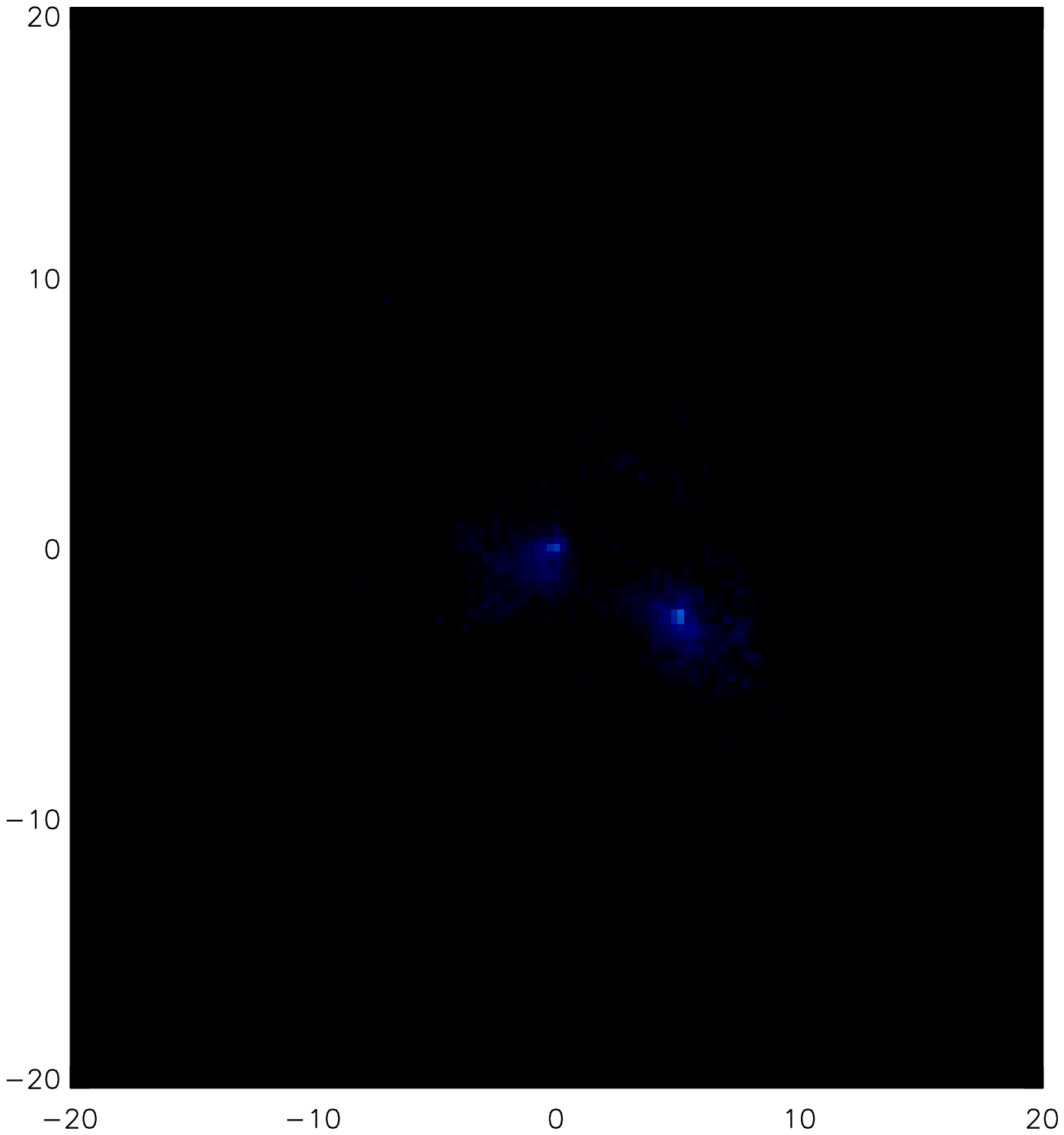,width=2.in}
\psfig{file=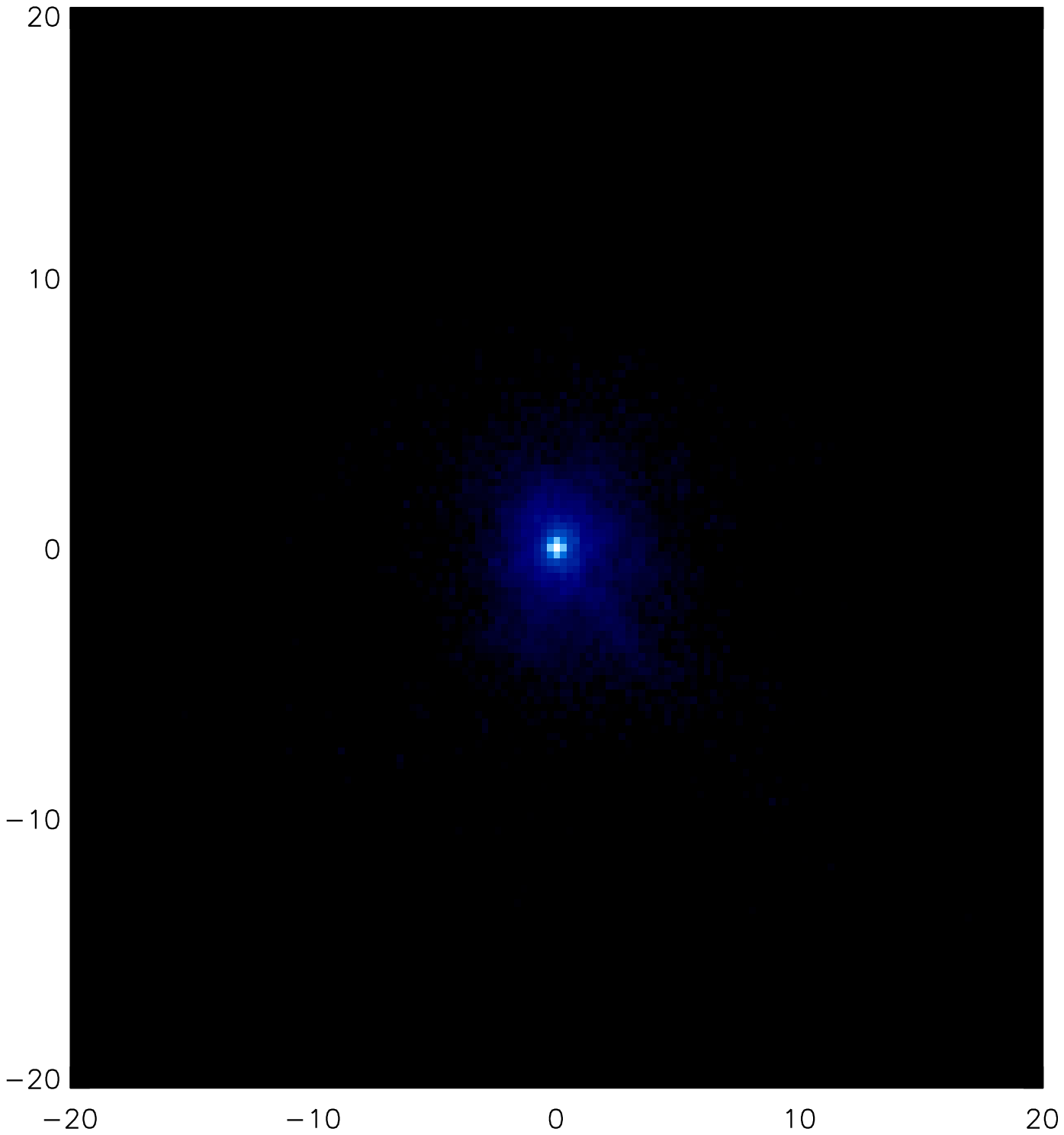,width=2.in}
{\psfig{file=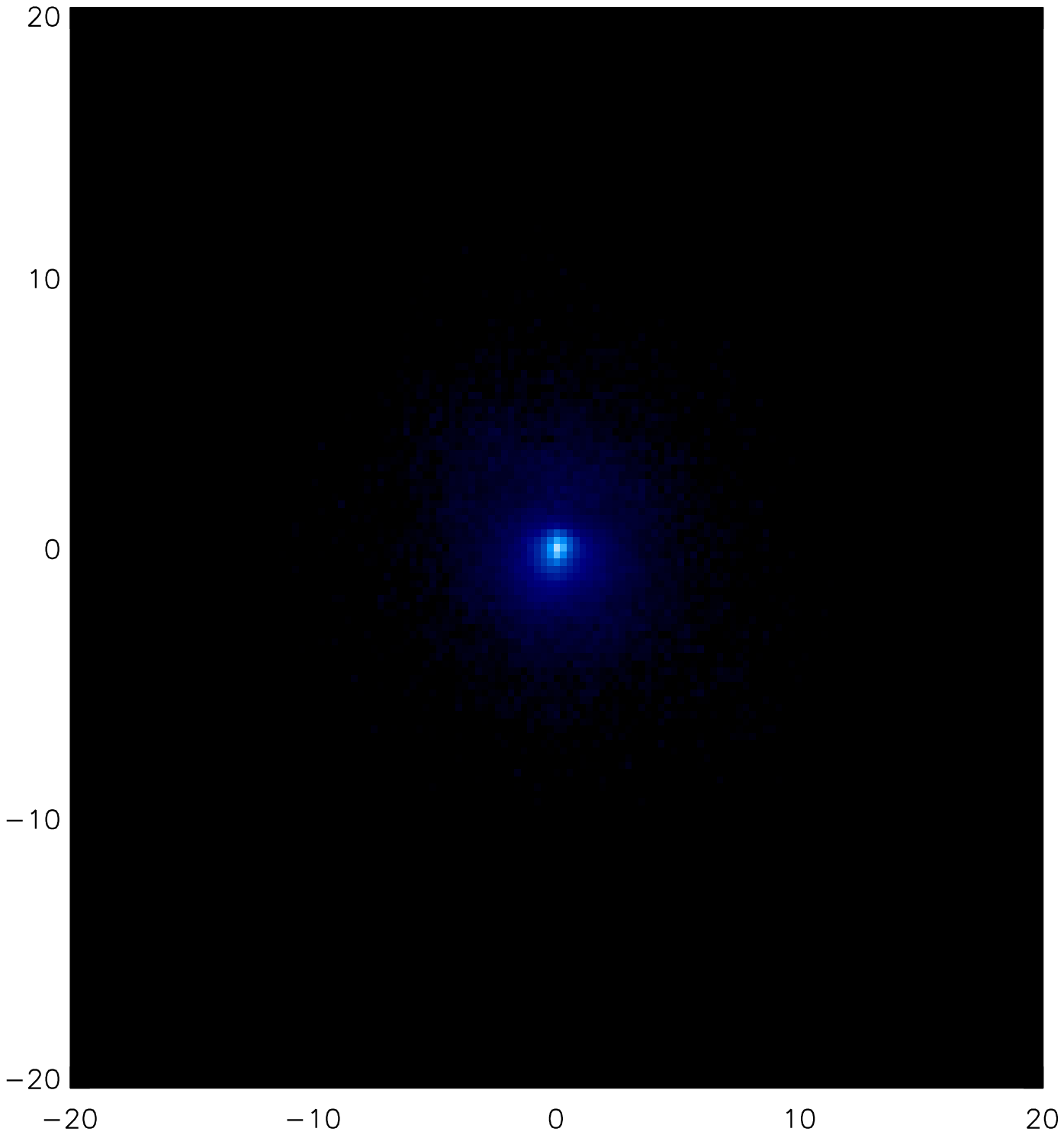,width=2.in}}}
\end{center}
\caption{Photo Album of the Lifetime of an SMG produced of a $z=3$ scaled merger (z3e226) in observed $3.6~\micron$ band, (a) Pre-merger phase, (b) Close to Main Feedback Phase, (c) After Main Feedback Phase.}\label{fig:18}
\end{figure*}

\begin{figure*} \begin{center}
\centerline{\psfig{file=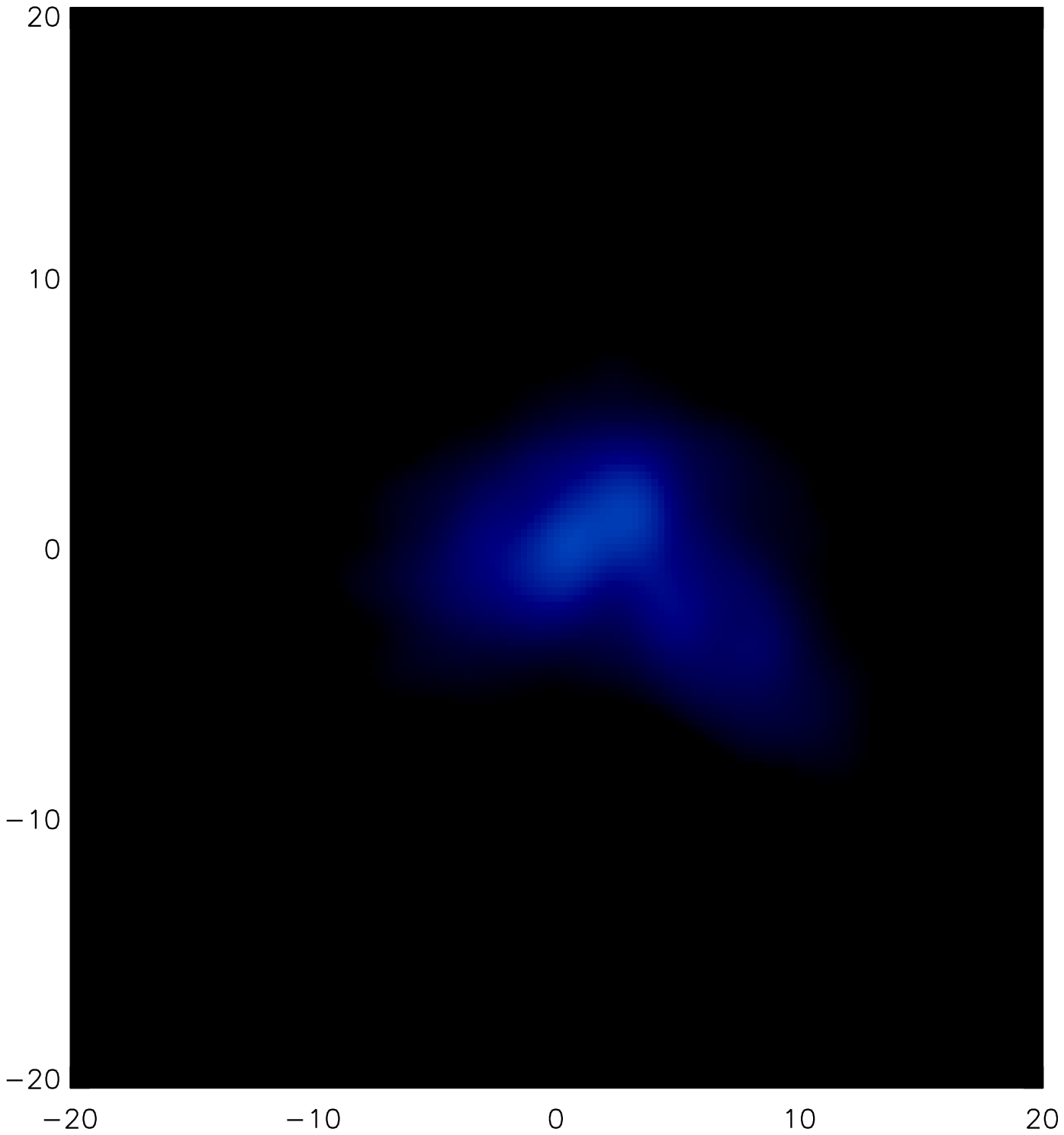,width=2.in}
\psfig{file=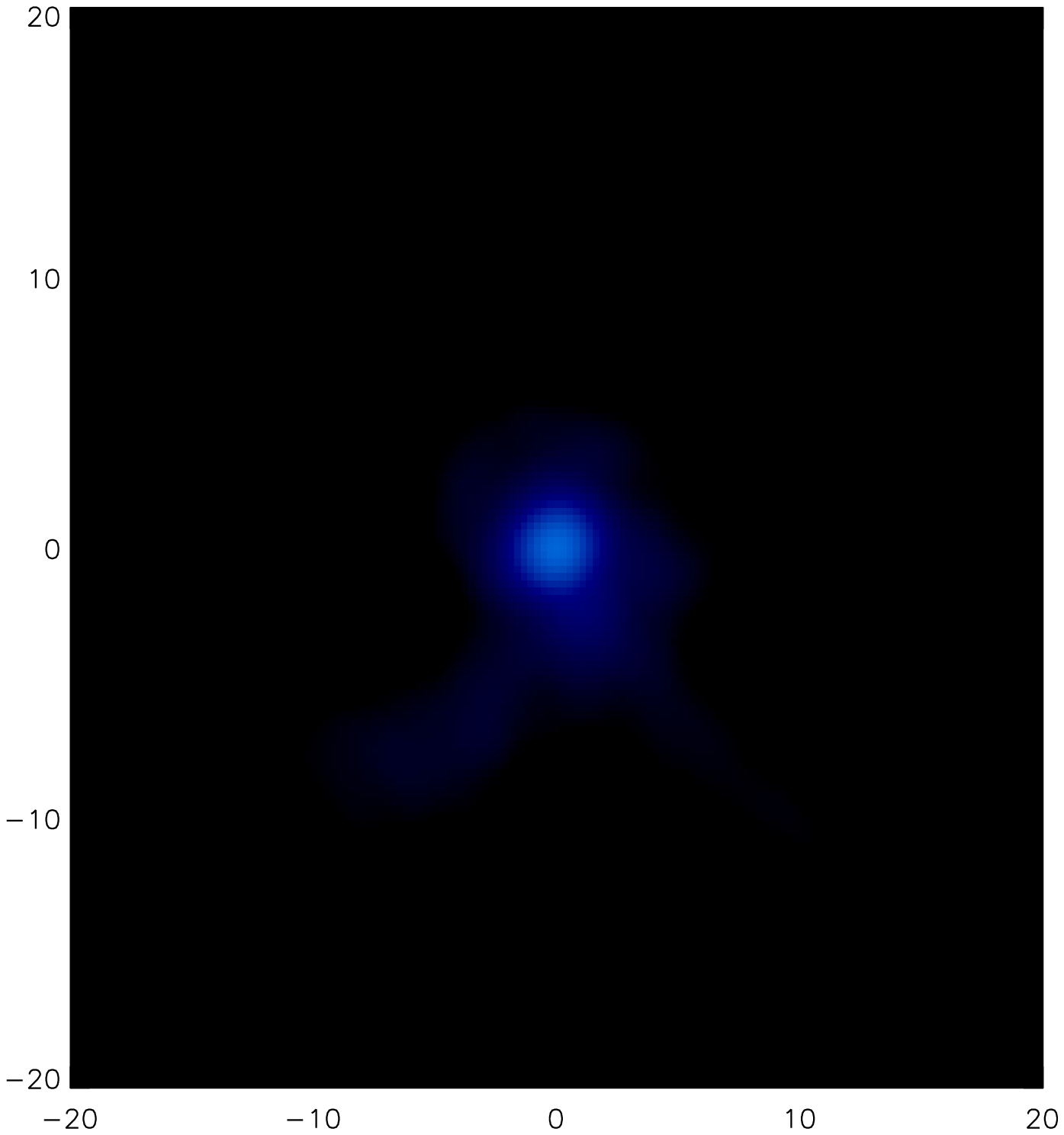,width=2.in}
{\psfig{file=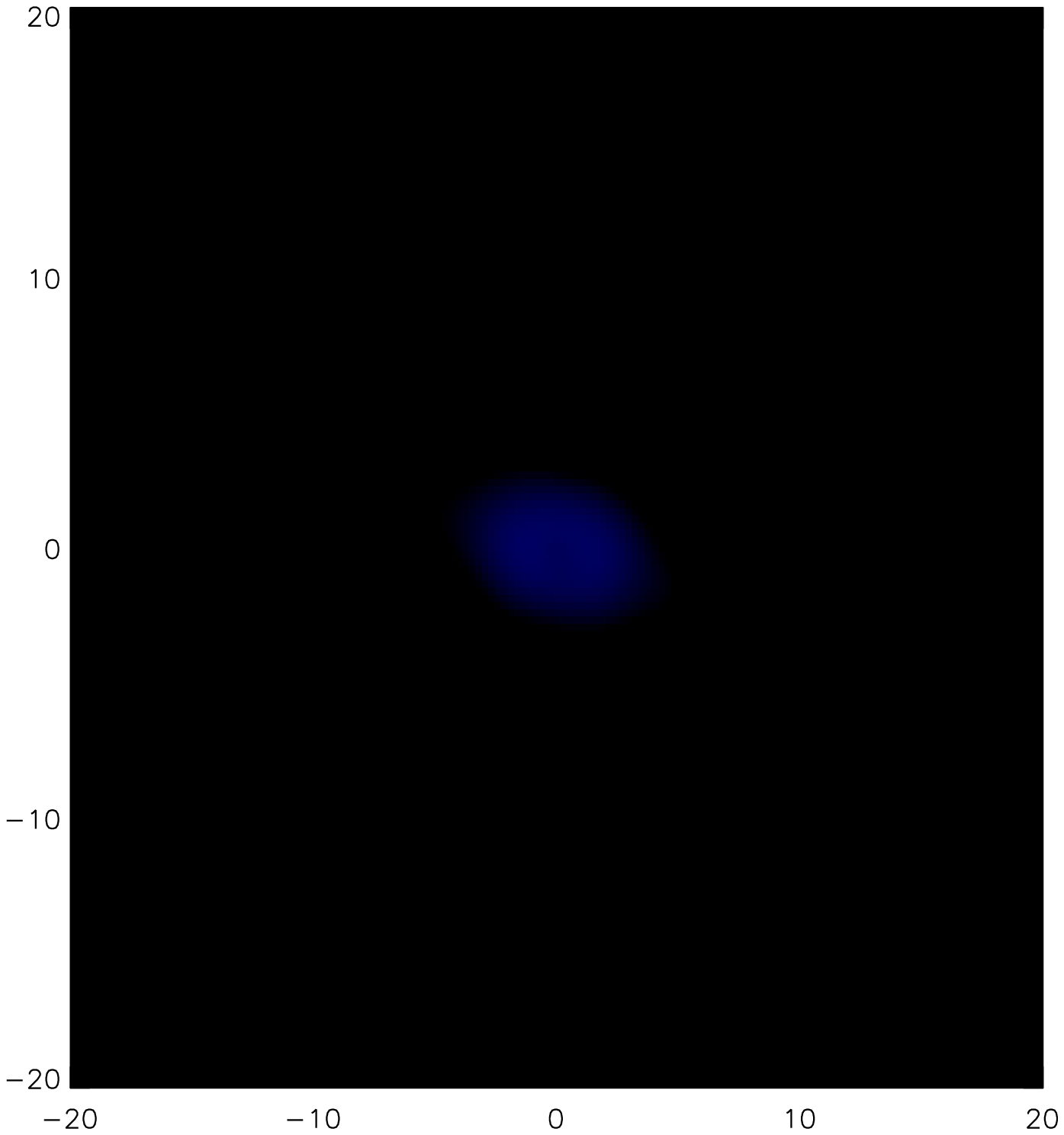,width=2.in}}}
\end{center}
\caption{Photo Album of the Lifetime of an SMG produced of a $z=3$ scaled merger (z3e226) in observed $450~\micron$ band.  Times beyond the last phase would be black; i.e., would show no emission in this band.}\label{fig:19}
\end{figure*}

Chapman et al. (2003a) obtained HST images of about a dozen SMGs ($z \sim 2$), and found that the morphologies were irregular and complex, suggestive of a merger origin.  Chapman et al. (2004) compared optical and radio imaging of this sample, and using the radio as a tracer of the star forming regions, argued for extended (about 10 kpc, but not much larger), obscured starbursts in about 70 \% of their sample, while in the rest of the sample, the radio extends to about 1 kpc.  While it is tempting to speculate about the differences in morphology between our $z=0$ and $z=3$ mergers and the potential to witness such differences in morphology in high resolution images, it is not clear that we can use images alone (even of HST resolution) to determine whether our models for SMGs with disk properties of $z=0$ systems would
be favored over those with properties of $z=3$ systems.  As we have noted earlier, a merger with progenitors representative of $z=3$ galaxies in the Mo, Mao \& White (1998) formalism, is less massive, even if it has the same virial velocity as a $z=0$ scaled merger.  A more massive progenitor, even with disk properties of $z=3$ simulations, can have a larger extent - hence, this mass degeneracy prohibits direct comparison between morphological structure seen in high resolution images and the the prevalence of mergers with progenitors at a given redshift.  We note also that Genzel et al. (2003) studied a bright, lensed SMG at $z=2.8$ to find a rotation velocity in excess of $400~\rm km/s$, and millimeter emission in a disk-like structure extending to about 10 kpc.  These properties seem to be akin to our h320, h500 simulations of co-planar mergers.  It is important to note here that a disk-like image is a natural outcome of co-planar mergers (though such a orbital inclination is expected to be rare); i.e., the appearance of a disk-like structure does not imply that the system had not undergone a major merger in the past.  While there have been a number of recent papers (Pope et al. 2006; Iono et al. 2006) reporting high resolution imaging of SMGs, detailed modeling of individual sources requires kinematic information along the lines obtained by Genzel et al. (2006) for a $z \sim 3$ optically bright galaxy inferred to have a star formation rate of $\sim 100~\rm M_{\odot}/yr$ and a circular velocity of $\sim 230~\rm km/s$.  
  
\section{Discussion}

In comparing our predictions to observed data, there are several important points of note.  First, we have depicted the IRAC and infrared X-ray correlations in the rest-frame, which leads to a clear dynamical interpretation, as we have discussed previously.  To test these predictions will require a large number of observations in a narrow redshift slice.  Second, in comparing the $850~\micron$ fluxes (which have been shown at $z=2$, unless otherwise noted) to observations, it will be necessary to disentangle the effects of lensing, and of multiple counterparts contributing to the wide SCUBA field $850~\micron$ flux.  Moreover, a detailed study of current observational biases is needed.  We defer this study to a future paper.  Here, we point out a few basic observational comparisons - the IRAC color-color plot for galaxies placed at $z=0.3-2$ in the $\it{observed}$ frame; the time-evolution of a rest-frame SED; a comparison of our SEDs to recent data obtained by Kovacs et al. (2006); the $850~\micron$ flux (shown at $z=2$) as function of the ratio of black hole luminosity to total luminosity, and as a function of the X-ray luminosity.  

\begin{figure} \begin{center}
\centerline{\psfig{file=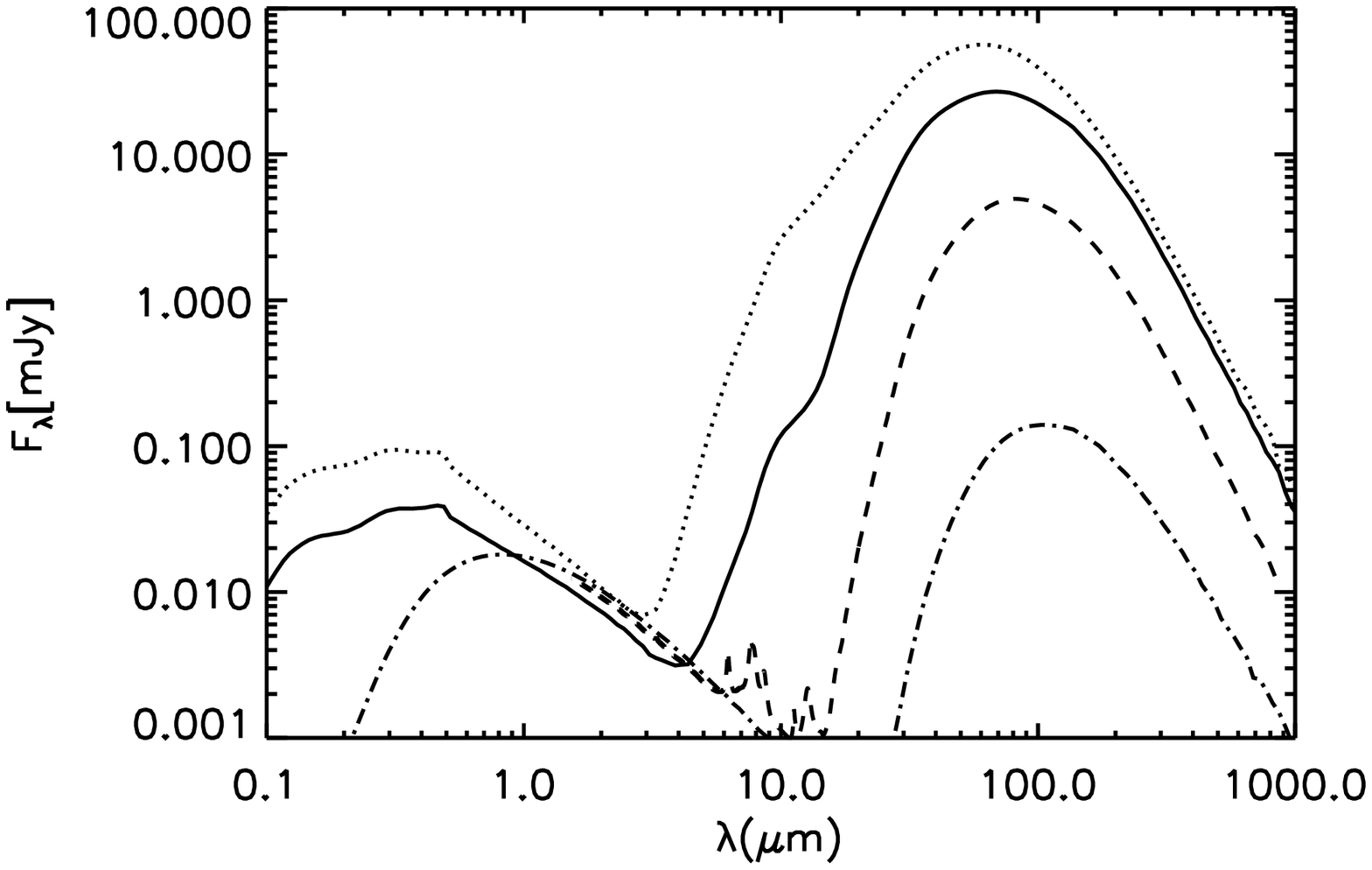,width=3.7in}}
\end{center}
\caption{Time Evolution of a Rest-Frame SED for e320.  As the AGN feedback clears out the obscuring material, more of the emitted energy shifts from the longer wavelengths to the shorter wavelengths (compare the solid line which is the pre-merger phase to the dash-dotted line which is from the post-feedback phase).}\label{fig:20}
\end{figure}

\begin{figure} \begin{center}
\centerline{\psfig{file=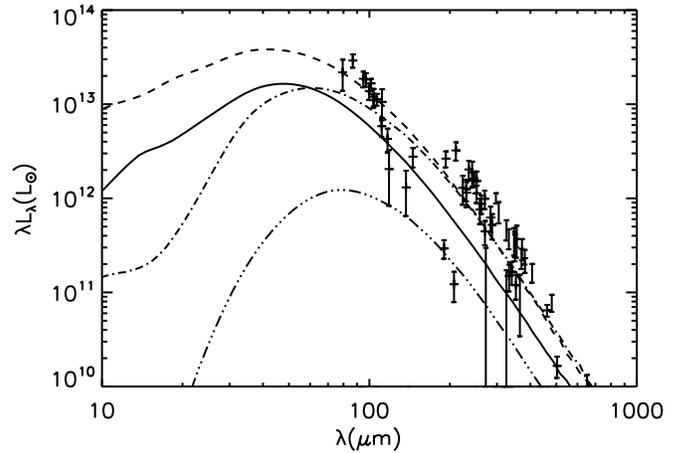,width=3.7in}}
\end{center}
\caption{Comparison to SHARC-2 and SCUBA data reported in Kovacs et al., astro-ph/0604591.  Dash-double-dotted line is the h160 simulation, solid line is e226, dashed line is e320, and dash-dotted line is e500; all are shown close to the peak of their sub-mm bright phase.}\label{fig:21}
\end{figure}

\begin{figure}\begin{center}
\centerline{\psfig{file=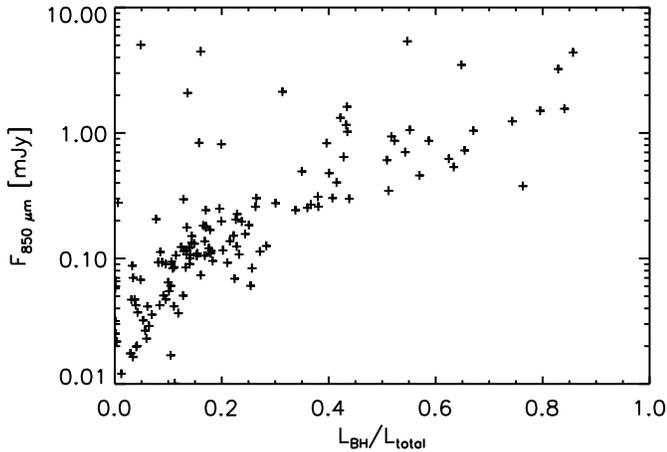,width=3.7in}}
\end{center}
\caption{The observed $F_{850~\micron}$ flux (mJy) as a function of the ratio of black hole luminosity to total luminosity.}\label{fig:22}
\end{figure}

\begin{figure} \begin{center}
\centerline{\psfig{file=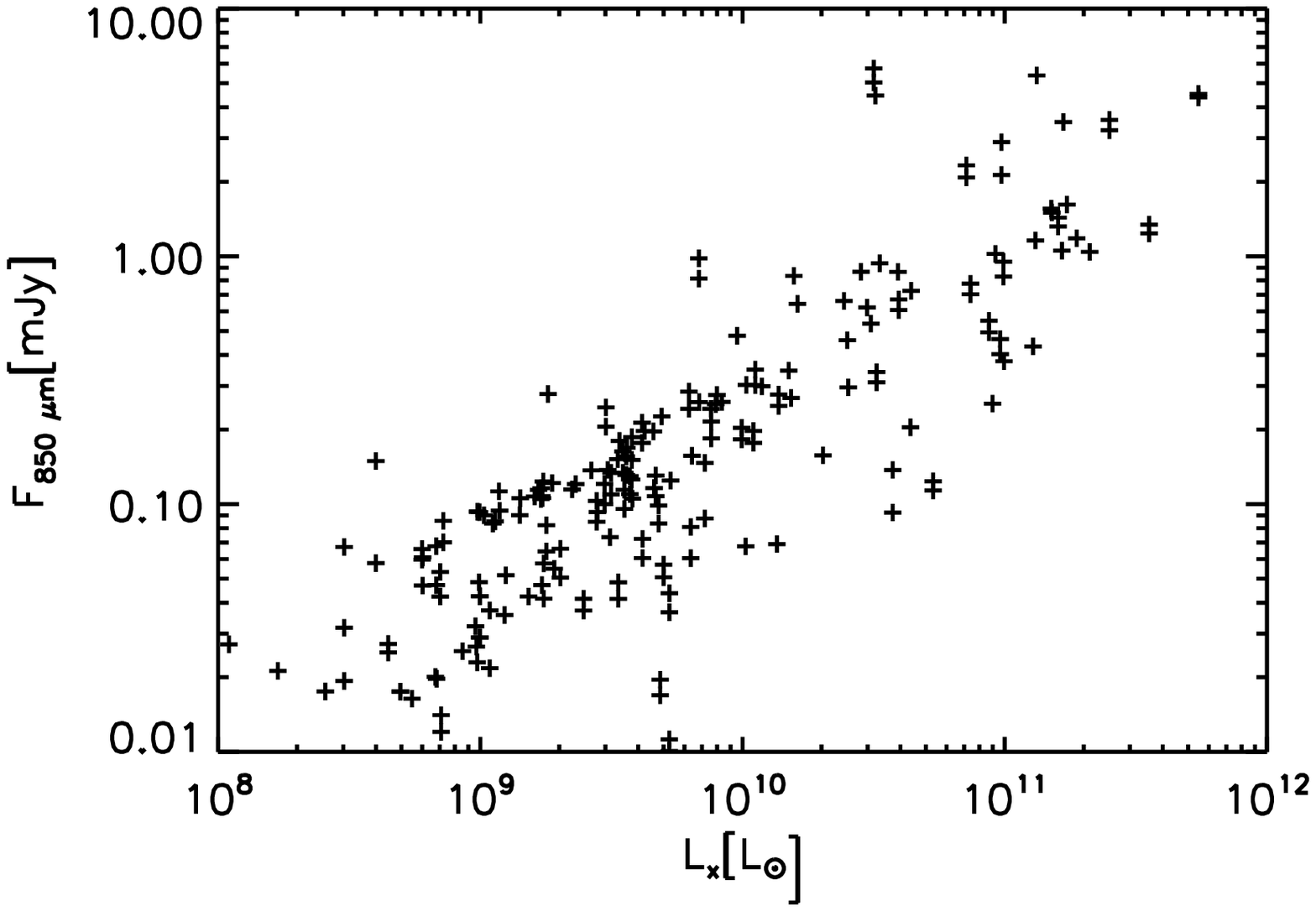,width=3.7in}}
\end{center}
\caption{The observed $F_{850~\micron}$ flux (mJy) as a function of the X-ray luminosity.}\label{fig:23}
\end{figure}

Figure 20 shows the time evolution of the rest-frame SED for the A5e simulation.  As the simulation progresses, more of the energy is distributed to shorter wavelengths, in a fashion similar to the time evolution of the SEDs for local ULIRGs as discussed by Chakrabarti et al. (2006a).  Figures 14-19 depict this same trend in the surface brightness maps, as the apparent brightness in the $3.6~\micron$ band increases as the brightness in the longer wavelength $450~\micron$ band decreases.  In Figure 21, we compare the SEDs from the simulations h160, e226, e320, and e500 during the bright sub-mm phase, to observed SMG data obtained by Kovacs et al. (2006), and find reasonable agreement.  In our simulations, except for mergers of very massive systems ($V_{\rm vir} \ga 400 ~\rm~km/s$), the brighter SMGs ($F_{850~\micron} \ga 1 ~\rm mJy$) at $z \sim 2$, do have non-negligible contribution from the black hole to the total bolometric luminosity.  Kovacs et al. (2006) use the radio far-infrared correlation to find that the SMGs in their sample would not have a significant contribution from the black hole luminosity.  Their sample is biased towards radio-loud sources, and does not include radio-quiet AGN that would contribute to the far-infrared luminosity, increasing the derived $q_{L}$ parameter, which is a measure of the far-infrared to radio luminosity.  To better quantify our prediction for the variation of the $850~\micron$ fluxes as a function of the black luminosity, we show in Figures 22 and 23 the $850~\micron$ flux as a function of the ratio of the black hole luminosity to the total bolometric luminosity and as a function of the X-ray luminosity.   Figure 22 shows that there is a general trend for the $850~\micron$ flux to increase as a function of the ratio of the black hole luminosity to the total bolometric luminosity.  The brighter SMGs which have low $L_{\rm BH}/L_{\rm total}$ are the more massive systems.  A significant caveat to interpreting observations however, is a Compton thick population, which the current simulations cannot model in detail.  

\begin{figure*} \begin{center}
\centerline{\psfig{file=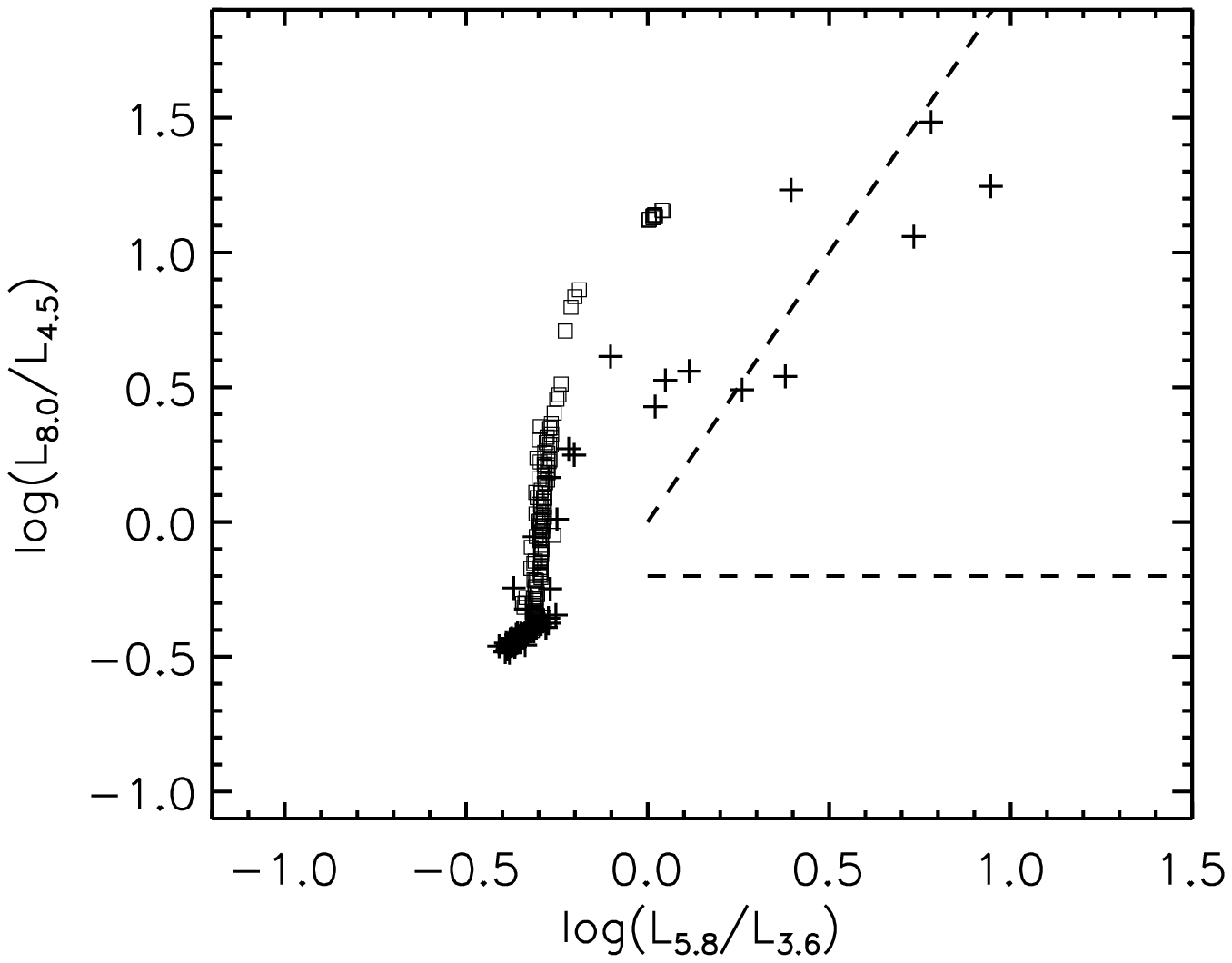,width=2.5in}
\psfig{file=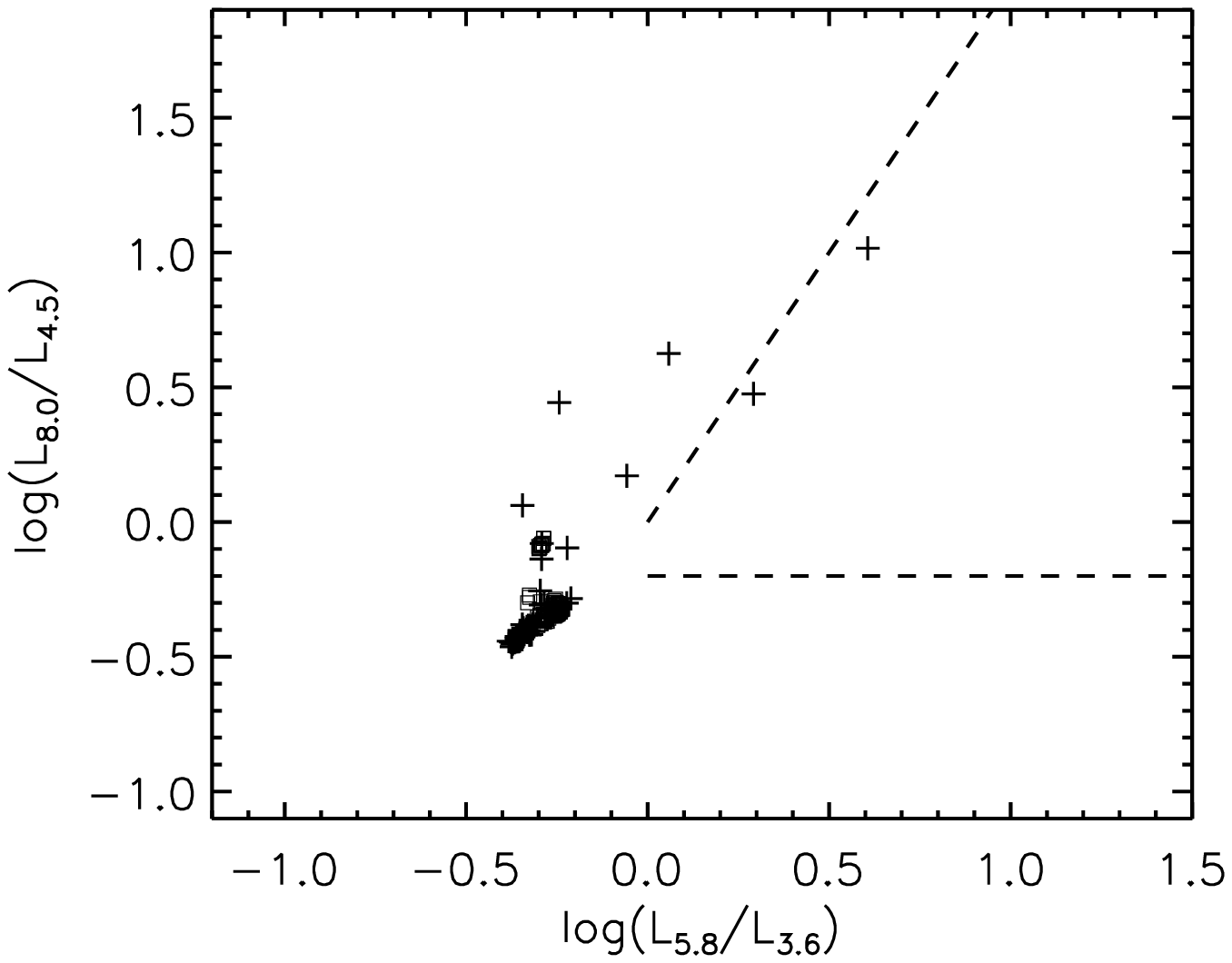,width=2.5in}
{\psfig{file=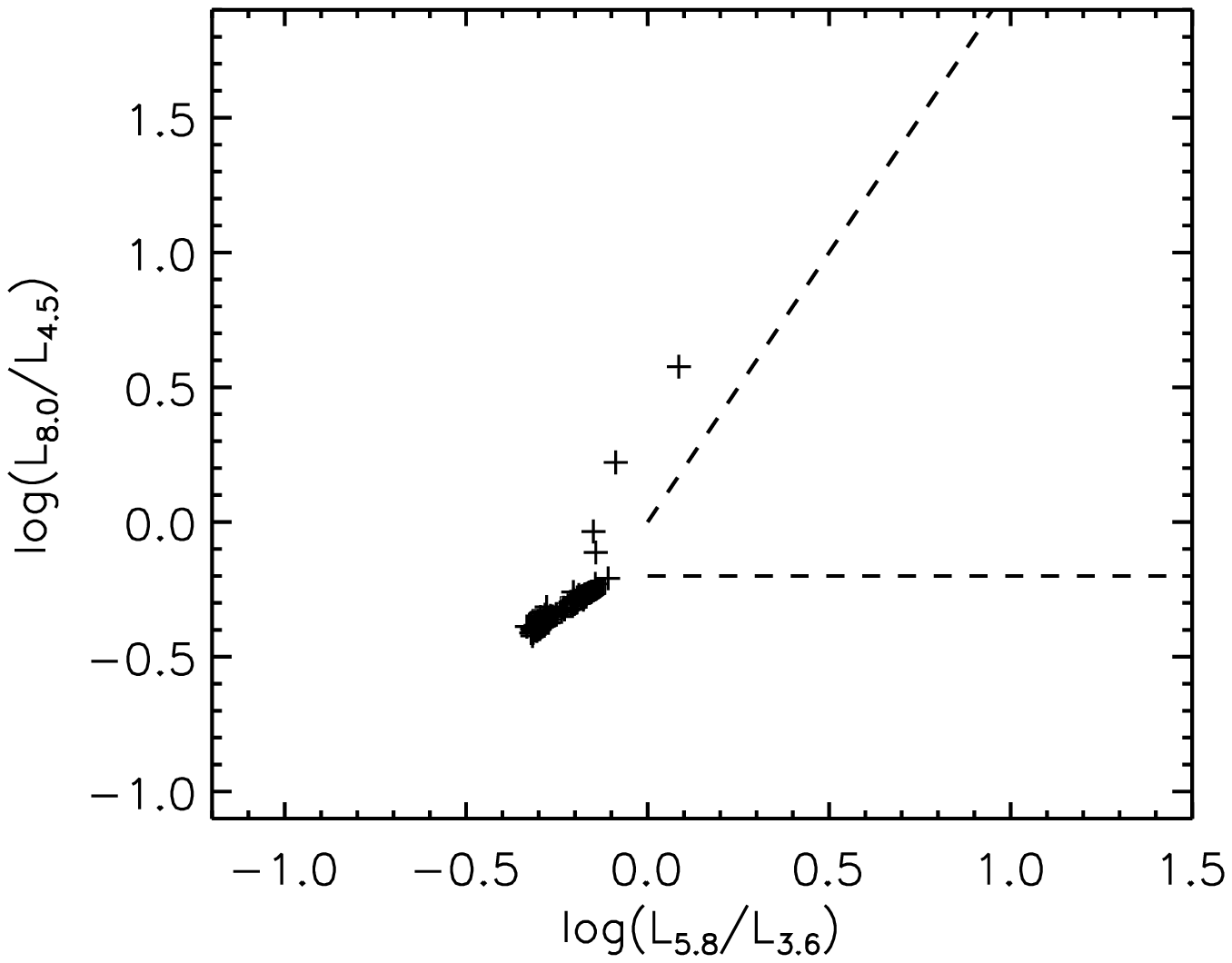,width=2.5in}}}
\end{center}
\caption{IRAC color-color plot in $\it{observed}$ frame, with all simulations placed at a) $z=0.3$, b) $z=1$, c) $z=2$.  Crosses designate the continuum and squares the combined spectrum, i.e., the continuum and the PAH model, as before.  Note the slightly ``bunny-ear'' shape of the IRAC color-color plot for the $z=0.3$ slice.}\label{fig:24}
\end{figure*}

We show in Figure 24 the IRAC color-color plot in the $\it{observed}$ frame for simulations placed at $z=0.3$, 1, and 2.   The PAH spectral features redshift out of the IRAC bands by $z \sim 0.5$, and the relative increase in $L_{8~\micron}/L_{4.5~\micron}$ for energetically active AGN also can no longer be seen.  However, the transition from a roughly spherical distribution of colors for low-redshift objects to a flattened elliptical distribution of colors for $z\sim 2$ objects may be useful for interpreting observations.  As mentioned previously, the $z=0.3$ IRAC color-color plot displays the characteristic ``bunny-ear'' shape seen in the observed FLS color-color plot.  We show the $z=0.3$ case here as Lacy et al. (2004) note that the median photometric redshift of their candidate AGN is $\sim 0.3$.  This contrasts with the rest-frame IRAC color-color plot shown in Figure 11 primarily owing to the redshifting of the PAH template, which is a more sensitive function of redshift than the continuum.  Ultimately, it will be useful to simulate observed color-color plots for a particular population which probe a range of redshifts.  Work is underway to analyze current observational biases to realize this.

Since the far-infrared ($L_{\rm IR}$, $L_{\rm 70}~\micron$) and to a lesser extent $L_{\rm 24}~\micron$, are well correlated with the X-ray luminosity, we review them here and give relations between these quantities.  The red dots in Figure 25 identify cases which have $F_{850~\micron} > ~\rm 1 mJy$.  The diagonal lines show constant $L_{i}/L_{\rm x}$; the best fit lines (the solid lines in these figures) are $L_{70~\micron}=40~L_{\rm x}$ and $L_{24~\micron}=10~L_{\rm x}$ respectively; there is nearly an order of magnitude of scatter relative to this line.            

\begin{figure*} \begin{center}
\centerline{\psfig{file=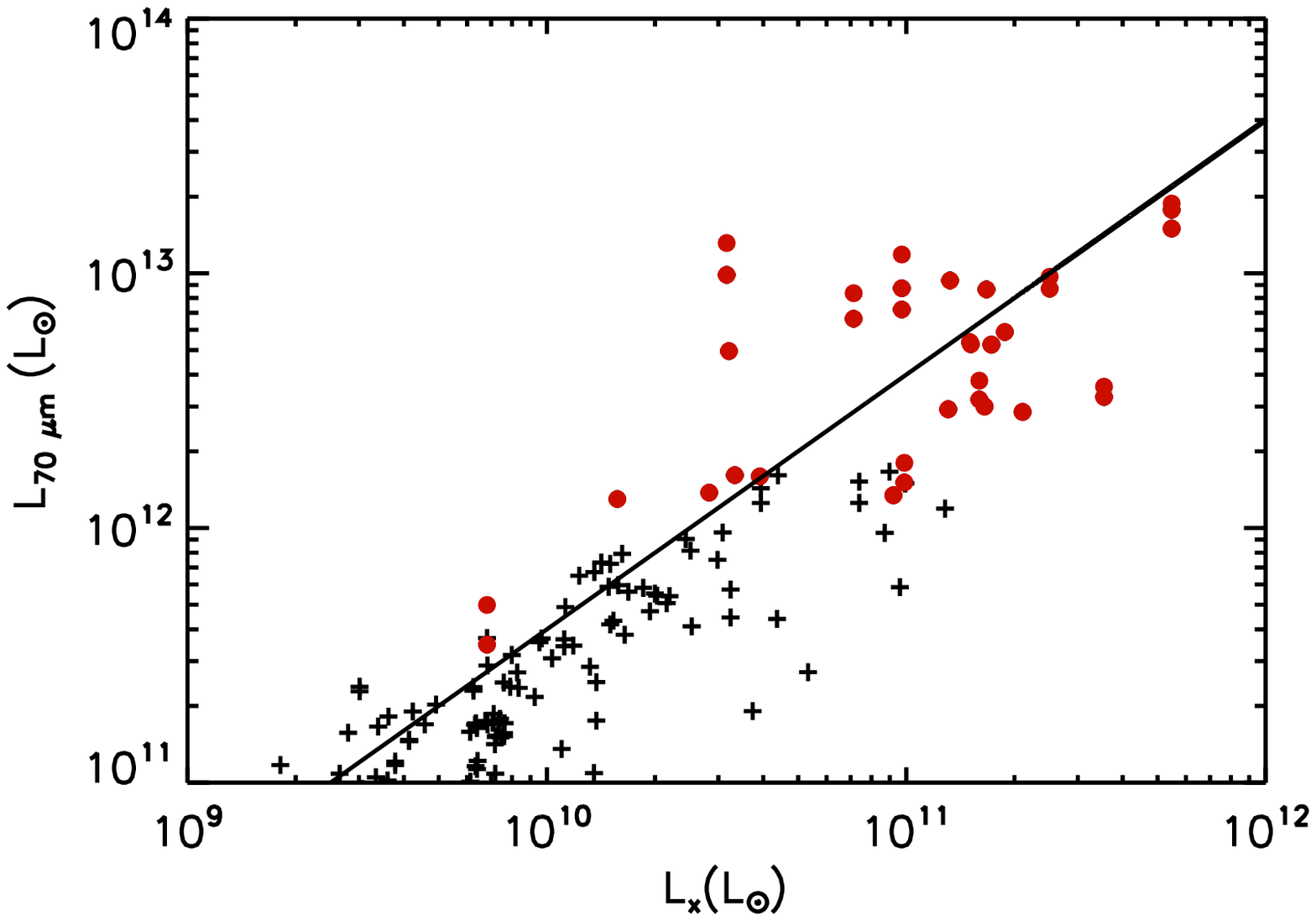,width=3.7in}
\psfig{file=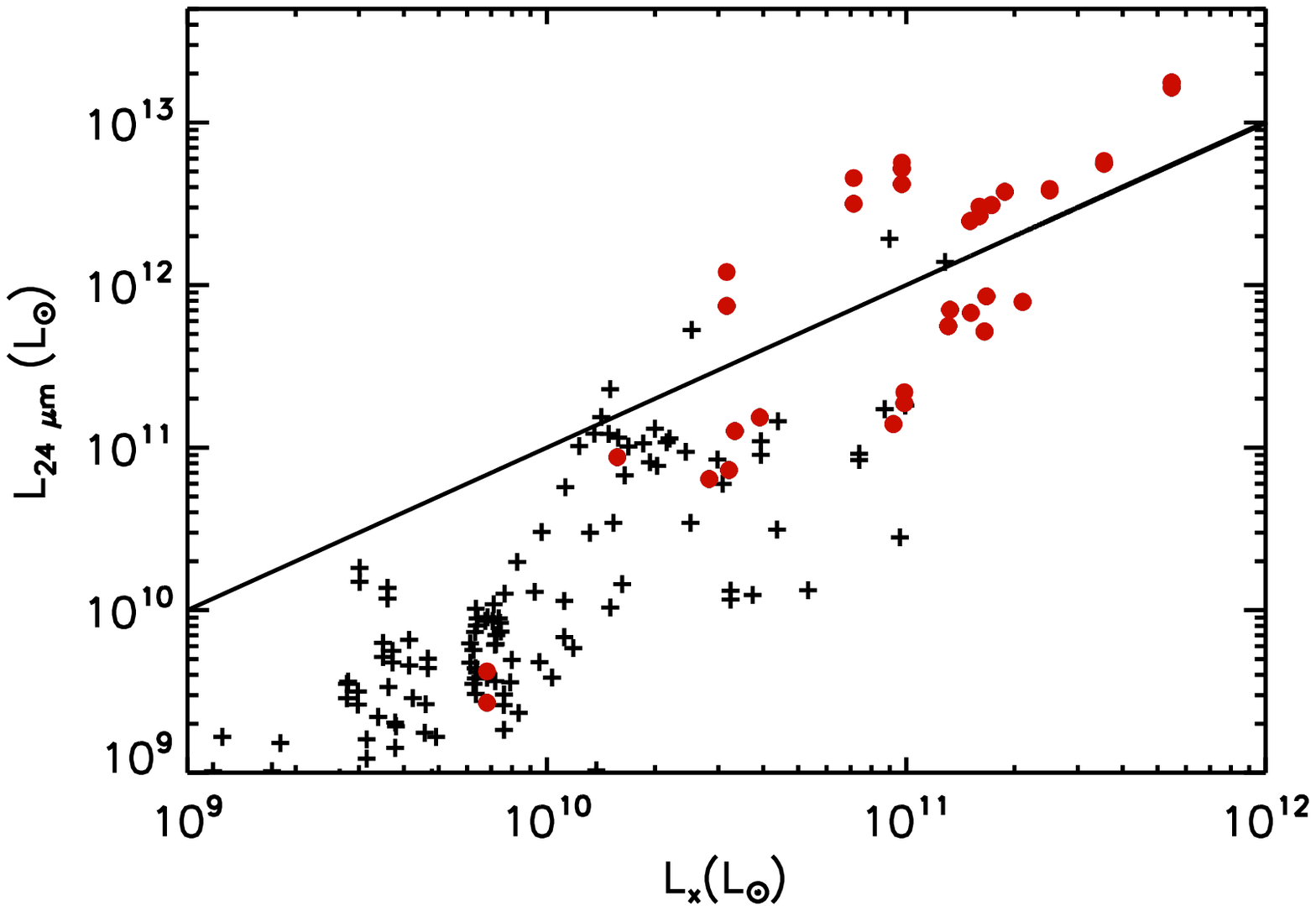,width=3.7in}}
\end{center}
\caption{(a) $L_{70~\micron}$ (rest-frame) vs. the hard X-ray luminosity, and (b) $L_{24~\micron}$ (rest-frame) vs. the hard X-ray luminosity.  Red dots denote cases which have $F_{850~\micron} > 1 \rm mJy$.  The solid lines correspond to $L_{70~\micron}=40~L_{\rm x}$ and $L_{24~\micron}=10~L_{\rm x}$.}\label{fig:25}
\end{figure*}

\begin{figure*} \begin{center}
\centerline{\psfig{file=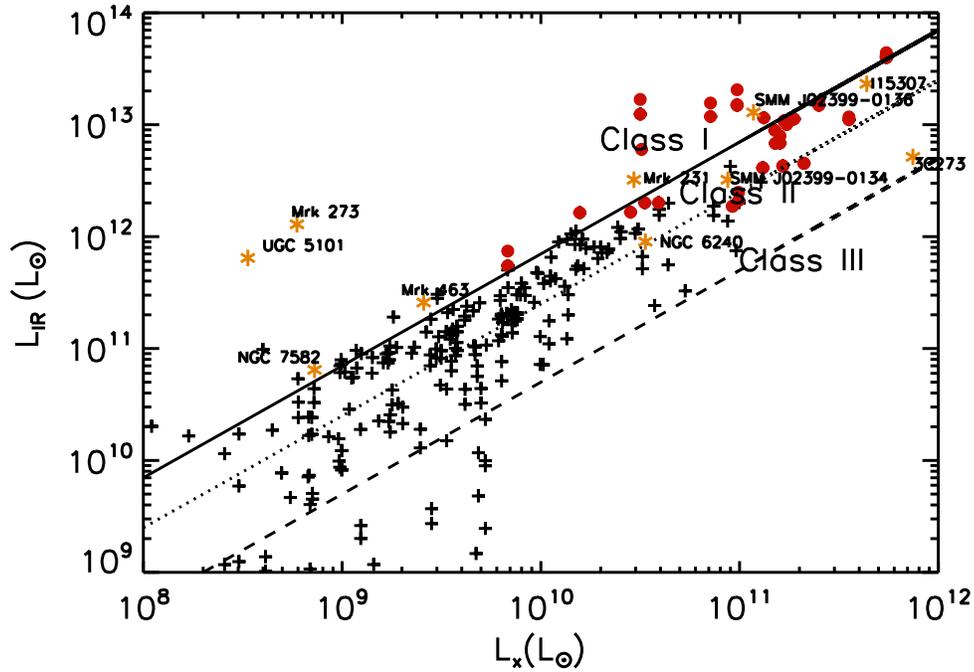,width=5.5in}}
\end{center}
\caption{The Class I, Class II, Class III designation scheme for SMGs: $L_{\rm IR} \ga 70~ L_{\rm x}$ is Class I, $L_{\rm IR} \ga 25~ L_{\rm x}$ is Class II, $L_{\rm IR} \la 5 ~L_{\rm x}$ is Class III.  Points shown in red are those that have $F_{850~\micron} \ga 1~\rm mJy$ at $z=2$; i.e., would be empirically designated as SMGs.  Yellow asterisks show observed sources from the literature.}\label{fig:26}
\end{figure*}

\begin{figure*} \begin{center}
\centerline{\psfig{file=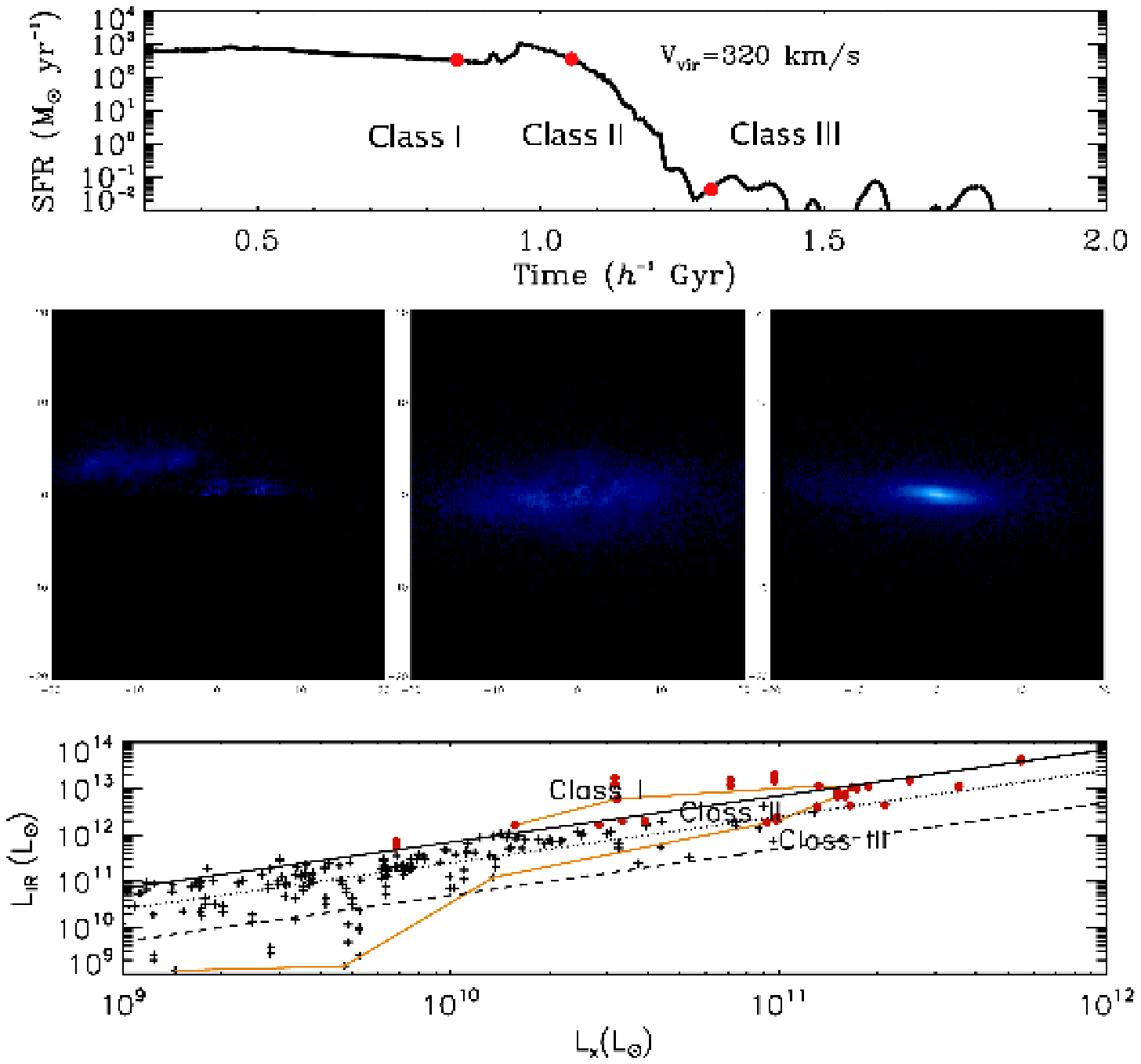}}
\end{center}
\caption{An Evolutionary Model for SMGs: (a) top panel shows the time evolution of the star formation rate, with the Class I $\rightarrow$ Class II $\rightarrow$ Class III time evolution marked.  Red dots mark the times of the images shown below.  (b) middle panel shows the $3.6~\micron$ images of h320 along this time sequence.  (c) bottom panel shows the time evolution on the $L_{\rm IR}-L_{\rm x}$ plot in yellow for this simulation.  As time progress, $L_{\rm IR}/L_{\rm x}$ increases first as the galaxies merge and the black hole becomes energetically active, and decreases as black hole feedback expels the obscuring material, ultimately producing a merger remnant.}\label{fig:27}
\end{figure*}

The total integrated infrared luminosity and the hard X-ray luminosity are essentially global properties of these simulations.  Hence, we investigate where SMGs from our merger simulations would lie on a $L_{\rm IR}-L_{\rm x}$ plane.  Figure 26 marks in red the location of SMGs (i.e., $F_{850~\micron} > 1~\rm mJy$ for simulations placed at $z=2$, at a luminosity distance of 15.5 Gpc) on the $L_{\rm IR}-L_{\rm x}$ plane, with divisions into Class I, Class II, and Class III, which correspond to $L_{\rm IR} \ga 70~ L_{\rm x}$ (the best-fit line that goes through our simulated data), $L_{\rm IR} \ga 25~ L_{\rm x}$ (the traditional quasar line, e.g., Elvis et al. 1994), and $L_{\rm IR} \la 5 ~L_{\rm x}$ respectively.  Most of the SMGs studied by Alexander et al. (2005a) were found to have $L_{\rm IR} \sim 200 L_{\rm x}$, on the basis of the radio far-IR correlation.  Indeed, quite a few from our simulations have similar $L_{\rm IR}/L_{\rm x}$ values, though most have lower values.  Interestingly, samples of quasars (e.g. Page et al. 2001) have similar $F_{850~\micron}$ values as SCUBA selected galaxies, which suggests that SMGs as defined by their $850~\micron$ fluxes are a diverse and broad class of objects, as already highlighted by observational studies on the basis of their radio and optical properties (Ivison et al. 2000).  Figure 26 demonstrates the diversity of SMGs, which traverse the Class I-Class II divide - some SMGs would fall on the traditional quasar line ($L_{\rm IR} \ga 25 L_{\rm x}$), yet many have higher values of $L_{\rm IR}/L_{\rm x}$.  This shows explicitly that SMGs are a broader class of systems than quasars or starbursts (which have higher $L_{\rm IR}/L_{\rm x}$), as selecting on the basis of the $F_{850~\micron}$ values allows these objects to span a broad range in $L_{\rm IR}/L_{\rm x}$.  We also show in Figure 26 a few observed sources from the literature (shown in yellow asterisks), comprised of local LIRGs (Mrk 463), ULIRGs (e.g., Mrk 231), quasars (e.g., 3C273), and well-known SMGs (e.g., SMM J02399-0136) (as reported by Sanders \& Mirabel 1996; Hughes et al. 1997; Bautz et al. 2000; Genzel et al. 2003; Alexander et al. 2005a).  Since we do not do a detailed comparison to observations in this paper, this plot certainly should not be interpreted to imply an exact correspondence between our simulated data points and observed data in cases of overlap.  Nonetheless, it is clear that $z \sim 2$ systems such as SMM J02399-1036 (as studied in by Genzel et al. 2003) do have similar $L_{\rm IR}/L_{\rm x}$ ratios as the systems in our simulations, which suggests that the formation of such objects at $z \sim 2$ may be described by major mergers, whose time evolution is influenced by feedback from the central AGN.  As mentioned previously, the current merger simulations do not resolve the inner few parsecs, which may well be Compton thick and attenuate even the hard X-rays.  However, Figure 26 does show that quasars like 3C273 in elliptical (McLeod \& Rieke 1994b) host galaxies (though see recent work by Martel et al. 2003 which identifies intricate features in the host galaxy not captured in early HST images), which are presumably more evolved, do lie closer to the Class III region than Mrk 463 or UGC 5101, which are interpreted to be in an earlier stage of a merger (Sanders \& Mirabel 1996).    

A natural question to ask at this point is whether this classification scheme may be interpreted also as an evolutionary scheme within the context of the merger simulations studied here.  Figure 27 presents the time evolution of the star formation rate, with the three classes marked, along with the time evolution of the $3.6~\micron$ image, and shows the track that a particular simulation (h320) would follow on the $L_{\rm IR}-L_{\rm x}$ diagram (marked in yellow in the bottom panel of the $L_{\rm IR}/L_{\rm x}$ diagram).  As shown, the system starts out at high values of $L_{\rm IR}/L_{\rm x} \sim 100$, and then after reaching a peak in $L_{\rm x}$, progressively moves down in $L_{\rm IR}/L_{\rm x}$ towards the bottom portion of this diagram, which we have marked as Class III, and which we associate with elliptical merger remnants.  (Other simulations evolve in a generally similar manner.) The $L_{\rm IR}/L_{\rm x}$ values for ellipticals are not well determined, unfortunately.  Extrapolating from SCUBA observations of a sample of elliptical galaxies (Di Matteo et al. 1999), we suggest that $L_{\rm IR} \sim 5 L_{\rm x}$ may represent an upper bound for ellipticals.  We emphasize that the Class III designation is highly uncertain owing to the lack of a determination for $L_{\rm IR}$ for merger remnants (though see Temi et al. 2005 for a discussion of a correlation between the mid-IR emission and age of ellipticals), as well as the wide range in X-ray luminosities (Mathews et al. 2006; O'Sullivan et al. 2001, 2004; Allen et al. 2000; Fabbiano 1989) and complex range of phenomena that produce X-ray luminosities in ellipticals, i.e., X-ray emission from hot gas, black hole, or residual star formation.  Nonetheless, the increase in $L_{\rm x}/L_{\rm B}$ with age (Mackie \& Fabbiano 1997; O'Sullivan et al. 2001) does suggest that the amount of X-ray emitting hot gas increases with time relative to the amount of cold (infrared emitting) gas.  It will also be useful to directly compare correlations between photometric and kinematic correlations for merger remnants, as explored recently by Rothberg \& Joseph (2006) to better quantify this tentative evolutionary scheme.  We study correlations between kinematic and multiwavelength photometric properties in a future paper, and analyze a larger set of simulations to quantitatively address the transition from SMGs to ellipticals.  From our preliminary analysis here, Figure 27 does suggest that the evolution of SMGs, from the pre-merger phase to the merger remnant, may be qualitatively understood in a relatively simple manner, wherein the system occupies some characteristic region of the $L_{\rm IR}-L_{\rm x}$ diagram during each phase of its lifetime.        

\section{Conclusion}

$\bullet$ Our simulations of gas-rich major mergers with black hole feedback naturally lead to the production of SMGs, which match photometric observations.  The SMGs formed in these simulations have star formation
rates of $\sim 500-1000 ~\rm M_{\odot}/yr$, infrared luminosities of $\sim 1-5 \times 10^{12} L_{\odot}$, and virial velocities of $\sim 300-400 ~\rm km/s$.

$\bullet$ We comment on the $M_{\rm star}- M_{\rm BH}$ relation for SMGs, and its inference from observations.  We do find that the black holes in SMGs are in a rapidly growing phase, and grow by 
factors $\sim 5$ between $z \sim 2$ and the present day.  However, we do not find that SMGs at redshift 2 would lie two orders of magnitude below the local relation which is an extreme lower limit that follows from deriving the black hole masses under the assumption of Eddington-limited accretion.

$\bullet$ We demonstrate that clustering in IRAC color-color space can be naturally
explained within the context of the merger simulations studied here; clustering
in this context translates to the system spending more of its lifetime
in a given region of color-color space; it is also positively correlated with
the stars dominating in their contribution to the bolometric luminosity.  We 
recover a similar percentage of sources occupying the AGN-demarcated region
as in the Lacy et al. (2004) plot.  We use a phenomenological model for the PAH emission to demonstrate the change in color-color plot when a PAH template is included in calculating the colors, both in the rest-frame, and in the observed frame for galaxies from $z=0.3-2$.

$\bullet$  Our models would predict an inherent similarity between color (not fluxes)
 evolution
for local ULIRGs and high redshift luminous galaxies, which is established
largely by the dynamical effect of feedback from the AGN in these simulations.
The extension of this model to high redshifts ($z\ga 6$) would suggest 
that these correlations would also hold for the highest redshift
galaxies known.

$\bullet$ We predict a correlation between the rest-frame infrared, $70~\micron$, and $24~\micron$ luminosity
and the hard X-ray luminosity for SMGs, with an increase in scatter at low X-ray luminosities.  To aid observational studies, we quantify the far-infrared X-ray correlations.  These predictions will be directly testable by future instruments, such 
as Herschel, and possibly through source stacking analysis using current Spitzer data.  

$\bullet$ Our photo albums of the lifetimes of SMGs visually illustrate the 
differential variation in surface brightness between the SCUBA and IRAC bands.
Of particular note is the increase in apparent brightness in the IRAC bands,
which is concomitant with a decrease in brightness in the SCUBA bands, towards
the late phases of the merger.  This shows that more and more of the emitted
energy shifts to the shorter wavelengths as AGN feedback disperses
the obscuring material.  We also demonstrate that the morphology as
seen in these bands is partly a function of orbital inclination, with co-planar
mergers producing disk-like morphologies in the active phase.  As such, even galaxies with 
observed disk-like morphologies may have experienced major (co-planar) mergers.  The simulations of merger with progenitors at high redshifts lead to more compact morphologies than progenitors of lower redshift systems, which generally have emission extending to $\sim 10 ~\rm kpc$ in the active phase in the IRAC and SCUBA bands.

$\bullet$ We find that our predicted SEDs are in good agreement with recent
multiwavelength photometry.  We show the IRAC color-color plot 
for galaxies in a narrow redshift slice, with $z \sim 2$, which results 
in a flattened elliptical distribution of colors unlike the nearly spherical distribution 
at low redshifts.   

$\bullet$ We depict the variation of the observed $850~\micron$ flux as a function of 
both the X-ray luminosity and the ratio of the black hole luminosity to the total bolometric 
luminosity.  There is a general trend for SMGs on the bright end at $z \sim 2$ to have a significant ($\ga 50\%$) contribution from the black hole to their total bolometric luminosity.  

$\bullet$ We stress that the correlations and photometric properties of SMGs listed here are predictions of our model and can be observationally tested - by obtaining a large sample ($\sim 100$) of observations in a narrow redshift range.

$\bullet$ We find that SMGs are a broader class of systems than quasars or starbursts.  We introduce a simple, heuristic classification scheme for SMGs on the basis of their $L_{\rm IR}/L_{\rm x}$ ratios,  $L_{\rm IR} \ga 70~ L_{\rm x}$ is Class I, $L_{\rm IR} \ga 25~ L_{\rm x}$ is Class II, $L_{\rm IR} \la 5 ~L_{\rm x}$ is Class III.  We suggest that this may also be interpreted as an evolutionary scheme as SMGs transit from the pre-merger stage through the quasar phase to merger remnants.

\bigskip
\bigskip
\acknowledgements 

We thank George Rybicki and Barbara Whitney for helpful discussions on
scattering processes.  We thank Giovanni Fazio, Pauline Barmby and Tiziana Di Matteo for
helpful discussions.  We especially thank Jiasheng Huang for many
helpful discussions on the observations of (and interpretations of)
SMGs.  We also thank Volker Springel for constructive feedback on the simulations
discussed in this paper.  SC is supported by an NSF Postdoctoral Fellowship.  The
simulations were performed on the Institute for Theory and
Computation Cluster.


\begin{references}

\reference{} Allamandola, L.J., Sandford, S.A., et al., 1992, ApJ, 399, 134

\reference{} Alexander, D.M., Bauer, F.E., Chapman, S.C., et al., 2005a, ApJ, 632, 736

\reference{} Alexander, D.M., et al., 2005b, Nature, 434, 738

\reference{} Allen, S.W., Di Matteo, T., \& Fabian, A.C., 2000, MNRAS, 311, 493

\reference{} Alonso-Herrero et al. 2004, ApJS, 154, 155A

\reference{} Baugh, C.M., et al., 2005, MNRAS, 356, 1191

\reference{} Barmby, P., Alonso-Herroro, A., Donley, J.L., et al., 2006, ApJ, 642, 126

\reference{} Barnes, J.E. \& Hernquist, L.E., 1991, ApJ, 370L, 65

\reference{} Barnes, J.E., \& Hernquist, L.E., 1996, ApJ, 471, 115

\reference{} Blain, A.W., Smail, I., Ivison, R.J., et al., 2002, PhR, 369, 111

\reference{} Bautz, M.W., et al., 2000, ApJ, 543L, 119B

\reference{} Blain, A.W., et al., 2004a, ApJ, 611, 725

\reference{} Blain, A.W., et al., 2004b, ApJ, 611, 52

\reference{} Blitz, L., Fukui, Y., Kawamura, A., et al., astro-ph/0602600

\reference{} Bjorkman, J.E., \& Wood, K., 2001, ApJ, 554, 615

\reference{} Borys, C., Smail, I., Chapman, S.C., et al. 2005, ApJ, 635, 853

\reference{} Bullock, J., et al., 2001, MNRAS, 321, 559

\reference{} Chakrabarti, S., \& McKee, C.F.M., 2005, ApJ, 631, 792

\reference{} Chakrabarti, S., Cox, T.J., Hernquist, L., Hopkins, P.F., et al., 2006a, accepted to ApJ, astro-ph/0605652

\reference{} Chakrabarti, S., \& McKee, C.F.M., 2006, in preparation

\reference{} Chakrabarti, S., \& Whitney, B.A., 2006, in preparation

\reference{} Chapman, S.C., Windhorst, R., et al., 2003a, ApJ, 599, 92

\reference{} Chapman, S.C., Blain, A.W., et al., 2003b, Nature, 422, 695

\reference{} Chapman, S.C., Smail, I., et al., 2004, ApJ, 611, 732

\reference{} Chapman, S.C., Blain, A.W., et al., 2005, ApJ, 622, 772

\reference{} Comastri, A., Setti, G., et al., 1995, A\&A, 296, 1

\reference{} Cox, T.J., Chakrabarti, S., Di Matteo, T., et al., 2006a, in preparation

\reference{} Cox, T.J., et al., 2006b, ApJ, 643, 692

\reference{} Croom, S.M., Boyle, B.J., Shanks, T., et al., 2005, MNRAS, 356, 415C

\reference{} Dale, D.A., \& Helou, G., 2002, ApJ, 576, 159

\reference{} Dasyra, K.M, Tacconi, L.J., et al. 2006, ApJ, 638, 745

\reference{} De Grijp et al. 1985, Nature, 314, 240

\reference{} Di Matteo, T., Springel, V., \& Hernquist, L., 2005, Nature, 433, 604

\reference{} Di Matteo, T., Fabian, A.C., et al., 1999, MNRAS, 305, 492

\reference{} Dopita, M.A., Groves, B.A., et al., 2005, ApJ, 619, 755

\reference{} Dunne, L., Eales., S. et al., 2000, MNRAS, 315, 115

\reference{} Dunne, L., \& Eales, S.A., 2001, MNRAS, 327, 697

\reference{} Efstathiou, A., \& Rowan-Robinson, M., 2003, MNRAS, 343, 322

\reference{} Elvis, M., et al., 1994, ApJS, 95,1

\reference{} Fabbiano, G., 1989, Annual Reviews of Astronomy \& Astrophysics, 27:87-138

\reference{} Farrah, D., et al., 2002, MNRAS, 335, 1163

\reference{} George, I.M., Turner, T.J., et al. 1998, ApJS, 114, 73

\reference{} Genzel, R., Baker, A.J., et al., 2003, ApJ, 584, 633

\reference{} Genzel, R., Tacconi, L.J., et al., 2006, Nature, 442, 786

\reference{} Gilfanov, M., et al., 2004, MNRAS, 347L, 57G

\reference{} Goldader, J.D., Meurer, G., Heckman, T.M, et al. 2002, ApJ, 568, 651

\reference{} Greve, T.R., Ivison, R.J., \& Papadopoulos, P.P., 2004, A\& A, 419, 99

\reference{} Hopkins, P., et al. 2005a, ApJL, 625, L71

\reference{} Hopkins, P., et al. 2005b, ApJ, 630, 716

\reference{} Hopkins, P., et al. 2005c, ApJ, 632, 81

\reference{} Hopkins, P., et al. 2006a, ApJS, 163, 1

\reference{} Hopkins, P., et al. 2006b, ApJS, 163, 50

\reference{} Hopkins, P., et al. 2006c,  ApJ, submitted, astro-ph/0602290

\reference{} Hopkins, P., et al. 2006d,  ApJ, 639, 700

\reference{} Hopkins, P. \& Hernquist, L. 2006,  ApJS, 166, 1

\reference{} Hopkins, P., Richards, G., \& Hernquist, L. 2006,  ApJ, in press [astro-ph/0605678] [HRH]

\reference{} Hughes, D.H., et al., 1997, MNRAS, 289, 766H

\reference{} Iono, D., Peck, A.B., Pope, A., et al., 2006, ApJ, 640L, 1

\reference{} Ivison, R.J., Smail, I., Le Borgne, J.-F., et al., 1998, MNRAS, 298, 583

\reference{} Ivison, R.J., Smail, I., Barger, A.J., et al., 2000, MNRAS, 315, 209

\reference{} Kennicutt, R.C., 1998, ApJ, 498, 541

\reference{} Klaas, U., Haas, M., et al., 2001, A\&A, 379, 823

\reference{} Kovacs, A., Chapman, S.C., et al., 2006, accepted to ApJ, astro-ph/0604591

\reference{} Lacy, M., Storrie-Lombardi, L.J., Sajina, A., et al., 2004, ApJS, 154, 166

\reference{} Leitherer, C., Schaerer, S., et al., 1999, ApJS, 123, 3

\reference{} Lidz, A., Hopkins, P.F., et al., 2006, ApJ, 641, 41

\reference{} Lilly, S., 1999, $\it{The~Hy~Redshift~Universe}$, ASP Conference Series, Vol., 193

\reference{} Mackie, G., \& Fabbiano, G., 1997, in Arnaboldi M., Da Costa G.S., Saha P., eds, ASP Conf. Ser. Vol., 116, The Nature of Elliptical Galaxies; 2nd Stromlo Symposium. Astron. Soc. Pac., San Francisco, p. 401

\reference{} Magdziarz, P., \& Zdziarski, A.A., 1995, MNRAS, 273, 837

\reference{} Magorrian, J., Tremaine, S., Richstone, D., et al., 1998, AJ, 115, 2285M

\reference{} Marconi, A., Risaliti, G., et al., 2004, MNRAS, 351, 169

\reference{} Martel, A.R., et al., 2003, AJ, 125, 2964M

\reference{} Mathews, W.G., Brighenti, F., et al., astro-ph/0610694

\reference{} McLeod, K.K., \& Rieke, G.H., 1994,ApJ, 431, 137M

\reference{} Mihos, C.J., \& Hernquist, L., 1994, ApJ, 431L, 9

\reference{} Mihos, C.J., \& Hernquist, L., 1996, ApJ, 464, 641

\reference{} Mo, H.J., Mao, S., \& White, S.D.M., 1998, MNRAS, 295, 319

\reference{} Narayanan, D. et al. 2006, ApJ, submitted

\reference{} Neri, R., Genzel, R., Ivison, R.J., et al., 2003, ApJ, 597L, 113

\reference{} O'Sullivan, E., Forbes, D., A., \& Ponman, T.J., 2001, MNRAS, 2001, 420-426

\reference{} O'Sullivan, E., \& Ponman, T.J., 2004, MNRAS, 349, 535

\reference{} Page, M.J., Stevens, J.A., et al. 2001, Science, 294, 2516

\reference{} Perola, G.C., et al., 2002, A\&A, 389, 802

\reference{} Plume, R., et al., 1997, ApJ, 476, 730P

\reference{} Pollack, J.B., Hollenbach, D., et al., 1994, ApJ, 421, 615

\reference{} Pope, A., Scott, D.., Dickinson, M., et al., 2006, MNRAS, 370, 1185

\reference{} Rigby, J.R., Rieke, G.H., et al., 2004, ApJS, 154, 160

\reference{} Robertson, B., Cox., T.J., Hernquist, L., et al., 2006, ApJ, 641, 21R

\reference{} Rothberg, B., \& Joseph, R.D., 2006, AJ, 132, 976

\reference{} Rowan-Robinson, M., 2000, MNRAS, 316, 885

\reference{} Sadler, E.M., Jackson, C.A., et al., 2002, MNRAS, 329, 227S

\reference{} Sajina, A., Lacy, M., \& Scott, D., 2005, ApJ, 621, 256

\reference{} Sanders, D.B., \& Mirabel., I.F., 1996, ARA\&A, 34, 749S

\reference{} Scoville, N.Z., Evans, A.S., et al., 2000, AJ, 119, 991

\reference{} Silva, L., et al., 1998, ApJ, 509, 103

\reference{} Smail, I., Ivison, R.J., \& Blain, A.W., 1997, ApJL, 490, p. L5

\reference{} Soifer, B.T. et al. 2000, AJ, 119, 509

\reference{} Solomon, P.M., et al., 1997, ApJ, 478, 144S

\reference{} Spoon, H.W.W., Keane, J.V., Tielens, A.G.G.M., et al., 2002, A\&A, 385, 1022

\reference{} Springel, V., 2005, MNRAS, 364, 1105

\reference{} Springel, V. \& Hernquist, L, 2002, MNRAS, 333, 649

\reference{} Springel, V. \& Hernquist, L, 2003a, MNRAS, 339, 289 [SH03]

\reference{} Springel, V. \& Hernquist, L, 2003b, MNRAS, 333, 312

\reference{} Springel, V., di Matteo, T., \& Hernquist, L, 2005b, MNRAS, 361, 776

\reference{} Springel, V., di Matteo, T., \& Hernquist, L, 2005a, ApJ, 620, L79

\reference{} Swinbank, A.M., Smail., I, Chapman, S.C., et al., 2004, ApJ, 617, 64

\reference{} Tacconi, L.J., Neri, R., Chapman, S.C., et al., 2006, ApJ, 640, 228

\reference{} Telfer, R.C., et al., 2002, ApJ, 565, 773

\reference{} Temi, P., et al., 2005, ApJ, 635L, 25T

\reference{} Ueda, Y., Akiyama, M., et al., 2003, ApJ, 598, 886

\reference{} Vazquez, G.A., \& Leitherer, C., 2005, ApJ, 621, 695

\reference{} Vignali, C., Brandt, W.N., \& Schneider, D.P., 2003, AJ, 125, 433

\reference{} Weingartner, J., \& Draine, B.T., 2001, ApJ, 548, 296 [WD01]

\reference{} Whitney, B.A., Wood, K., et al., 2003, ApJ, 598, 1079

\reference{} Whitney, B.A., \& Wolff, M.J. 2002, ApJ, 574, 205:231

\reference{} Wood, K., \& Reynolds, R.J., 1999, ApJ, 525, 799

\reference{} Yan, L., Chary, R., Armus, L, et al. 2005, ApJ, 628, 604

\reference{} Yusef-Zadeh, F., Morris, M., \& White, R.L, 1984, ApJ, 278, 186


\end{references}
\end{document}